%% file: main.tex
\documentclass[letterpaper,twocolumn,10pt]{article}
\usepackage{usenix}

\usepackage[T1]{fontenc}
\usepackage{lmodern}
\usepackage{graphicx,color}
\usepackage{algorithm}
\usepackage{algpseudocode}
\usepackage{subcaption}
\usepackage{xcolor}
\usepackage{glossaries}
\usepackage{listings}

\usepackage{todonotes}
\usepackage{pgfplots}
\usepackage{tikz}
\usepackage{pgfplotstable}
\usepackage{amsmath,amsfonts}

\usepackage{booktabs}
\usepackage{multirow}
\usepackage{multicol}
\usetikzlibrary{shapes.geometric, arrows.meta, shapes.multipart, positioning, fit, backgrounds, shapes.misc, calc, patterns, colorbrewer}

\usepgfplotslibrary{statistics, groupplots}
\pgfplotsset{compat=1.8}
\usepackage{comment}

\definecolor{color0}{RGB}{27,158,119} 
\definecolor{color1}{RGB}{217,95,2}   
\definecolor{color2}{RGB}{117,112,179} 
\definecolor{color3}{RGB}{231,41,138} 
\definecolor{color4}{RGB}{102,166,30} 

\usetikzlibrary{patterns}
\usetikzlibrary{colorbrewer}

\pgfdeclarelayer{background}
\pgfdeclarelayer{foreground}
\pgfsetlayers{background,main,foreground}

\definecolor{cornflowerblue121150222}{HTML}{0000FF} 
\definecolor{darkseagreen12518595}{HTML}{F8AB5D}
\definecolor{darkslategray38}{RGB}{38,38,38}
\definecolor{darkslategray66}{RGB}{66,66,66}
\definecolor{lightgray204}{RGB}{204,204,204}
\definecolor{palevioletred227108143}{RGB}{227,108,143}
\definecolor{peru21513172}{RGB}{215,131,72}
\definecolor{sandybrown22319372}{RGB}{223,193,72}

\definecolor{cblue1}{HTML}{377eb8}
\definecolor{cred1}{HTML}{e41a1c}
\definecolor{cgreen1}{HTML}{4daf4a}

\definecolor{color0}{HTML}{f77189}
\definecolor{color1}{HTML}{ae9d31}
\definecolor{color2}{HTML}{33b07a}
\definecolor{color3}{HTML}{38a9c5}
\definecolor{color4}{HTML}{cc7af4}

\definecolor{static-rho-color}{HTML}{0000FF} 
\definecolor{adaptive-rho-color}{HTML}{F8AB5D}

\input{nsections/acronyms}

\begin{document}

\date{}


\title{\Large \bf Synapse: Virtualizing Match Tables in Programmable Hardware} 

\author{
{\rm Seyyidahmed Lahmer}\\
University of Padova\\
seyyidahmed.lahmer@unipd.it
\and
{\rm Angelo Tulumello}\\
University of Rome Tor Vergata\\
tulumello@uniroma2.it
\and
{\rm Alessandro Rivitti}\\
University of Rome Tor Vergata\\
alessandro.rivitti@uniroma2.it
\and
{\rm Giuseppe Bianchi}\\
University of Rome Tor Vergata\\
giuseppe.bianchi@uniroma2.it
\and
{\rm Andrea Zanella}\\
University of Padova\\
andrea.zanella@unipd.it
}

\maketitle

\begin{abstract}

Efficient network packet processing increasingly demands dynamic, adaptive, and run-time resizable match table allocation to handle the diverse and heterogeneous nature of traffic patterns and rule sets. Achieving this flexibility at high performance in hardware is challenging, as fixed resource constraints and architectural limitations have traditionally restricted such adaptability.

In this paper, we introduce Synapse, an extension to programmable data plane architectures that incorporates the \gls{vmt} framework, drawing inspiration from virtual memory systems in \glspl{os}, but specifically tailored to network processing. This abstraction layer allows logical tables to be elastic, enabling dynamic and efficient match table allocation at runtime. Our design features a hybrid memory system, leveraging on-chip associative memories for fast matching of the most popular rules and off-chip addressable memory for scalable and cost-effective storage. Furthermore, by employing a sharding mechanism across physical match tables, Synapse ensures that the power required per key match remains bounded and proportional to the key distribution and the size of the involved shard. To address the challenge of dynamic allocation, we formulate and solve an optimization problem that dynamically allocates physical match tables to logical tables based on pipeline usage and traffic characteristics at the millisecond scale. We prototype our design on FPGA and develop a simulator to evaluate the performance, demonstrating its effectiveness and scalability.
\end{abstract}

\input{nsections/introduction}

\vspace{-0.1cm}
\input{nsections/motivation}
\vspace{-0.1cm}
\input{nsections/overview}
\vspace{-0.1cm}
\input{nsections/design}
\vspace{-0.1cm}
\input{nsections/synapse-opt}

\vspace{-0.1cm}
\input{nsections/evaluation}
\vspace{-0.2cm}
\input{nsections/rwork}

\vspace{-0.2cm}
\input{nsections/limitations}
\vspace{-0.2cm}
\input{nsections/conclusion}

\clearpage
\bibliographystyle{plain}
\bibliography{references}

\clearpage
\appendix
\input{nsections/appendix}


\end{document}

%% file: nsections/acronyms.tex
\newacronym{cam}{CAM}{Content Addressable Memory}
\newacronym{tcam}{TCAM}{Ternary CAM}
\newacronym{mu}{MU}{Matching Unit}
\newacronym{mus}{MUs}{Matching Units}
\newacronym{mkr}{MKR}{Missed Key Resolver}
\newacronym{rp}{RP}{Replacement Policy}
\newacronym{rv}{RV}{Random Variable}
\newacronym{rmt}{RMT}{Reconfigurable Match-action Table}
\newacronym{drmt}{dRMT}{Disaggregated Reconfigurable Match-action Table}
\newacronym{mt}{MT}{Match Table}
\newacronym{vmt}{VMT}{Virtual Matching Table}
\newacronym{vmu}{VMU}{Virtual Matching Unit}
\newacronym{pmt}{PMT}{Physical Matching Table}
\newacronym{pmu}{PMU}{Physical Match Unit}
\newacronym{emu}{EMU}{External matching unit}
\newacronym{elu}{ELU}{External Lookup Unit}
\newacronym{dma}{DMA}{Direct Memory Access}
\newacronym{ilp}{ILP}{Integer Linear Program}
\newacronym{orb}{ORB}{Outstanding Request Buffer}
\newacronym{sram}{SRAM}{Static Random Access Memory}
\newacronym{dram}{DRAM}{Dynamic Random Access Memory}
\newacronym{hbm}{HBM}{High Bandwidth Memory}
\newacronym{lpm}{LPM}{Longest Prefix Match}
\newacronym{sdn}{SDN}{Software Defined Networking}
\newacronym{fpga}{FPGA}{Field Programmable Gate Arrays }
\newacronym{bst}{BST}{Binary Search Tree}
\newacronym{cfg}{CFG}{Control Flow Graph}
\newacronym{tdg}{TDG}{Table Dependency Graph}
\newacronym{dag}{DAG}{Directed Acyclic Graph}
\newacronym{phv}{PHV}{Packet Header Vector}
\newacronym{upf}{UPF}{User Plane Function}
\newacronym{ras}{RAS}{Runtime Auto-Scaler}
\newacronym{ps}{PS}{Processing System}
\newacronym{pc}{PC}{Pseudo Channel}
\newacronym{pl}{PL}{Processing Logic}
\newacronym{pim}{PIM}{Processing In Memory}
\newacronym{dpu}{DPU}{Data Processing Unit}
\newacronym{cpu}{CPU}{Central Processing Unit}
\newacronym{nic}{NIC}{Network Interface Card}
\newacronym{os}{OS}{Operating System}
\newacronym{iaas}{IaaS}{Infrastructure as a Service}
\newacronym{paas}{PaaS}{Platform as a Service}
\newacronym{usl}{USL}{Universal Scalability Law}
\newacronym{tss}{TSS}{Tuple Space Search}
\newacronym{ovs}{OVS}{Open vSwitch}

%% file: nsections/introduction.tex
\section{Introduction}
\gls{sdn} has revolutionized network control and management by decoupling the control plane from the data plane, enabling centralized management of network traffic using high-level policies and rules. Protocols like OpenFlow~\cite{Nick.Openflow} facilitate this dynamic and flexible management by allowing the control plane to program the data plane. Additionally, advancements in data plane programmability have enabled custom packet processing logic through languages such as P4\cite{Bosshart.P4} and architectures like \gls{rmt} and \gls{drmt}\cite{Bosshart.RMT,Chole.dRMT}. 

In \gls{sdn} architectures, maintaining high performance relies on the efficient division of tasks between the fast path and the slow path. The fast path, implemented in hardware, handles high-speed packet processing, while the slow path, managed by \glspl{cpu}, oversees complex decision-making and protocol management. Along this line, the limited size of hardware match tables has been so-far mainly managed centrally by presenting a larger logical table to the control plane and dynamically placing essential matching rules~\cite{Zulfiqar.Slowpath}. However, modern networks require finer granularity in resource management to handle varying workloads and traffic patterns. The static nature of current data plane resource allocation struggles to adapt to dynamic traffic conditions, resulting in performance bottlenecks, particularly with large rulesets. This necessitates larger \glspl{pmt} to maintain high-speed packet processing, increasing hardware resources and power consumption. Therefore, there is a need for dynamic allocation of \glspl{pmt} to logical tables, allowing for the shrinking and extending of logical tables based on current needs and enabling the sharing of \glspl{pmt} among logical tables over time without requiring hardware re-synthesis.

To address these limitations, we introduce Synapse, a \gls{vmt} framework designed to enhance the flexibility and efficiency of programmable data planes. Synapse virtualizes match tables, thus permitting to handle large, abstracted, logical tables directly in the hardware-based fast path. Specifically, by employing \gls{dma} for external lookups and leveraging \gls{hbm} for scalable storage, Synapse enables dynamic resizing of logical tables at runtime, allowing for adaptive resource allocation without the need for hardware re-synthesis. Key features of Synapse include:

\begin{itemize} 
\item \textbf{Elasticity}: The \gls{vmt} abstraction allows the runtime association of \glspl{pmt} to logical match tables, dynamically adapting to changing network demands, such as varying numbers of active rules or entries, without requiring re-synthesis.
\item \textbf{Scalability}: By offloading the logical table abstraction to hardware and utilizing an \gls{elu} with \gls{hbm} for slow path lookups, Synapse can accommodate significantly larger rulesets, ensuring efficient hardware management without relying on \gls{cpu} intervention. 
\item \textbf{Dynamic Allocation Abstraction}: A key strength of the system is its ability to abstract dynamic allocation, allowing the \gls{vmt} to associate with \glspl{pmt} at runtime. This eliminates the need for the programmer to manually adjust resources, while ensuring efficient resource usage.
\item \textbf{Power and Energy Efficiency}: Synapse can also improve power efficiency through a sharding mechanism and consistent hashing for effective key distribution. By replacing tranitional multicast lookups---particularily power-hungry on \gls{tcam} blocks---with unicast lookups and minimizing external memory accesses, Synapse ensures bounded power usage per key match and increases hit rates. \end{itemize}

In summary, our contributions include:
\begin{itemize}
    \item the design of the Synapse architecture and its hardware components;
    \item an FPGA prototype of Synapse and the implementation of a cycle-accurate simulator carefully representing the hardware design;
    \item an extensive evaluation of the Synapse components with real traffic traces.
\end{itemize}

%% file: nsections/motivation.tex
\section{Background and Motivation}


In modern programmable network infrastructures, achieving elasticity is a crucial objective, particularly when it comes to the dynamic adaptability of packet matching capabilities. While the control plane offers runtime flexibility by adjusting traffic policies and configurations, the dataplane limited physical resources pose significant challenges. Matching tables, essential for packet classification and flow management, are typically allocated at compile time, making it difficult to adjust to changing network conditions. This inflexibility can result in inefficient resource utilization when traffic patterns shift or when rule sets need rapid updates.

What we aim to achieve is a system where the matching capabilities can dynamically scale and adapt in real-time, without requiring the programmer to manually reconfigure or reallocate resources. This would allow the dataplane to handle fluctuating traffic loads efficiently, ensuring that match tables are sized according to current demands while maintaining high lookup performance.

In this section, we first overview the traditional mechanisms for matching and packet classification, the desired characteristics for achieving dynamic adaptability in match table allocations, and how the Synapse framework seeks to bring this elasticity to programmable dataplanes through abstraction interfaces.

\begin{figure*}[htbp]
    \centering
    \includegraphics[width=0.9\linewidth]{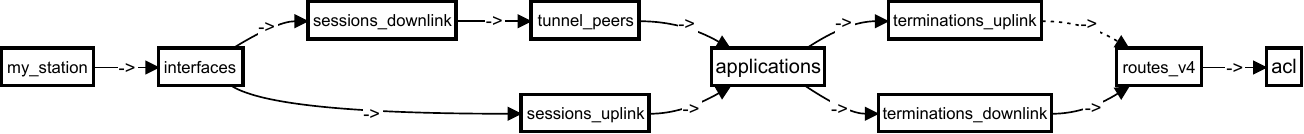}
    \caption{Figure illustrating the UPF.p4 \gls{cfg}, where a \gls{phv} follows one of two paths. For downlink traffic (from the internet to the user), the packet is processed through the tables on the upper path, while uplink traffic (from the user to the internet) is handled by the tables on the bottom path.} 
    \label{fig:upf-cfg}
    \vspace{-0.3cm}
\end{figure*}

\subsection{Packet Classification Strategies}

In high-performance networks, packet classification must match incoming packets to predefined rules at wire speed with a minimum-sized packet. Achieving this requires a focus on the \textit{lookup speed} as the dominant metric, while update speed--the time it takes to insert, delete, or modify a rule—is important in certain contexts, it is often secondary to lookup efficiency. However, in systems utilizing caching, rapid updates become critical when frequent rule replacements are required.

\noindent\textbf{Simple Approaches.}
A foundational approach to packet classification is linear search, which checks each packet sequentially against every rule. This method is straightforward but quickly becomes impractical due to its O(N) complexity, especially as rule sets grow.

A more structured and efficient alternative is \gls{tss}~\cite{Srinivasan.TupleSpace}. \gls{tss} groups rules into tuple spaces based on prefix size, allowing the classifier to search within relevant groups rather than the entire rule set. This reduces the complexity of the search. \gls{tss} is employed in \gls{ovs} for macroflows, where rules matching multiple packet fields are organized into tuple spaces. In contrast, microflows—which handle individual packet streams—use a hash table for faster lookups~\cite{Pfaff.OVS}. \gls{tss} is effective for complex rule sets but can suffer from overhead as the number of tuple spaces increases.

\noindent\textbf{Caching Mechanisms.}
Caching plays a pivotal role in accelerating packet classification, particularly by storing frequently matched rules to avoid repeated lookups in the main rule set. In the Linux routing subsystem, a hash table was historically used as a first-level cache to expedite routing lookups. This cache was designed for quick retrievals, but after kernel 3.6, it was replaced with a Trie-based structure to address security concerns~\cite{LinuxKernelNetworking,UnderstandingLinuxNetworkInternals}.

Caching techniques, such as simple hashing, map packet fields to cache entries, offering fast lookups. However, traditional hash tables often face collision issues, which degrade performance. Cuckoo hashing~\cite{Pagh2001CuckooHashing} mitigates these collisions by using multiple hash functions and allowing entries to displace one another across multiple locations. This ensures constant-time lookups, though cache updates become more complex under heavy load.

\noindent\textbf{Trie Structures.}
Especially for IP lookups, Tries~\cite{Fredkin1960Trie} offer a powerful method for handling hierarchical data. A Trie organizes IP prefixes in a tree, with each node representing a bit in the prefix. Multi-bit Tries~\cite{Gupta1999MultibitTrie} optimize this by reducing the number of tree levels, improving lookup speeds, while Set-pruning Tries~\cite{Eatherton2004SetPruningTries} conserve memory by sharing nodes between rules. Grid of Tries~\cite{Waldvogel2000GridOfTries} further enhances performance by handling different fields in parallel.

Hierarchical Tries~\cite{Srinivasan1999HierarchicalTrie} build on this by organizing multiple levels to handle multi-field classification. This approach is well-suited to environments with complex rules but can become memory-intensive as the number of rules increases.

\noindent\textbf{Hardware Solutions.}
Hardware-based solutions such as \gls{tcam} provide extremely fast lookups. \gls{tcam} allows for wildcard matches and range lookups, which are critical for multi-field classifications, such as those involving IP prefixes and port ranges. Despite their speed, \gls{tcam}'s high power consumption and cost limit their scalability in large deployments~\cite{Panigrahy.Power,Newman.Gigabit,Lu.LargeCAM,Chang.PowerPerf}.

 \noindent For a more thorough discussion on state-of-the-art packet classification approaches, the reader is referred to \cite{Varghese.NA}.

\subsection{The Need for Flexible Memory in Network Dataplanes}

Current programmable dataplane architectures leverage hardware-based matching mechanisms by partitioning tables according to the programmer's needs. However, this partitioning is fixed at compile time, limiting the ability to change them during execution. Meanwhile, modern networking demands have outpaced the capabilities of traditional data plane architectures, driving the need for dynamic adaptation to the variability of traffic. Consequently, these architectures struggle to leverage runtime characteristics, such as packet dynamics, and dynamically adapt physical resource allocation in response to these characteristics. 


Our key objective is to enhance support for Rule-Set Activity and \gls{cfg} Dynamicity. This involves optimizing match table utilization through the efficient use of both on-chip and off-chip memory, particularly in systems utilizing FPGAs and SmartNICs. Moreover, dynamic adaptability in packet traversal paths (\gls{cfg} Dynamicity) provides opportunities for real-time resource optimization, improving efficiency and reducing latency in data plane processing.

\textbf{Rule-Set Activity.}
With the huge growth in the number of devices connected to the internet, it becomes imperative to support expansive network policies encompassing a huge set of rules. Despite the necessity to accommodate such extensive rule sets, not all rules are actively used at any given moment \cite{MacDavid.P4UPF}. This spotlights a significant inefficiency in how match tables are currently utilized, suggesting a pivotal opportunity for optimization.

Memory\footnote{Throughout this document, the term "memory" refers to "on-chip memory" components such as LUTs , BRAM , and URAM in the context of FPGA architectures. "Off-chip memory" is used interchangeably to describe external memory components such as DDR-like or HBM, unless explicitly stated otherwise.} utilization becomes increasingly important and relevant with the advent of \gls{fpga} and SmartNICs and their integration into the networking infrastructure. These technologies, which often house a variety of applications even beyond networking task implying a further constrains on the on-chip resources particularly memory. While on-chip memory is known for its speed, its high cost and scarcity necessitate a more strategic approach to memory usage, namely, by leveraging both on-chip and off-chip memory solutions.


\textbf{\gls{cfg} Dynamicity.}
The traversal path of packets through the pipeline is predominately defined by a \gls{cfg}, typically represented as a \gls{dag}, which is determined by the data plane program, with some exceptions, such as cycles that can be introduced due to packet re-circulation. State of the art compilers leverage the \gls{tdg} to optimize stage usage within the data plane pipeline. By identifying and co-locating independent tables within the same processing stage, these methodologies achieve two main objectives: (1) reducing the latency, (2) enhancing the efficiency of resource utilization per stage (i.e., memory utilization, turn off unused stages). However, these strategies overlook one crucial dimension -- runtime variability.

\noindent\textbf{The UPF example.} The 5G  \gls{upf} provides an illustrative example of where these features are essential. We observe that the path a packet traverses within a pipeline has semantics. As shown in Fig.~\ref{fig:upf-cfg}, a packet traversing the UPF pipeline has two primary paths: uplink and downlink. These paths experience varying levels of utilization based on real-time network demands—such as increased downlink activity during live streaming or heightened uplink activity during cloud backups. The utilization of these paths, and thus the resource demands, are not static but exhibit temporal fluctuations influenced by various network activities. This suggests a significant opportunity: by leveraging runtime information, the UPF can efficiently manage its resources without manual intervention, thus optimizing efficiency across resources of the pipeline. 


\subsection{Elastic Match Tables for Next-generation networking}

Our research addresses the above mentioned issues by introducing elasticity to the hardware match tables, enabling them to respond to the changing behavior of traffic in real time, thus enhancing the adaptability and efficiency of programmable dataplanes.
To this end, enabling virtualization and the efficient integration of SmartNICs and programmable switches within cloud infrastructures requires robust abstraction layers for the physical resources embedded within these devices. These abstraction layers facilitate operations such as Create, Read, Update, and Delete (CRUD) for managing resources dynamically, akin to the capabilities seen in orchestration systems like OpenStack for \gls{iaas} and OpenShift for \gls{paas} -- natively provided by the \gls{os}. The transformative trends in network function integration — from fixed-function devices to general-purpose \glspl{cpu}, and subsequently to programmable switches and Smart\glspl{nic} - highlight a significant evolution. This evolution, facilitated by software-defined data planes using languages like P4, as well as software frameworks such as eBPF XDP~\cite{Brunella.hXDP,Rivitti.eHDL} and DPDK, has empowered cloud infrastructures to offload all or part of their network functions (NFs) to specialized executors, significantly enhancing both efficiency and performance. While network programmers are increasingly accustomed to utilizing software interfaces that offer comprehensive abstraction features, this level of abstraction has not yet been fully realized in hardware dataplanes.

However, restructuring resources at runtime is challenging because it often involves reconfiguring the program or firmware in the programmable hardware. The process of generating bitstreams for reconfiguration can be time-consuming, complicating the dynamic management of resources.
Thus, we argue that abstraction interfaces are crucial for next-generation networking as they enable efficient resource sharing, fast instantiation, and robust isolation of network functions. By providing these interfaces, network operators can dynamically manage and allocate resources, optimizing performance and resource utilization without frequent hardware re-configuration.

%% file: nsections/overview.tex
\section{Synapse Overview}
\begin{figure}[t]
    \centering
    \includegraphics[width=\columnwidth]{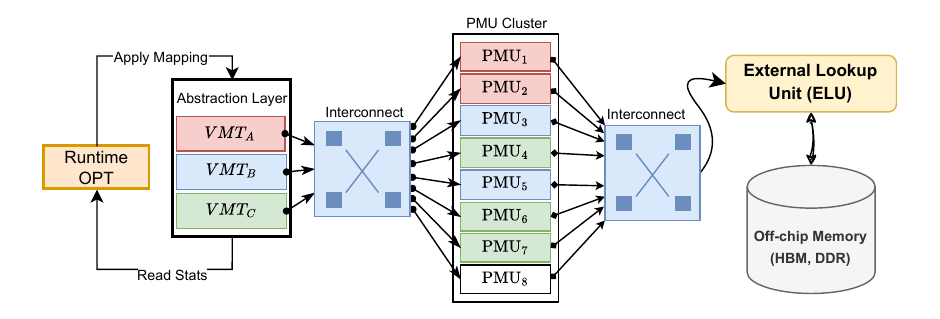}
    \caption{Synapse -- Closed loop control system}
    \label{fig:synapse-overview}
    \vspace{-0.5cm}
\end{figure}

Synapse introduces an architecture extension that facilitates \gls{pmt} abstraction through virtualization. This abstraction layer shares physical match tables among multiple concurrent logical tables on a programmable data plane dynamically at runtime. Figure~\ref{fig:synapse-overview} shows a general schematic of the proposed architecture, which includes three main components: (1)~a runtime optimizer (OPT), (2)~a \gls{vmt} serving as an abstraction layer and (3)~a~set~of \glspl{pmt} along with the external memory.

\noindent\textbf{Virtual Matching Unit.} Synapse implements the \gls{vmt} as an abstraction layer that simplifies the dynamic allocation of match tables to the different logical tables defined by the programmer. This concept can be integrated with both pipeline- or processor-based architectures defined by \gls{rmt} and \gls{drmt}, respectively, although with some considerations, which will be detailed Sec~\ref{sec:design}. This abstraction layer acts as a northbound interface for the \gls{ps}, allowing it to make updates at runtime. The \gls{vmt} abstraction provides a flexible mechanism for managing match tables, facilitating the allocation of resources based on application requirements. By decoupling the logical representation of match tables from their physical implementation, Synapse enhances adaptability and resource utilization, enabling efficient utilization of hardware resources in diverse network scenarios. This abstraction layer is crucial for accommodating dynamic changes in network traffic patterns (i.e., \gls{cfg} paths activity ), ensuring optimal performance and scalability in evolving network environments.

\noindent\textbf{Physical Matching Unit.} It serves as a fundamental component of the architecture. It implements a \textit{non-blocking}, \textit{asynchronous} match table that can be associated with any \gls{vmt} (i.e., logical table) at runtime. It acts as an independent cache, by storing popular rules and their corresponding actions; available \glspl{pmu} are interconnected with a simple bus with a simple external lookup module described in the following section. The \glspl{pmu} play an important role in accelerating packet processing by leveraging cached rules for rapid and efficient matching. Their non-blocking and asynchronous nature ensures that the long latency of missed keys is hidden from the main pipeline, although bounded by the \gls{elu} throughput. When a key is not found, the \gls{pmu} sends an early miss notification, allowing the pipeline to store and proceed with subsequent keys without waiting for the long-latency lookup to complete. This mechanism ensures that the pipeline remains active and efficient despite the latency of missed keys. Moreover, the \glspl{pmu} can scale down their frequencies independently of the main pipeline, leading to more power-efficient operation. This decoupling of frequency scaling enables finer-grained power management, allowing the system to dynamically adjust performance based on workload demands while optimizing energy consumption.

\noindent\textbf{Synapse OPT.} Synapse supports dynamic \gls{pmu} allocation through the runtime OPT. This enables closed-loop control for the resources within our architecture, as the OPT: monitors the different logical tables through simple counters; estimates \glspl{phv} routes through the pipeline, and applies a mapping of each \gls{vmt} to a set of \glspl{pmt}. Runtime-OPT addresses two major challenges: firstly, it estimates how each edge in the \gls{cfg} is being used, providing insights into resource utilization and potential bottleneck tables. Secondly, it dynamically evaluates a per-table utility function at runtime, determining how adding or removing \glspl{pmu} from a given \gls{vmu} would affect the main objective. By continuously optimizing \gls{pmu} allocation based on real-time traffic patterns and application demands, Synapse ensures efficient resource utilization and maximizes pipeline throughput, ultimately enhancing network performance and scalability in dynamic environments.


%% file: nsections/design.tex
\section{Design}
\label{sec:design}
In this section, we focus on the details of the fundamental hardware components of the Synapse architecture: the \gls{vmu}, the  \gls{pmu} and the \gls{elu}, outlining their hardware design, operations and challenges. The runtime optimizer component will be discussed in Section \ref{sec:opt}.

\subsection{Virtualized Matching-Unit}
\begin{figure}[t!]
    \centering
    \includegraphics[width=0.8\linewidth]{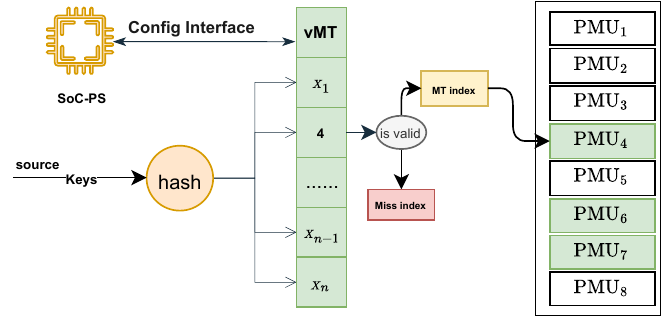}
    \caption{\gls{vmt} Lookup table}
    \label{fig:synapse-vmt}
    \vspace{-0.5cm}
\end{figure}

In Synapse, each logical table corresponds to a \gls{vmt} that serves to abstract the base match tables. Associated within each \gls{vmt} is a lookup table designed to direct each key to a specific \gls{pmu}. We utilize a hashing mechanism to enable a stateless\footnote{Stateless here refers  to the fact that we do not store pairs (key, pmu\_id)} mapping of keys to \glspl{pmu}, ensuring that keys requests associated with the same flow \textit{consistently} reach the same \gls{pmu} for processing. As depicted in Fig~\ref{fig:synapse-vmt}, the key undergoes a hash function that assigns it to an interval $[0..v-1]$, where $v$ represents the size of the lookup table. If this process results in an invalid entry, the system’s architecture determines the subsequent action, typically a default action. Conversely, if the entry is valid, it specifies the \gls{pmu} ID responsible for executing the lookup request. The configuration of the lookup table is managed by the CPU through a configuration interface. The matching process is non-deterministic and governed by the dynamics of caching, this introduces four core challenges that the \gls{vmu} must overcome: (1) hiding the long latency of missed keys, (2) ensuring accurate policy execution, (3) maintaining deadlock-free operations , and (4) minimal re-ordering.

\noindent\textbf{Lookup Table.}
Employing a hash function to map keys to \glspl{pmu} can lead to performance degradation during \gls{pmu} reallocation, which occurs when the CPU adds or removes \glspl{pmu} to/from a \gls{vmu}. Such  modifications often increase cache misses as they alter the key-to-PMU mappings. To mitigate this, we implement consistent~Hashing~\cite{Karger.ConsistentHash} across each \gls{vmt} to minimize the likelihood of keys migrating between different \glspl{pmu} during reconfigurations. Standard consistent hashing utilizes a \gls{bst} to assign a key to its nearest node—in our case, the nearest \gls{pmu}. We take advantage of our relatively small lookup table domain, allowing the CPU to pre-compute the lookup table. It performs a \gls{bst} search across all possible entries, populates them accordingly, and then updates only those \gls{vmt} lookup table entries that are affected by the change.

\noindent\textbf{Key Matching.}
Synapse implements the virtual match unit in an asynchronous, non-blocking way; the process by which \gls{vmt} is implemented is divided into main routines (refer to Appendix \ref{appendix:implementation}). A~key request is first received along with the corresponding \gls{pmu} id by the Request Producer, where it generates a unique identifier for the key, and then sends a key lookup request to the corresponding \gls{pmu} -- via the request interconnect; following, the \gls{phv} reference to the corresponding key is then passed to the Response Consumer. It is interconnected with a response network that delivers responses from the \glspl{pmu} to the \gls{vmt}, in case of invalid match result, this signifies that the current key was not found in the cache -- explicit early cache miss notification --, therefore, the Response Consumer will buffer the \gls{phv} in a local FIFO buffer. In the case of valid response, the response might correspond to the current \gls{phv} at hand\footnote{We note that the \gls{phv} buffer is shared between Request Producer and Response Consumer; This can be seen as a message queue of limited size, the writer blocks until there is at least one empty space}, or to front element of the buffer -- ensured by the re-order buffer in the \gls{elu} + priority enforcing in the \gls{pmu}; in the first case, the \gls{phv} is forwarded along with the response to the action unit; in the latter case, the \gls{vmt} pulls the \gls{phv} reference from the local buffer and forward it to the action unit along with the received response.

\subsection{Physical Matching-Unit}

\begin{figure}[t]
    \centering
    \includegraphics[width=\columnwidth]{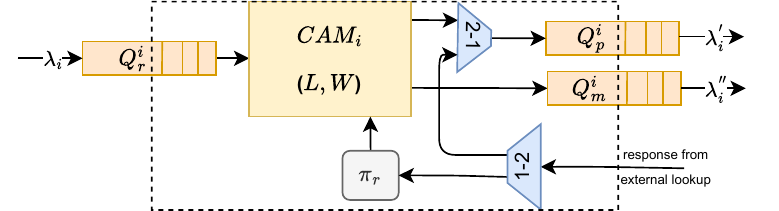}
    \caption{Schematic of the $i$th \gls{pmt}}
    \label{fig:pmu}
    \vspace{-0.4cm}
\end{figure}

The \gls{pmu}, depicted in Fig~\ref{fig:pmu}, serves as the main interface responsible for handling lookup requests coming from the main pipeline. It performs non-blocking, asynchronous lookup operations and interfaces with the \gls{elu}, which resolves lookups within an external data structure residing in the off-chip memory via \gls{dma}.

Therefore, \gls{pmu} architecture has the following key characteristics:

\begin{itemize}
    \item \textbf{Asynchronous Operation:} Any \gls{pmu} $i$ operates independently of the main pipeline's clock, processing incoming keys as long as there are requests in the \(Q_{r}^{i}\) queue. Additionally, with the \(Q_{p}^{i}\) queue for domain-crossing frequency, the \gls{pmu} can dynamically adjust its operating frequency relative to the main pipeline. \(Q_{r}^{i}\) and \(Q_{p}^{i}\) deliver requests to the \gls{pmu} and responses to the main pipeline, respectively; \(Q_{m}^{i}\) forward missed keys to the \gls{elu} for an external match request.
    
    \item \textbf{Reconfigurability:} The \gls{pmu} functions as an independent shard\footnote{The term shard is used interchangeably with \gls{pmu} throughout the document} that caches rules. Each delivered key is accompanied by a byte-level mask indicating the valid bytes of the key. This enables the \gls{pmu} to serve different logical tables at different configuration times.

\end{itemize}

Synapse employs a fully associative data structure to implement the cache within each shard. The \gls{cam} performs a parallel search across its entries to determine if the incoming key matches any existing cached rule entry. The request mask bytes are used to disable the masked bytes within all entries in the \gls{cam}. The matching process occurs within a deterministic number of clock cycles, denoted as  \(\tau_{c}\), and $\text{II}_c$ (initiation interval). A hit occurs when the action corresponding to the key is found, and the result, along with the key, is pushed into  \( Q_{\text{m}}^{i}\) to be delivered back to the action unit. If the \gls{cam} search yields a miss, indicating that the requested key does not match any existing rule in the local cache, the key is pushed into the \( Q_{\text{m}}^{i}\) queue. Simultaneously, the same key is duplicated and pushed into \( Q_{\text{p}^{i}}\) along with a special action \texttt{not\_found} (i.e., a pointer to \texttt{nops} code segment) to the main action unit. This allows the action unit to store the corresponding \gls{phv} in a special awaiting buffer with a FIFO discipline, preventing the corresponding processor/stage from blocking on missed keys.

The \gls{pmu} serves as a runtime-allocatable shard, acting as the primary bridge between the \gls{vmt} and the \gls{elu}. It is essential for the \gls{pmu} to ensure that no outstanding requests are pending during migration periods, specifically when changing \gls{pmu} associations from \gls{vmt} \(i\) to \gls{vmt} \(j\). To manage this, the operation of the \gls{pmu} is governed by a three-state finite state machine consisting of \textit{free}, \textit{transient}, and \textit{associated} states. In the free state, the \gls{pmu} is available for allocation to any \gls{vmt}. Upon allocation, it transitions to the associated state. If a rescheduling is required, moving from \gls{vmt} \(i\) to \gls{vmt} \(j\), the \gls{pmu} first enters the transient state. At this point, it completes servicing all outstanding requests for the currently associated \gls{vmt} before reverting to the free state, ready to associate with a new~\gls{vmt}.

\subsection{External Lookup Unit}

\begin{figure}[t]
    \centering
    \includegraphics[width=\linewidth]{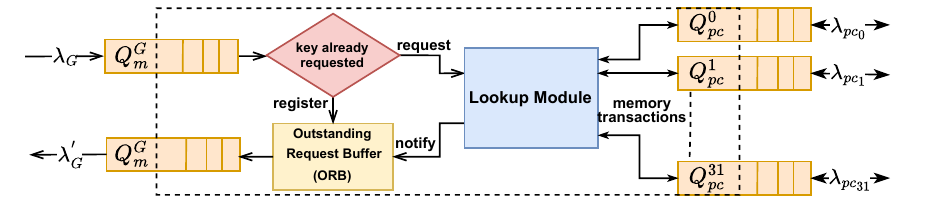}
    \caption{A high level schematic of the \gls{elu}}
    \label{fig:elu}
\end{figure}

The \gls{elu} serves as the interface with the off-chip memory. As illustrated in Figure~\ref{fig:elu}, the \gls{elu} manages communication between the \gls{pmu} and the \gls{hbm} used to store large rulesets and handle lookup operations that cannot be accommodated within the on-chip memory. The memory is accessed via \gls{dma}.

The \gls{elu} architecture includes two main queues \(Q_{m}^{G}\) responsible for receiving missed requests from all available \glspl{pmu}, and \(Q_{l}^{G}\), for communicating responses back, each to its corresponding \gls{pmu}. On Xilinx UltraScale+ devices such as the Alveo U50 board\cite{XilinxU50}, the \gls{fpga} is coupled with two \gls{hbm} stacks.

\noindent The lookup module within the \gls{elu} implements a multi-bit hierarchical trie data structure for each policy, with the policy being populated via the control plane. The lookup module delegates the following pre-processing tasks to the control plane:

\begin{itemize}
    \item \textbf{Rule Expansion.} The control plane converts any non-compatible trie rule (e.g., range match) into a set of equivalent trie-compatible rules, e.g. LPM.

    \item \textbf{Spine-pruning.} The control plane eliminates the need for backtracking during lookups by duplicating backtracked nodes for all possible paths.
\end{itemize}

We employ a static scheduling strategy which simplifies the design and allows the HLS compiler to efficiently reuse hardware resources. During a trie lookup, a request traverses from the root node through edges based on the key value until it finds the corresponding rule. This process cannot be pipelined due to dependency between iterations. However, by carefully designing the system to exploit different banks of external memory, we can pipeline different lookup requests. This requires pre-locating each node bank at compile time.

Using horizontal trie partitioning, we place each level in separate memory banks. A high N-bit trie reduces external memory lookups and benefits from low latency and high bandwidth due to sequential access, though it increases memory usage due to rule expansion. While other trie methods like Grid of Trie reduce memory footprint and eliminate backtracking, they complicate static scheduling by using pointers instead of duplication.

To enhance lookup efficiency, methods like trie duplication across ports can reduce memory contention, and hybrid approaches can be used. A dynamic scheduler could also be considered, but designing a high-performance \gls{elu} is beyond the scope of our current research and is proposed for future work.

\noindent The \gls{orb} is implemented as a reorder buffer similar to Tomasulo Algorithm~\cite{Tomasulo}, where each entry contains an identifier for an outstanding request generated by the matching unit, along with an initially unset action pointer and a validity flag. Incoming requests are first registered in the \gls{orb} and marked as invalid, with responses pushed into a local FIFO queue with unset action fields. The \gls{orb} uses two pointers, \texttt{commit} and \texttt{issue}, to track the newest and oldest pending requests, respectively.

Requests are forwarded to the Lookup module for external data structure lookups without guaranteed order. When a new lookup response is received, the \gls{orb} marks the corresponding entries as valid. The \gls{orb} then pulls responses from the FIFO queue as long as the \texttt{commit} points to a valid entry, sending responses back via \(Q_m^G\). This setup enables efficient batched external lookups and ensures that responses are processed out of order but replied in-order.

\subsection{Interconnect Design Choices}
The interconnection network is central to the efficiency of Synapse’s architecture. In this design, two key interconnects are responsible for communication: one between the \glspl{vmt} and the \glspl{pmt}, and the other between the \glspl{pmt} and the \gls{elu}.

\noindent\textbf{Interconnect Between \glspl{vmt} and \glspl{pmt}.}

One possible design for the interconnect is a fully connected \textit{many-to-many} architecture, where each \gls{vmt} is connected to every \gls{pmu} through dedicated buses. This would allow any \gls{vmt} to communicate with any \gls{pmu} simultaneously, offering maximum parallelism.

While this design maximizes parallel communication, the hardware complexity grows quadratically as \(O(V \times P)\), where \(V\) is the number of \gls{vmt}s and \(P\) is the number of \gls{pmu}s. Such complexity requires extensive \gls{fpga} resources and increases power consumption due to the large number of active connections, making it impractical for large-scale systems 

At the other extreme, the interconnect could be implemented using a \textit{single-shared-bus} design, where all \gls{vmt}s share a single communication bus to issue requests to the \gls{pmu}s. For each cycle, only one \gls{vmt} can issue a lookup request, limiting system performance but significantly reducing hardware complexity. Although this design minimizes the number of connections and hardware complexity, it severely limits performance. Only one \gls{vmt} can issue a request per clock cycle, resulting in low \gls{pmu} utilization. The maximum throughput is reduced to \(1/V\), where \(V\) is the number of \gls{vmt}s, drastically limiting system performance as the number of \gls{vmt}s increases. 

A more natural choice is the \textit{segmented-channel} design, where \gls{pmu}s are divided into channels. Each channel can process requests independently, allowing multiple \gls{vmt}s to issue lookup requests in parallel, provided they target different channels. This approach is similar to modern memory systems, where memory banks within the same channel cannot be accessed simultaneously, but banks in different channels can. 

The segmented channel design balances the performance and hardware complexity. By dividing the \gls{pmu}s into independent channels, multiple \gls{vmt}s can issue requests to different \gls{pmu}s without contention, maximizing throughput. If two \gls{vmt}s attempt to access \gls{pmu}s within the same channel, one request is buffered and processed in the next clock cycle, introducing a slight latency. However, this latency is offset by the overall increase in parallelism and throughput. The complexity of the interconnect grows as \(O(V  \frac{P}{C} )\), making it more scalable than the many-to-many design while avoiding the performance limitations of the shared bus approach. Synapse utilize this design.

\vspace{-0.3cm}
\subsection{Hardware Implementation}
Our hardware implementation is centered around the FPGA-based prototype, deployed and tested on the Alveo U50 FPGA-based SmartNIC. Various modules were designed using a combination of High-Level Synthesis (HLS) and Hardware Description Language (HDL) to achieve low latency and high throughput in packet processing.

We extended the open-source Xilinx project for CAM-based packet processing \cite{XilinxHLS} to incorporate an efficient LRU replacement policy. The CAM was modified to include a linked list structure that tracks the least recently used (LRU) entries, enabling O(1) updates. Each CAM entry stores a tuple \texttt{<action, LRU pointer>}, ensuring constant-time updates on access, while the linked list also maintains pointers to the corresponding CAM entries for fast LRU replacement.

The consistent\_lb module, responsible for managing load balancing, is implemented using a single-port BRAM for the lookup table, which is updated by the processor as needed.

For interconnect communication between the \glspl{vmt} and \glspl{pmt}), we utilized the Xilinx AXI Stream (AXIS) interconnect \cite{xilinxaxis}, enabling efficient packet handling with minimal latency.

The HLS implementation of both the consistent load balancing and action execution unit can be found in the corresponding listings in Appendix E. These listings provide the details of the implementation.

In addition, we developed a discrete-time simulator to model the FPGA prototype's behavior. This simulator accounts for key hardware parameters such as module latencies and initiation intervals, providing a cycle-accurate representation of the system's performance.

%% file: nsections/synapse-opt.tex
\section{Synapse Runtime OPT}
\label{sec:opt}

The \textbf{Synapse Runtime OPT} is critical for ensuring efficient use of resources in our architecture. By monitoring traffic patterns and system performance in real-time, the optimizer dynamically allocates \textit{\gls{pmt}} to \textit{\gls{vmt}}. The goal is to maximize throughput by adjusting how resources are distributed across the system as conditions change.

\noindent\textbf{System Model Overview.} 
We represent the problem of allocating \glspl{pmt} to \glspl{vmt} as a \textit{\gls{cfg}} with $G = (V, E)$, where each node $v_i \in V$ represents a \gls{vmt} handling packet classification, and each edge $e(i,j) \in E$ shows how packets flow between \glspl{vmt}. The performance of each \gls{vmt} depends on how many \glspl{pmt} it has available. Our task is to decide how to assign \glspl{pmt} to maximize throughput, while keeping the total number of \glspl{pmt} within available limits.


We chose to model the optimization problem using the \textit{Maximum Flow Problem} framework because it closely mirrors how packets flow through the network~\footnote{The term network refers to logical representation of the \gls{cfg} (i.e. P4 representation), abstracting away the actual physical implementation which could be a sequential representation (e.g.  in the \gls{rmt})}. \textit{\glspl{phv}} traverse the \glspl{vmt}, and each \gls{vmt}'s processing capacity depends on the number of \glspl{pmt} assigned to it. However, our scenario introduces several key challenges:

\begin{itemize}
    \item \textbf{Uncertain Packet Paths}: \glspl{phv} don’t always follow the same path. We capture this uncertainty using a \textit{stochastic matrix} $P$, where $P_{ij}$ represents the probability of a \gls{phv} moving from \gls{vmt} $v_i$ to \gls{vmt} $v_j$. Real-time traffic conditions are constantly changing, and our model must adapt to this variability.
    
    \item \textbf{Nonlinear Capacity}: The capacity of each edge is not fixed. Instead, it depends on how many \glspl{pmt} are allocated to each \gls{vmt}. We use the \textit{\gls{usl})} to model this nonlinear relationship between the number of \glspl{pmt} and the resulting throughput.
    
    \item \textbf{Resource Allocation Limits}: It is essential to ensure that the total number of \glspl{pmt} allocated to all \glspl{vmt} remains within the available resources and avoids overallocation.
\end{itemize}

This problem formulation is well-suited for capturing the key characteristics of \gls{phv} flows in data plane, particularly under dynamic and variable traffic conditions. Although a similar approach utilizing a stochastic matrix to model \gls{phv} traversal was introduced in a different context \cite{xing.UnleasingSmartnic}, our work focuses on leveraging the maximum flow model to efficiently allocate \glspl{pmt} and optimize system throughput in real-time environments.

\noindent\textbf{Problem Formulation}

Our optimization problem can be summarized as:

\begin{equation}
\max_{f, n} \sum_{(s, v_j) \in E} f(s, v_j)
\end{equation}

Subject to the following constraints:

1. \textbf{Flow conservation}:
\begin{equation}
f(v_j, v_k) \leq P_{jk} \cdot \sum_{v_i : (v_i, v_j) \in E} f(v_i, v_j), \forall v_k \neq \{s, t\}, (v_j, v_k) \in E
\end{equation}

2. \textbf{Capacity limits}:
\begin{equation}
f(v_j, v_k) \leq s(j) \cdot P_{jk}, \forall (v_j, v_k) \in E
\end{equation}

3. \textbf{Resource constraints}:
\begin{equation}
\sum_{v_i \in V} n_i \leq N
\end{equation}

4. \textbf{Nonlinear capacity function} modeled by the USL:
\begin{align}
s(j) = &\frac{\sum_{v_i : (v_i, v_j) \in E} f(v_i, v_j)}{1 + (n_j \alpha_{j0} + \alpha_{j1}) \cdot (\sum_{v_i : (v_i, v_j) \in E} f(v_i, v_j) - 1)} \nonumber \\
&+ (n_j \beta_{j0} + \beta_{j1}) \cdot \sum_{v_i : (v_i, v_j) \in E} f(v_i, v_j)
\end{align}
This formulation ensures efficient PMU allocation to maximize throughput while adapting to changes in network traffic in real-time. By solving this optimization problem continuously, we dynamically adjust resources to maintain optimal network performance.

%% file: nsections/evaluation.tex
\section{Evaluation}

\begin{figure*}[ht]
    \centering
    \input{figures/eval-results/shared-legend}\\
    \begin{subfigure}[t]{0.24\textwidth}
        \centering
        \input{figures/eval-results/latency/latency-trace-1} 
        \caption{Latency for Trace 1}
        \label{fig:lat1}
    \end{subfigure}
    \hfill
    \begin{subfigure}[t]{0.24\textwidth}
        \centering
        \input{figures/eval-results/latency/latency-trace-2}
        \caption{Latency for Trace 2}
        \label{fig:lat2}
    \end{subfigure}
    \hfill
    \begin{subfigure}[t]{0.24\textwidth}
        \centering
        \input{figures/eval-results/bandwidth/bandwidth-trace-1}
        \caption{Memory BW for Trace 1}
        \label{fig:bw1}
    \end{subfigure}
    \hfill
    \begin{subfigure}[t]{0.24\textwidth}
        \centering
        \input{figures/eval-results/bandwidth/bandwidth-trace-2}
        \caption{Memory BW for Trace 2}
        \label{fig:bw2}
    \end{subfigure}

    \begin{subfigure}[b]{\columnwidth}
        \centering
        \input{figures/eval-results/hit-rate/hit-rate-trace-1} 
        \vspace{-10pt}        
        \caption{Hit Rate for Trace 1}
        \label{fig:hit-rate-1}
    \end{subfigure}
    \begin{subfigure}[b]{\columnwidth}
        \centering
        \input{figures/eval-results/hit-rate/hit-rate-trace-2}
        \vspace{-10pt}
        \caption{Hit Rate for Trace 2}
        \label{fig:hit-rate-2}
    \end{subfigure}
    
    \caption{[Sim]~Latency, Memory Bandwidth and Hit Rate results varying the block size of the \glspl{pmu}}
    \label{fig:lat-bw-hitrate}
    \vspace{-0.3cm}
\end{figure*}
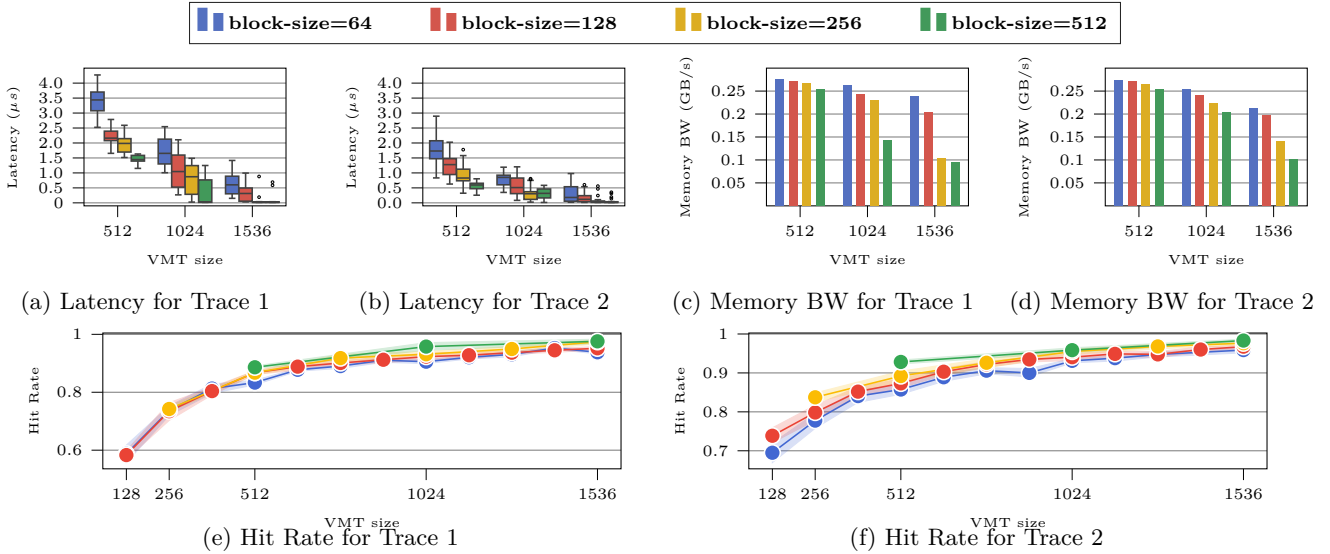

In this section, we describe the setup used for our evaluation, detailing the traffic generation, policy rule set generation, simulation environment, and performance metrics.


\noindent\textbf{Traffic Generation.}
We used two CAIDA traffic traces (2019 and 2014) to derive the flow size distribution for our simulation. Flow sizes were sampled from these traces to ensure realistic traffic patterns. Each flow's rate followed a Poisson distribution, with the rate being linearly proportional to its size. The arrival of flows was uniformly distributed during the simulation. Increasing the input rate was achieved by increasing the number of sampled flows, allowing the generation of synthetic traffic with realistic flow sizes and adjustable rates.
\textbf{Policy Rule Set Generation.}
For generating the policy rule sets, we employed the ClassBench-ng framework, which is widely used for benchmarking and generating common policy tables such as ACLs and routing tables. In addition, we employed flow-bench~\cite{Chen.FlowBench} to generate other non-supported fields.
\noindent\textbf{Performance Metrics}
The primary performance metrics in our evaluation include hit rate, latency distribution of per-key matches, and external memory bandwidth usage. We also conducted a stress test to assess \gls{vmt} throughput relative to input rate and compared the performance of the elastic \gls{vmt} configuration with a static one, where \gls{vmt} size is allocated based on an oracle matching the maximum required size.
\vspace{-0.2cm}
\noindent\subsection*{Performance Evaluation}

\subsubsection*{\textbf{\gls{vmt} and \gls{pmu} size impact}}
The proposed architecture makes it possible to assign multiple \glspl{pmu} to the same \gls{vmt} in a more dynamic manner. Each \gls{pmu} is characterized by its block-size, i.e., the size of the \gls{cam} associated with the \gls{pmu}. The nominal capacity of a \gls{vmt} is given by the sum of the block-sizes of the associated \glspl{pmu}. For instance, to create a \gls{vmt} size of 1536 entries, using \glspl{pmu} with 64 entries each would require 24 \glspl{pmu}, whereas using \glspl{pmu} with 512 entries each would need only 3 \glspl{pmu}. Therefore, the block-size determines the granularity with which we can tune the capacity of \gls{vmt}. Moreover, the larger the block-size, the higher the energy consumption and cost of the \gls{pmu}. It is therefore interesting to investigate the impact of the \gls{pmu} block-size on the system performance for a given aggregate capacity. 

To investigate these aspects, we report in Fig.~\ref{fig:lat-bw-hitrate} the average hit-rate, the box-plot distribution of the latency, and the mean bandwidth utilization to reach the external memory when varying the \gls{vmt} capacity. For each \gls{vmt} capacity value, moreover, we considered four different configurations of the \gls{pmu} block-size, as indicated in the figure's legend. The upper and lower graphs have been obtained using the flow-size distribution extracted from Trace 1 and Trace 2, respectively. 

We can immediately observe that, as expected, the larger the \gls{vmt} capacity the better the hit rate and, consequently, the lower the latency and the external memory bandwidth. These observations are consistent across different traffic traces, indicating that the results are qualitatively the same regardless of the specific traffic patterns considered. Looking at the effect of the block-size, we notice that smaller blocks for the same \gls{vmt} size yield significantly lower results. This is due to the non-uniform key distribution as well as the potential collisions associated with the consistent hashing.

The latency distribution for per-key matching is shown in Fig.s~\ref{fig:lat1}~and~\ref{fig:lat2}. It is mainly affected by the likelihood of key mismatch events, which require access to the slower external memory. Therefore, the latency distribution reflects the behavior of the hit rate, improving with larger \gls{vmt} capacity. For a given capacity, moreover, the average latency increases for smaller \gls{pmu} block-sizes, because of the higher risk of bottlenecks is some \glspl{pmu}. Figures~\ref{fig:lat1}~and~\ref{fig:lat2} show that an appropriately sized \gls{vmt} can achieve average latency in the order of 20 nanoseconds, with minimal variance in the latency distribution. 

Finally, Figs~\ref{fig:bw1}~and~\ref{fig:bw2} illustrate the memory bandwidth usage (in GB/s). Again, we observe a clear correlation with \gls{vmt} size and \gls{pmu} block-size, as for the other metrics. This is no surprise, considering that the access to the external memory occurs in case of key mismatch and, hence, depends directly on the hit rate. Once again, smaller block sizes exhibit higher bandwidth usage for the same \gls{vmt} size because they have fewer entries per a single \gls{pmu}, increasing the likelihood of over-utilizing a single \gls{pmu}. 

We can hence conclude that having smaller blocks allows for finer-grained allocation of \glspl{pmu} and requires less power per key match, proportional to the size of the \gls{pmu}. However, this advantage comes at the cost of potentially moving the bottleneck to the interconnection. Conversely, having larger tables is more feasible in terms of reducing interconnection complexity and balancing the load more effectively across \glspl{pmu}, ensuring overall system efficiency.

\begin{figure}[t]
    \centering
    \begin{subfigure}[t]{0.49\linewidth}
        \centering
        \input{figures/eval-results/throughput/throughput-trace-1}
        \vspace{-11.5pt}
        \caption{}
        \label{fig:throughput-t1}
    \end{subfigure}
    \begin{subfigure}[t]{0.49\linewidth}
        \centering
        \input{figures/eval-results/throughput/throughput-trace-2}
        \caption{}
        \label{fig:mem-usage-t1}
    \end{subfigure}
    \caption{[U50]~(a) System average throughput and (b) average external memory bandwidth usage (exposing a single pseudo channel (128 MB)) as a function of the input rate, for three distinct VMT sizes with a basic block-size of 256.}
    \label{fig:sys-throughput}
\end{figure}
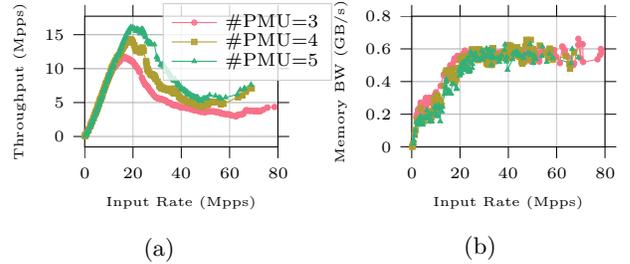

\noindent\textbf{Stress Test} Fig.~\ref{fig:throughput-t1} represents the input rate on the x-axis in millions of packets per second (Mpps) and the \gls{vmt} throughput on the y-axis. We observe that for each configuration with 3, 4, and 5 \glspl{pmu} with block size 256 attached to the \gls{vmt}, there is a linear increase in throughput until a saturation point is reached. This saturation point occurs when the arrival packet rate exceeds the \gls{vmt}'s processing capacity, and it increases with the \gls{vmt} size. Beyond this saturation point, we observe a decline in performance, which can be better explained by observing Fig.~\ref{fig:mem-usage-t1} that shows the external memory usage resulting from the same experiment. Here, we note a linear increase in memory usage with the incoming traffic, with a lower slope for \glspl{vmt} with more active entries, or equivalently, with a higher number of associated \glspl{pmu}. As the input rate increases, the external memory usage continues to rise, indicating an increase in the miss rate. Beyond  the saturation point, the external memory bandwidth usage becomes approximately constant. In this region, the system cannot amortize the miss rate, leading to a larger number of keys getting queued, which in turn causes a pipeline stall, and consequently decreases the throughput.

\begin{figure*}[t]
    \centering
    \begin{subfigure}[t]{0.24\textwidth}
        \centering
        \input{figures/eval-results/adapt-vs-static/adapt-static-t1}  
        \caption{Traffic Profile 1.}
    \end{subfigure}
    \hfill
    \begin{subfigure}[t]{0.24\textwidth}
        \centering
        \input{figures/eval-results/adapt-vs-static/adapt-static-t2}
        \caption{Traffic Profile 2.}
    \end{subfigure}
    \hfill
    \begin{subfigure}[t]{0.24\textwidth}
        \centering
        \input{figures/eval-results/adapt-vs-static/adapt-static-t3}
        \caption{Traffic Profile 3.}
    \end{subfigure}
    \hfill
    \begin{subfigure}[t]{0.24\textwidth}
        \centering
        \input{figures/eval-results/adapt-vs-static/adapt-static-t4}
        \caption{Traffic Profile 4.}
    \end{subfigure}
    \caption{[Sim]~The figure shows 4 different traffic profiles with varying flow arrival rates. The blue line represents the static approach, which provisions the maximum number of entries beforehand, while the red line is runtime-opt, which dynamically scales the number of active PMUs. Synapse closely matches the static allocation with minimal throughput drop. In figure (d), Synapse efficiently handles high traffic rates with a low number of active PMUs.}
    \label{fig:sys-prov}
    \vspace{-0.3cm}
\end{figure*}
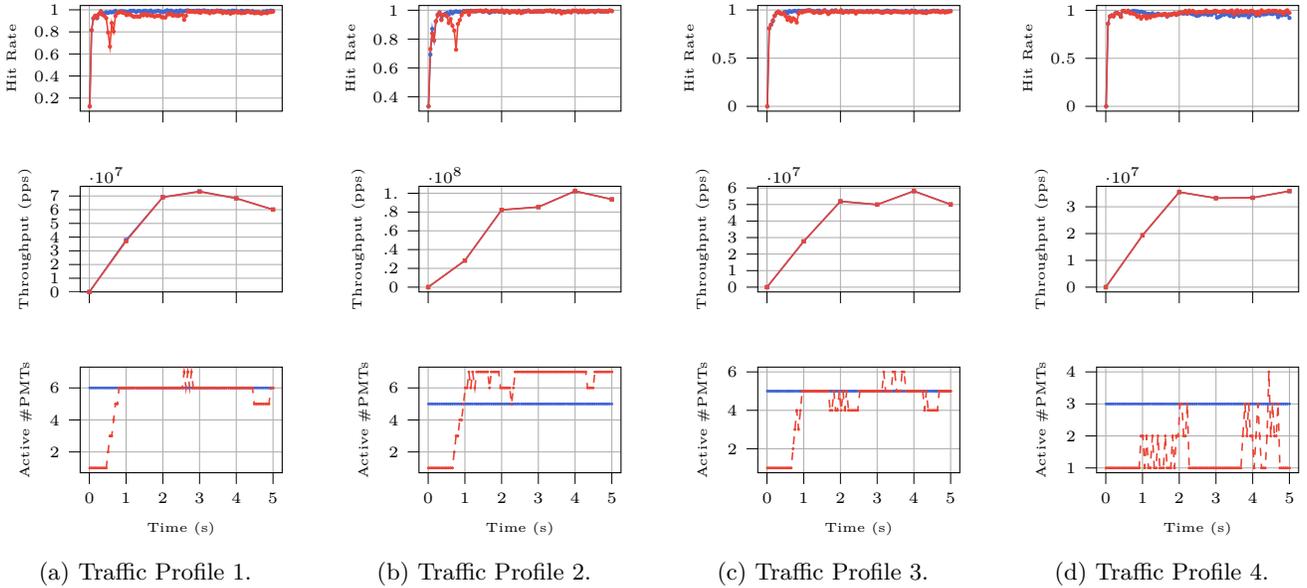

It is worth noting that the memory saturation point serves as a reference, indicating that the \gls{elu} implementation is not fully utilizing the memory port bandwidth. This limitation arises from data dependencies on per-key external lookups, non-sequential reads, and the static scheduler, which constrain performance. Two potential optimization strategies can be utilized. Firstly, as demonstrated, with a single port, we can utilize roughly 0.64 GB/s bandwidth with a single channel exposed. By duplicating the trie on different channels, we can easily double the utilizable memory bandwidth, effectively doubling the achievable external match rate. However, this comes at the cost of increased external~memory~usage. 

\noindent Another approach is to minimize the number of memory accesses. Similar to a B-tree structure, we could use a higher N-bit per node, which would result in more sequential reads but would require additional memory due to the prefix-rule expansion. All these approaches delay the decline point in the achievable throughput. Finally, the proposed architecture can optimize performance by extending or adding more cache blocks, essentially associating more \glspl{pmu} with the \gls{vmu} at hand.


\noindent\textbf{Runtime OPT effect on \gls{vmt} performance}  
We compared the static allocation scheme, where an oracle provisions the needed number of entries and keeps them fixed (blue), with the adaptive scheme (red), which relies on Runtime-OPT. Across all traffic profiles, Runtime-OPT achieves comparable throughput to the static allocation with only minor throughput drops. Even when hit rates decrease, the architecture compensates by hiding the external lookup latency, maintaining similar throughput. Furthermore, Runtime-OPT demonstrates its efficiency by achieving the same throughput and hit rate with fewer active PMUs (Fig.~\ref{fig:sys-prov}d), enabling the system to power down unused PMUs. In Fig.~\ref{fig:sys-prov}b, Runtime-OPT over-provisions two extra PMUs, as our model does not perfectly capture the system behavior (see Fig.~10, Appendix). In subfigures (a) and (c), the number of PMUs allocated by Runtime-OPT aligns with the static allocation on average.


\noindent\textbf{Hardware Cost}
The \gls{vmt} implementation in Synapse can be realized using a single lookup table from a hardware perspective, e.g., utilizing BRAM and LUTs in the FPGA.  An additional hardware cost is introduced by the \gls{phv} buffering FIFO queue, which can also be implemented using FPGA resources such as BRAM and LUTs. The number of FIFO queues required depends on the architecture. In the \gls{rmt} architecture, exactly \(S\) FIFO queues are needed, where \(S\) represents the number of stages. In contrast, the \gls{drmt} processor-based architecture required \(P * V\) FIFO queues, where \(P\) denotes the number of processors. Each processor processes the packet until the end, necessitating a private queue for each \gls{vmt}.

We acknowledge the complexity and additional cost introduced by the interconnection network between the \glspl{vmt} and \glspl{pmu}, as well as between the \glspl{pmu} and the \gls{elu}. While a full crossbar is an option, it may not scale efficiently with a large number of \glspl{pmu}. A solution to this is to divide the \glspl{pmu} into memory clusters, as traditionally done in data plane architectures, and implement a segment crossbar as proposed in \cite{Chole.dRMT}. This might lead to congestion when different \glspl{vmt} request different \glspl{pmu} located within the same memory cluster. Since we do not have a multicast lookup but only unicast, and the OPT manages which \glspl{pmu} to allocate for a given \gls{vmt}, namely, maximizing the intra-cluster \gls{pmu} allocation for each \gls{vmt}, this approach leads to less congestion. The interconnection from \glspl{pmu} to the \gls{elu} is expected to have fewer demands in terms of throughput and, therefore, result in simpler interconnection requirements, such as a simple shared bus.

%% file: figures/eval-results/shared-legend.tex
\begin{tikzpicture}
	\definecolor{color0}{rgb}{0.329411764705882,0.438235294117647,0.752941176470588}
    \definecolor{color1}{rgb}{0.828921568627451,0.337745098039216,0.29656862745098}
    \definecolor{color2}{rgb}{0.863235294117647,0.677941176470588,0.136764705882353}
    \definecolor{color3}{rgb}{0.26078431372549,0.601960784313726,0.351960784313725}
\begin{axis}[
    width=0,
    height=0,
    at={(0,0)},
    scale only axis,
    xmin=0,
    xmax=0,
    ybar,
    ybar legend,
    xtick={},
    ymin=0,
    ymax=0,
    ytick={},
    axis background/.style={fill=white},
    legend style={area legend,fill, legend cell align=center, align=center, draw=white!15!black, font=\scriptsize, at={(0, 0)}, anchor=center, /tikz/every even column/.append style={column sep=2em}},
    legend columns=10,
]
\addplot [thick, color0,fill=color0]
table {%
0 1
};
\addlegendentry{\textbf{block-size=64}}

\addplot [thick, color1,fill=color1]
table {%
0 1
};
\addlegendentry{\textbf{block-size=128}}

\addplot [thick, color2,fill=color2]
table {%
0 1
};
\addlegendentry{\textbf{block-size=256}}

\addplot [thick, color3,fill=color3]
table {%
0 1
};
\addlegendentry{\textbf{block-size=512}}
\end{axis}

\end{tikzpicture}

%% file: figures/eval-results/latency/latency-trace-1.tex
\begin{tikzpicture}
\definecolor{color0}{rgb}{0.329411764705882,0.438235294117647,0.752941176470588}
\definecolor{color1}{rgb}{0.828921568627451,0.337745098039216,0.29656862745098}
\definecolor{color2}{rgb}{0.863235294117647,0.677941176470588,0.136764705882353}
\definecolor{color3}{rgb}{0.26078431372549,0.601960784313726,0.351960784313725}

\begin{axis}[
width=\linewidth,
height=0.8\linewidth,
legend cell align={left},
legend columns=4,
legend style={fill opacity=0.8, draw opacity=1, text opacity=1, at={(0.5,1)}, anchor=south, draw=none},
tick align=outside,
tick pos=left,
x grid style={white!69.0196078431373!black},
xlabel={VMT size},
xmin=-0.5, xmax=2.5,
xtick style={color=black},
xtick={0,1,2},
yticklabels={0,0.5,1.0,1.5,2.0,2.5,3.0,3.5,4.0},
ytick={0.0,0.5,1,1.5,2,2.5,3,3.5,4},
xticklabels={512,1024,1536},
y grid style={white!50.1960784313725!black},
ylabel={Latency ($\mu s$)},
ymajorgrids,
ymin=-0.1, ymax=4.5,
ytick style={color=black},
ylabel style={font=\fontsize{5}{5}\selectfont},
xlabel style={font=\fontsize{5}{5}\selectfont},
tick label style={font=\fontsize{6}{6}\selectfont}
]
\path [draw=white!25.8823529411765!black, fill=color0, semithick]
(axis cs:-0.398,3.07696415910571)
--(axis cs:-0.202,3.07696415910571)
--(axis cs:-0.202,3.70190825591292)
--(axis cs:-0.398,3.70190825591292)
--(axis cs:-0.398,3.07696415910571)
--cycle;
\path [draw=white!25.8823529411765!black, fill=color1, semithick]
(axis cs:-0.198,2.07850562066749)
--(axis cs:-0.002,2.07850562066749)
--(axis cs:-0.002,2.39471493600498)
--(axis cs:-0.198,2.39471493600498)
--(axis cs:-0.198,2.07850562066749)
--cycle;
\path [draw=white!25.8823529411765!black, fill=color2, semithick]
(axis cs:0.00200000000000003,1.69691351857356)
--(axis cs:0.198,1.69691351857356)
--(axis cs:0.198,2.14591962114042)
--(axis cs:0.00200000000000003,2.14591962114042)
--(axis cs:0.00200000000000003,1.69691351857356)
--cycle;
\path [draw=white!25.8823529411765!black, fill=color3, semithick]
(axis cs:0.202,1.39524274026351)
--(axis cs:0.398,1.39524274026351)
--(axis cs:0.398,1.58530119060914)
--(axis cs:0.202,1.58530119060914)
--(axis cs:0.202,1.39524274026351)
--cycle;
\path [draw=white!25.8823529411765!black, fill=color0, semithick]
(axis cs:0.602,1.30004391601203)
--(axis cs:0.798,1.30004391601203)
--(axis cs:0.798,2.1275088822486)
--(axis cs:0.602,2.1275088822486)
--(axis cs:0.602,1.30004391601203)
--cycle;
\path [draw=white!25.8823529411765!black, fill=color1, semithick]
(axis cs:0.802,0.52546584941566)
--(axis cs:0.998,0.52546584941566)
--(axis cs:0.998,1.59629705098805)
--(axis cs:0.802,1.59629705098805)
--(axis cs:0.802,0.52546584941566)
--cycle;
\path [draw=white!25.8823529411765!black, fill=color2, semithick]
(axis cs:1.002,0.285316803534162)
--(axis cs:1.198,0.285316803534162)
--(axis cs:1.198,1.24477637872755)
--(axis cs:1.002,1.24477637872755)
--(axis cs:1.002,0.285316803534162)
--cycle;
\path [draw=white!25.8823529411765!black, fill=color3, semithick]
(axis cs:1.202,0.0169225144776801)
--(axis cs:1.398,0.0169225144776801)
--(axis cs:1.398,0.766422749505134)
--(axis cs:1.202,0.766422749505134)
--(axis cs:1.202,0.0169225144776801)
--cycle;
\path [draw=white!25.8823529411765!black, fill=color0, semithick]
(axis cs:1.602,0.302649927364534)
--(axis cs:1.798,0.302649927364534)
--(axis cs:1.798,0.889987234299788)
--(axis cs:1.602,0.889987234299788)
--(axis cs:1.602,0.302649927364534)
--cycle;
\path [draw=white!25.8823529411765!black, fill=color1, semithick]
(axis cs:1.802,0.0551624766087464)
--(axis cs:1.998,0.0551624766087464)
--(axis cs:1.998,0.495807574681548)
--(axis cs:1.802,0.495807574681548)
--(axis cs:1.802,0.0551624766087464)
--cycle;
\path [draw=white!25.8823529411765!black, fill=color2, semithick]
(axis cs:2.002,0.0161317120440869)
--(axis cs:2.198,0.0161317120440869)
--(axis cs:2.198,0.0346874167589494)
--(axis cs:2.002,0.0346874167589494)
--(axis cs:2.002,0.0161317120440869)
--cycle;
\path [draw=white!25.8823529411765!black, fill=color3, semithick]
(axis cs:2.202,0.0159935208181884)
--(axis cs:2.398,0.0159935208181884)
--(axis cs:2.398,0.0339862668450473)
--(axis cs:2.202,0.0339862668450473)
--(axis cs:2.202,0.0159935208181884)
--cycle;
\draw[draw=white!25.8823529411765!black,fill=color0,line width=0.3pt] (axis cs:0,0) rectangle (axis cs:0,0);

\draw[draw=white!25.8823529411765!black,fill=color1,line width=0.3pt] (axis cs:0,0) rectangle (axis cs:0,0);

\draw[draw=white!25.8823529411765!black,fill=color2,line width=0.3pt] (axis cs:0,0) rectangle (axis cs:0,0);

\draw[draw=white!25.8823529411765!black,fill=color3,line width=0.3pt] (axis cs:0,0) rectangle (axis cs:0,0);

\addplot [semithick, white!25.8823529411765!black, forget plot]
table {%
-0.3 3.07696415910571
-0.3 2.52289848688697
};
\addplot [semithick, white!25.8823529411765!black, forget plot]
table {%
-0.3 3.70190825591292
-0.3 4.27231846444985
};
\addplot [semithick, white!25.8823529411765!black, forget plot]
table {%
-0.349 2.52289848688697
-0.251 2.52289848688697
};
\addplot [semithick, white!25.8823529411765!black, forget plot]
table {%
-0.349 4.27231846444985
-0.251 4.27231846444985
};
\addplot [semithick, white!25.8823529411765!black, forget plot]
table {%
-0.1 2.07850562066749
-0.1 1.65408516042781
};
\addplot [semithick, white!25.8823529411765!black, forget plot]
table {%
-0.1 2.39471493600498
-0.1 2.78558912508285
};
\addplot [semithick, white!25.8823529411765!black, forget plot]
table {%
-0.149 1.65408516042781
-0.051 1.65408516042781
};
\addplot [semithick, white!25.8823529411765!black, forget plot]
table {%
-0.149 2.78558912508285
-0.051 2.78558912508285
};
\addplot [semithick, white!25.8823529411765!black, forget plot]
table {%
0.1 1.69691351857356
0.1 1.5147813381008
};
\addplot [semithick, white!25.8823529411765!black, forget plot]
table {%
0.1 2.14591962114042
0.1 2.59064004249077
};
\addplot [semithick, white!25.8823529411765!black, forget plot]
table {%
0.051 1.5147813381008
0.149 1.5147813381008
};
\addplot [semithick, white!25.8823529411765!black, forget plot]
table {%
0.051 2.59064004249077
0.149 2.59064004249077
};
\addplot [semithick, white!25.8823529411765!black, forget plot]
table {%
0.3 1.39524274026351
0.3 1.15170111420198
};
\addplot [semithick, white!25.8823529411765!black, forget plot]
table {%
0.3 1.58530119060914
0.3 1.63216730231894
};
\addplot [semithick, white!25.8823529411765!black, forget plot]
table {%
0.251 1.15170111420198
0.349 1.15170111420198
};
\addplot [semithick, white!25.8823529411765!black, forget plot]
table {%
0.251 1.63216730231894
0.349 1.63216730231894
};
\addplot [semithick, white!25.8823529411765!black, forget plot]
table {%
0.7 1.30004391601203
0.7 1.00673292125232
};
\addplot [semithick, white!25.8823529411765!black, forget plot]
table {%
0.7 2.1275088822486
0.7 2.54448500064852
};
\addplot [semithick, white!25.8823529411765!black, forget plot]
table {%
0.651 1.00673292125232
0.749 1.00673292125232
};
\addplot [semithick, white!25.8823529411765!black, forget plot]
table {%
0.651 2.54448500064852
0.749 2.54448500064852
};
\addplot [semithick, white!25.8823529411765!black, forget plot]
table {%
0.9 0.52546584941566
0.9 0.267003385255332
};
\addplot [semithick, white!25.8823529411765!black, forget plot]
table {%
0.9 1.59629705098805
0.9 2.10697574303612
};
\addplot [semithick, white!25.8823529411765!black, forget plot]
table {%
0.851 0.267003385255332
0.949 0.267003385255332
};
\addplot [semithick, white!25.8823529411765!black, forget plot]
table {%
0.851 2.10697574303612
0.949 2.10697574303612
};
\addplot [semithick, white!25.8823529411765!black, forget plot]
table {%
1.1 0.285316803534162
1.1 0.0256533432162138
};
\addplot [semithick, white!25.8823529411765!black, forget plot]
table {%
1.1 1.24477637872755
1.1 1.48867137574343
};
\addplot [semithick, white!25.8823529411765!black, forget plot]
table {%
1.051 0.0256533432162138
1.149 0.0256533432162138
};
\addplot [semithick, white!25.8823529411765!black, forget plot]
table {%
1.051 1.48867137574343
1.149 1.48867137574343
};
\addplot [semithick, white!25.8823529411765!black, forget plot]
table {%
1.3 0.0169225144776801
1.3 0.0156057029478458
};
\addplot [semithick, white!25.8823529411765!black, forget plot]
table {%
1.3 0.766422749505134
1.3 1.24783844130108
};
\addplot [semithick, white!25.8823529411765!black, forget plot]
table {%
1.251 0.0156057029478458
1.349 0.0156057029478458
};
\addplot [semithick, white!25.8823529411765!black, forget plot]
table {%
1.251 1.24783844130108
1.349 1.24783844130108
};
\addplot [semithick, white!25.8823529411765!black, forget plot]
table {%
1.7 0.302649927364534
1.7 0.151315020010005
};
\addplot [semithick, white!25.8823529411765!black, forget plot]
table {%
1.7 0.889987234299788
1.7 1.41671204033596
};
\addplot [semithick, white!25.8823529411765!black, forget plot]
table {%
1.651 0.151315020010005
1.749 0.151315020010005
};
\addplot [semithick, white!25.8823529411765!black, forget plot]
table {%
1.651 1.41671204033596
1.749 1.41671204033596
};
\addplot [semithick, white!25.8823529411765!black, forget plot]
table {%
1.9 0.0551624766087464
1.9 0.0320099552572707
};
\addplot [semithick, white!25.8823529411765!black, forget plot]
table {%
1.9 0.495807574681548
1.9 0.995195785070785
};
\addplot [semithick, white!25.8823529411765!black, forget plot]
table {%
1.851 0.0320099552572707
1.949 0.0320099552572707
};
\addplot [semithick, white!25.8823529411765!black, forget plot]
table {%
1.851 0.995195785070785
1.949 0.995195785070785
};
\addplot [semithick, white!25.8823529411765!black, forget plot]
table {%
2.1 0.0161317120440869
2.1 0.0153277304809721
};
\addplot [semithick, white!25.8823529411765!black, forget plot]
table {%
2.1 0.0346874167589494
2.1 0.0394066241940436
};
\addplot [semithick, white!25.8823529411765!black, forget plot]
table {%
2.051 0.0153277304809721
2.149 0.0153277304809721
};
\addplot [semithick, white!25.8823529411765!black, forget plot]
table {%
2.051 0.0394066241940436
2.149 0.0394066241940436
};
\addplot [black, mark=o, mark size=0.5, mark options={solid,fill opacity=0}, only marks, forget plot]
table {%
2.1 0.880662690562775
2.1 0.190597498749375
};
\addplot [semithick, white!25.8823529411765!black, forget plot]
table {%
2.3 0.0159935208181884
2.3 0.0153173042505593
};
\addplot [semithick, white!25.8823529411765!black, forget plot]
table {%
2.3 0.0339862668450473
2.3 0.0376684693877551
};
\addplot [semithick, white!25.8823529411765!black, forget plot]
table {%
2.251 0.0153173042505593
2.349 0.0153173042505593
};
\addplot [semithick, white!25.8823529411765!black, forget plot]
table {%
2.251 0.0376684693877551
2.349 0.0376684693877551
};
\addplot [black, mark=o, mark size=0.5, mark options={solid,fill opacity=0}, only marks, forget plot]
table {%
2.3 0.586973010034731
2.3 0.691609487045451
};
\addplot [semithick, white!25.8823529411765!black, forget plot]
table {%
-0.398 3.43796252190912
-0.202 3.43796252190912
};
\addplot [semithick, white!25.8823529411765!black, forget plot]
table {%
-0.198 2.16196222814068
-0.002 2.16196222814068
};
\addplot [semithick, white!25.8823529411765!black, forget plot]
table {%
0.00200000000000003 1.98555753355051
0.198 1.98555753355051
};
\addplot [semithick, white!25.8823529411765!black, forget plot]
table {%
0.202 1.44488421950304
0.398 1.44488421950304
};
\addplot [semithick, white!25.8823529411765!black, forget plot]
table {%
0.602 1.65331081342356
0.798 1.65331081342356
};
\addplot [semithick, white!25.8823529411765!black, forget plot]
table {%
0.802 1.04539368033118
0.998 1.04539368033118
};
\addplot [semithick, white!25.8823529411765!black, forget plot]
table {%
1.002 0.872441816489908
1.198 0.872441816489908
};
\addplot [semithick, white!25.8823529411765!black, forget plot]
table {%
1.202 0.0343349600533836
1.398 0.0343349600533836
};
\addplot [semithick, white!25.8823529411765!black, forget plot]
table {%
1.602 0.602173292141402
1.798 0.602173292141402
};
\addplot [semithick, white!25.8823529411765!black, forget plot]
table {%
1.802 0.313204607975567
1.998 0.313204607975567
};
\addplot [semithick, white!25.8823529411765!black, forget plot]
table {%
2.002 0.017408928804698
2.198 0.017408928804698
};
\addplot [semithick, white!25.8823529411765!black, forget plot]
table {%
2.202 0.017212701523125
2.398 0.017212701523125
};
\end{axis}

\end{tikzpicture}

%% file: figures/eval-results/latency/latency-trace-2.tex
\begin{tikzpicture}

\definecolor{color0}{rgb}{0.329411764705882,0.438235294117647,0.752941176470588}
\definecolor{color1}{rgb}{0.828921568627451,0.337745098039216,0.29656862745098}
\definecolor{color2}{rgb}{0.863235294117647,0.677941176470588,0.136764705882353}
\definecolor{color3}{rgb}{0.26078431372549,0.601960784313726,0.351960784313725}

\begin{axis}[
width=\linewidth,
height=0.8\linewidth,
legend cell align={left},
legend columns=4,
legend style={fill opacity=0.8, draw opacity=1, text opacity=1, at={(0.5,1)}, anchor=south, draw=none},
tick align=outside,
tick pos=left,
x grid style={white!69.0196078431373!black},
xlabel={VMT size},
xmin=-0.5, xmax=2.5,
xtick style={color=black},
xtick={0,1,2},
yticklabels={0.0,0.5,1.0,1.5,2.0,2.5,3.0,3.5,4.0},
ytick={0,0.5,1,1.5,2,2.5,3,3.5,4},
xticklabels={512,1024,1536},
y grid style={white!50.1960784313725!black},
ylabel={Latency ($\mu s$)},
ymajorgrids,
ymin=-0.1, ymax=4.5,
ytick style={color=black},
ylabel style={font=\fontsize{5}{5}\selectfont},
xlabel style={font=\fontsize{5}{5}\selectfont},
tick label style={font=\fontsize{6}{6}\selectfont}
]
\path [draw=white!25.8823529411765!black, fill=color0, semithick]
(axis cs:-0.398,1.4740902009783)
--(axis cs:-0.202,1.4740902009783)
--(axis cs:-0.202,2.07401622786429)
--(axis cs:-0.398,2.07401622786429)
--(axis cs:-0.398,1.4740902009783)
--cycle;
\path [draw=white!25.8823529411765!black, fill=color1, semithick]
(axis cs:-0.198,0.945905375762931)
--(axis cs:-0.002,0.945905375762931)
--(axis cs:-0.002,1.47700184416168)
--(axis cs:-0.198,1.47700184416168)
--(axis cs:-0.198,0.945905375762931)
--cycle;
\path [draw=white!25.8823529411765!black, fill=color2, semithick]
(axis cs:0.00200000000000003,0.736883157982643)
--(axis cs:0.198,0.736883157982643)
--(axis cs:0.198,1.12490209720486)
--(axis cs:0.00200000000000003,1.12490209720486)
--(axis cs:0.00200000000000003,0.736883157982643)
--cycle;
\path [draw=white!25.8823529411765!black, fill=color3, semithick]
(axis cs:0.202,0.473140492081216)
--(axis cs:0.398,0.473140492081216)
--(axis cs:0.398,0.65449750844869)
--(axis cs:0.202,0.65449750844869)
--(axis cs:0.202,0.473140492081216)
--cycle;
\path [draw=white!25.8823529411765!black, fill=color0, semithick]
(axis cs:0.602,0.603018669933369)
--(axis cs:0.798,0.603018669933369)
--(axis cs:0.798,0.924378499322272)
--(axis cs:0.602,0.924378499322272)
--(axis cs:0.602,0.603018669933369)
--cycle;
\path [draw=white!25.8823529411765!black, fill=color1, semithick]
(axis cs:0.802,0.30839493119455)
--(axis cs:0.998,0.30839493119455)
--(axis cs:0.998,0.829213008804522)
--(axis cs:0.802,0.829213008804522)
--(axis cs:0.802,0.30839493119455)
--cycle;
\path [draw=white!25.8823529411765!black, fill=color2, semithick]
(axis cs:1.002,0.115934328424114)
--(axis cs:1.198,0.115934328424114)
--(axis cs:1.198,0.372681596041284)
--(axis cs:1.002,0.372681596041284)
--(axis cs:1.002,0.115934328424114)
--cycle;
\path [draw=white!25.8823529411765!black, fill=color3, semithick]
(axis cs:1.202,0.165000987346754)
--(axis cs:1.398,0.165000987346754)
--(axis cs:1.398,0.464104233031259)
--(axis cs:1.202,0.464104233031259)
--(axis cs:1.202,0.165000987346754)
--cycle;
\path [draw=white!25.8823529411765!black, fill=color0, semithick]
(axis cs:1.602,0.0537284822152765)
--(axis cs:1.798,0.0537284822152765)
--(axis cs:1.798,0.538761859931347)
--(axis cs:1.602,0.538761859931347)
--(axis cs:1.602,0.0537284822152765)
--cycle;
\path [draw=white!25.8823529411765!black, fill=color1, semithick]
(axis cs:1.802,0.0389013468739071)
--(axis cs:1.998,0.0389013468739071)
--(axis cs:1.998,0.240394901914911)
--(axis cs:1.802,0.240394901914911)
--(axis cs:1.802,0.0389013468739071)
--cycle;
\path [draw=white!25.8823529411765!black, fill=color2, semithick]
(axis cs:2.002,0.0162405803313783)
--(axis cs:2.198,0.0162405803313783)
--(axis cs:2.198,0.0628265576476447)
--(axis cs:2.002,0.0628265576476447)
--(axis cs:2.002,0.0162405803313783)
--cycle;
\path [draw=white!25.8823529411765!black, fill=color3, semithick]
(axis cs:2.202,0.0145965296005909)
--(axis cs:2.398,0.0145965296005909)
--(axis cs:2.398,0.036794281102664)
--(axis cs:2.202,0.036794281102664)
--(axis cs:2.202,0.0145965296005909)
--cycle;
\draw[draw=white!25.8823529411765!black,fill=color0,line width=0.3pt] (axis cs:0,0) rectangle (axis cs:0,0);

\draw[draw=white!25.8823529411765!black,fill=color1,line width=0.3pt] (axis cs:0,0) rectangle (axis cs:0,0);

\draw[draw=white!25.8823529411765!black,fill=color2,line width=0.3pt] (axis cs:0,0) rectangle (axis cs:0,0);

\draw[draw=white!25.8823529411765!black,fill=color3,line width=0.3pt] (axis cs:0,0) rectangle (axis cs:0,0);

\addplot [semithick, white!25.8823529411765!black, forget plot]
table {%
-0.3 1.4740902009783
-0.3 0.8328935165307
};
\addplot [semithick, white!25.8823529411765!black, forget plot]
table {%
-0.3 2.07401622786429
-0.3 2.89543733254742
};
\addplot [semithick, white!25.8823529411765!black, forget plot]
table {%
-0.349 0.8328935165307
-0.251 0.8328935165307
};
\addplot [semithick, white!25.8823529411765!black, forget plot]
table {%
-0.349 2.89543733254742
-0.251 2.89543733254742
};
\addplot [semithick, white!25.8823529411765!black, forget plot]
table {%
-0.1 0.945905375762931
-0.1 0.628210575019154
};
\addplot [semithick, white!25.8823529411765!black, forget plot]
table {%
-0.1 1.47700184416168
-0.1 2.02516052638885
};
\addplot [semithick, white!25.8823529411765!black, forget plot]
table {%
-0.149 0.628210575019154
-0.051 0.628210575019154
};
\addplot [semithick, white!25.8823529411765!black, forget plot]
table {%
-0.149 2.02516052638885
-0.051 2.02516052638885
};
\addplot [semithick, white!25.8823529411765!black, forget plot]
table {%
0.1 0.736883157982643
0.1 0.32257833146597
};
\addplot [semithick, white!25.8823529411765!black, forget plot]
table {%
0.1 1.12490209720486
0.1 1.57887415556842
};
\addplot [semithick, white!25.8823529411765!black, forget plot]
table {%
0.051 0.32257833146597
0.149 0.32257833146597
};
\addplot [semithick, white!25.8823529411765!black, forget plot]
table {%
0.051 1.57887415556842
0.149 1.57887415556842
};
\addplot [black, mark=o, mark size=0.5, mark options={solid,fill opacity=0}, only marks, forget plot]
table {%
0.1 1.77657419728971
};
\addplot [semithick, white!25.8823529411765!black, forget plot]
table {%
0.3 0.473140492081216
0.3 0.255722835662395
};
\addplot [semithick, white!25.8823529411765!black, forget plot]
table {%
0.3 0.65449750844869
0.3 0.803767874123114
};
\addplot [semithick, white!25.8823529411765!black, forget plot]
table {%
0.251 0.255722835662395
0.349 0.255722835662395
};
\addplot [semithick, white!25.8823529411765!black, forget plot]
table {%
0.251 0.803767874123114
0.349 0.803767874123114
};
\addplot [semithick, white!25.8823529411765!black, forget plot]
table {%
0.7 0.603018669933369
0.7 0.348874897591805
};
\addplot [semithick, white!25.8823529411765!black, forget plot]
table {%
0.7 0.924378499322272
0.7 1.19153834311409
};
\addplot [semithick, white!25.8823529411765!black, forget plot]
table {%
0.651 0.348874897591805
0.749 0.348874897591805
};
\addplot [semithick, white!25.8823529411765!black, forget plot]
table {%
0.651 1.19153834311409
0.749 1.19153834311409
};
\addplot [semithick, white!25.8823529411765!black, forget plot]
table {%
0.9 0.30839493119455
0.9 0.0830160968833051
};
\addplot [semithick, white!25.8823529411765!black, forget plot]
table {%
0.9 0.829213008804522
0.9 1.20366582400914
};
\addplot [semithick, white!25.8823529411765!black, forget plot]
table {%
0.851 0.0830160968833051
0.949 0.0830160968833051
};
\addplot [semithick, white!25.8823529411765!black, forget plot]
table {%
0.851 1.20366582400914
0.949 1.20366582400914
};
\addplot [semithick, white!25.8823529411765!black, forget plot]
table {%
1.1 0.115934328424114
1.1 0.0243223783077426
};
\addplot [semithick, white!25.8823529411765!black, forget plot]
table {%
1.1 0.372681596041284
1.1 0.751066742010423
};
\addplot [semithick, white!25.8823529411765!black, forget plot]
table {%
1.051 0.0243223783077426
1.149 0.0243223783077426
};
\addplot [semithick, white!25.8823529411765!black, forget plot]
table {%
1.051 0.751066742010423
1.149 0.751066742010423
};
\addplot [black, mark=o, mark size=0.5, mark options={solid,fill opacity=0}, only marks, forget plot]
table {%
1.1 0.806374497354497
1.1 0.782878364202846
};
\addplot [semithick, white!25.8823529411765!black, forget plot]
table {%
1.3 0.165000987346754
1.3 0.0136219064454719
};
\addplot [semithick, white!25.8823529411765!black, forget plot]
table {%
1.3 0.464104233031259
1.3 0.584997807759449
};
\addplot [semithick, white!25.8823529411765!black, forget plot]
table {%
1.251 0.0136219064454719
1.349 0.0136219064454719
};
\addplot [semithick, white!25.8823529411765!black, forget plot]
table {%
1.251 0.584997807759449
1.349 0.584997807759449
};
\addplot [semithick, white!25.8823529411765!black, forget plot]
table {%
1.7 0.0537284822152765
1.7 0.0179626697213706
};
\addplot [semithick, white!25.8823529411765!black, forget plot]
table {%
1.7 0.538761859931347
1.7 0.979059928288465
};
\addplot [semithick, white!25.8823529411765!black, forget plot]
table {%
1.651 0.0179626697213706
1.749 0.0179626697213706
};
\addplot [semithick, white!25.8823529411765!black, forget plot]
table {%
1.651 0.979059928288465
1.749 0.979059928288465
};
\addplot [semithick, white!25.8823529411765!black, forget plot]
table {%
1.9 0.0389013468739071
1.9 0.014116873284239
};
\addplot [semithick, white!25.8823529411765!black, forget plot]
table {%
1.9 0.240394901914911
1.9 0.538445806614613
};
\addplot [semithick, white!25.8823529411765!black, forget plot]
table {%
1.851 0.014116873284239
1.949 0.014116873284239
};
\addplot [semithick, white!25.8823529411765!black, forget plot]
table {%
1.851 0.538445806614613
1.949 0.538445806614613
};
\addplot [black, mark=o, mark size=0.5, mark options={solid,fill opacity=0}, only marks, forget plot]
table {%
1.9 0.603451340415388
};
\addplot [semithick, white!25.8823529411765!black, forget plot]
table {%
2.1 0.0162405803313783
2.1 0.0130097328817048
};
\addplot [semithick, white!25.8823529411765!black, forget plot]
table {%
2.1 0.0628265576476447
2.1 0.105603498639049
};
\addplot [semithick, white!25.8823529411765!black, forget plot]
table {%
2.051 0.0130097328817048
2.149 0.0130097328817048
};
\addplot [semithick, white!25.8823529411765!black, forget plot]
table {%
2.051 0.105603498639049
2.149 0.105603498639049
};
\addplot [black, mark=o, mark size=0.5, mark options={solid,fill opacity=0}, only marks, forget plot]
table {%
2.1 0.24354657822741
2.1 0.55884044937485
2.1 0.4663769827653
};
\addplot [semithick, white!25.8823529411765!black, forget plot]
table {%
2.3 0.0145965296005909
2.3 0.0121629640585712
};
\addplot [semithick, white!25.8823529411765!black, forget plot]
table {%
2.3 0.036794281102664
2.3 0.0425027695351137
};
\addplot [semithick, white!25.8823529411765!black, forget plot]
table {%
2.251 0.0121629640585712
2.349 0.0121629640585712
};
\addplot [semithick, white!25.8823529411765!black, forget plot]
table {%
2.251 0.0425027695351137
2.349 0.0425027695351137
};
\addplot [black, mark=o, mark size=0.5, mark options={solid,fill opacity=0}, only marks, forget plot]
table {%
2.3 0.354156349990124
2.3 0.313302850274725
2.3 0.192974364592463
2.3 0.140887212451991
};
\addplot [semithick, white!25.8823529411765!black, forget plot]
table {%
-0.398 1.7320443232116
-0.202 1.7320443232116
};
\addplot [semithick, white!25.8823529411765!black, forget plot]
table {%
-0.198 1.27728963836793
-0.002 1.27728963836793
};
\addplot [semithick, white!25.8823529411765!black, forget plot]
table {%
0.00200000000000003 0.829401969562209
0.198 0.829401969562209
};
\addplot [semithick, white!25.8823529411765!black, forget plot]
table {%
0.202 0.596589472462167
0.398 0.596589472462167
};
\addplot [semithick, white!25.8823529411765!black, forget plot]
table {%
0.602 0.85027091649753
0.798 0.85027091649753
};
\addplot [semithick, white!25.8823529411765!black, forget plot]
table {%
0.802 0.510247969045359
0.998 0.510247969045359
};
\addplot [semithick, white!25.8823529411765!black, forget plot]
table {%
1.002 0.30361865828452
1.198 0.30361865828452
};
\addplot [semithick, white!25.8823529411765!black, forget plot]
table {%
1.202 0.317284840309847
1.398 0.317284840309847
};
\addplot [semithick, white!25.8823529411765!black, forget plot]
table {%
1.602 0.180595585730075
1.798 0.180595585730075
};
\addplot [semithick, white!25.8823529411765!black, forget plot]
table {%
1.802 0.118249170798003
1.998 0.118249170798003
};
\addplot [semithick, white!25.8823529411765!black, forget plot]
table {%
2.002 0.0229668180765243
2.198 0.0229668180765243
};
\addplot [semithick, white!25.8823529411765!black, forget plot]
table {%
2.202 0.0156265110634607
2.398 0.0156265110634607
};
\end{axis}

\end{tikzpicture}

%% file: figures/eval-results/bandwidth/bandwidth-trace-1.tex
\begin{tikzpicture}

\definecolor{color0}{rgb}{0.329411764705882,0.438235294117647,0.752941176470588}
\definecolor{color1}{rgb}{0.828921568627451,0.337745098039216,0.29656862745098}
\definecolor{color2}{rgb}{0.863235294117647,0.677941176470588,0.136764705882353}
\definecolor{color3}{rgb}{0.26078431372549,0.601960784313726,0.351960784313725}

\begin{axis}[
width=\linewidth,
height=0.8\linewidth,
legend cell align={left},
legend columns=3,
legend style={fill opacity=0.8, draw opacity=1, text opacity=1, at={(0.5,1)}, anchor=south, draw=none},
tick align=outside,
tick pos=left,
x grid style={white!69.0196078431373!black},
xlabel={VMT size},
xmin=-0.5, xmax=2.5,
xtick style={color=black},
xtick={0,1,2},
xticklabels={512,1024,1536},
ytick={0.05, 0.1, 0.15, 0.2, 0.25},
yticklabels={0.05, 0.1, 0.15, 0.2, 0.25},
y grid style={white!50.1960784313725!black},
ylabel={Memory BW (GB/s)},
ymajorgrids,
ymin=0, ymax=0.3,
ytick style={color=black},
ylabel style={font=\fontsize{5}{5}\selectfont},
xlabel style={font=\fontsize{5}{5}\selectfont},
tick label style={font=\fontsize{6}{6}\selectfont}
]
\draw[draw=none,fill=color0] (axis cs:-0.36,0) rectangle (axis cs:-0.24,0.27450925039893);

\draw[draw=none,fill=color0] (axis cs:0.64,0) rectangle (axis cs:0.76,0.261845260566963);
\draw[draw=none,fill=color0] (axis cs:1.64,0) rectangle (axis cs:1.76,0.237594825650004);
\draw[draw=none,fill=color1] (axis cs:-0.16,0) rectangle (axis cs:-0.04,0.270575332157998);

\draw[draw=none,fill=color1] (axis cs:0.84,0) rectangle (axis cs:0.96,0.243055065412074);
\draw[draw=none,fill=color1] (axis cs:1.84,0) rectangle (axis cs:1.96,0.203662199853754);
\draw[draw=none,fill=color2] (axis cs:0.04,0) rectangle (axis cs:0.16,0.265874847106013);

\draw[draw=none,fill=color2] (axis cs:1.04,0) rectangle (axis cs:1.16,0.229603552340389);
\draw[draw=none,fill=color2] (axis cs:2.04,0) rectangle (axis cs:2.16,0.104249699468677);
\draw[draw=none,fill=color3] (axis cs:0.24,0) rectangle (axis cs:0.36,0.25440065331244);

\draw[draw=none,fill=color3] (axis cs:1.24,0) rectangle (axis cs:1.36,0.143507094555931);
\draw[draw=none,fill=color3] (axis cs:2.24,0) rectangle (axis cs:2.36,0.0942081089713226);
\end{axis}

\end{tikzpicture}

%% file: figures/eval-results/bandwidth/bandwidth-trace-2.tex
\begin{tikzpicture}

\definecolor{color0}{rgb}{0.329411764705882,0.438235294117647,0.752941176470588}
\definecolor{color1}{rgb}{0.828921568627451,0.337745098039216,0.29656862745098}
\definecolor{color2}{rgb}{0.863235294117647,0.677941176470588,0.136764705882353}
\definecolor{color3}{rgb}{0.26078431372549,0.601960784313726,0.351960784313725}

\begin{axis}[
width=\linewidth,
height=0.8\linewidth,
legend cell align={left},
legend columns=3,
legend style={fill opacity=0.8, draw opacity=1, text opacity=1, at={(0.5,1)}, anchor=south, draw=none},
tick align=outside,
tick pos=left,
x grid style={white!69.0196078431373!black},
xlabel={VMT size},
xmin=-0.5, xmax=2.5,
xtick style={color=black},
xtick={0,1,2},
xticklabels={512,1024,1536},
ytick={0.05, 0.1, 0.15, 0.2, 0.25},
yticklabels={0.05, 0.1, 0.15, 0.2, 0.25},
y grid style={white!50.1960784313725!black},
ylabel={Memory BW (GB/s)},
ymajorgrids,
ymin=0, ymax=0.3,
ytick style={color=black},
ylabel style={font=\fontsize{5}{5}\selectfont},
xlabel style={font=\fontsize{5}{5}\selectfont},
tick label style={font=\fontsize{6}{6}\selectfont}
]
\draw[draw=none,fill=color0] (axis cs:-0.36,0) rectangle (axis cs:-0.24,0.272935698853655);

\draw[draw=none,fill=color0] (axis cs:0.64,0) rectangle (axis cs:0.76,0.254855458027373);
\draw[draw=none,fill=color0] (axis cs:1.64,0) rectangle (axis cs:1.76,0.211699183357645);
\draw[draw=none,fill=color1] (axis cs:-0.16,0) rectangle (axis cs:-0.04,0.270981865738168);

\draw[draw=none,fill=color1] (axis cs:0.84,0) rectangle (axis cs:0.96,0.241509069233378);
\draw[draw=none,fill=color1] (axis cs:1.84,0) rectangle (axis cs:1.96,0.196615425058853);
\draw[draw=none,fill=color2] (axis cs:0.04,0) rectangle (axis cs:0.16,0.264735021229935);

\draw[draw=none,fill=color2] (axis cs:1.04,0) rectangle (axis cs:1.16,0.223151243541694);
\draw[draw=none,fill=color2] (axis cs:2.04,0) rectangle (axis cs:2.16,0.139457886619927);
\draw[draw=none,fill=color3] (axis cs:0.24,0) rectangle (axis cs:0.36,0.253688900479712);

\draw[draw=none,fill=color3] (axis cs:1.24,0) rectangle (axis cs:1.36,0.203026826425326);
\draw[draw=none,fill=color3] (axis cs:2.24,0) rectangle (axis cs:2.36,0.100424076107805);
\end{axis}

\end{tikzpicture}

%% file: figures/eval-results/hit-rate/hit-rate-trace-1.tex
\begin{tikzpicture}

\definecolor{color0}{rgb}{0.258823529411765,0.403921568627451,0.823529411764706}
\definecolor{color1}{rgb}{0.917647058823529,0.262745098039216,0.207843137254902}
\definecolor{color2}{rgb}{0.984313725490196,0.737254901960784,0.0156862745098039}
\definecolor{color3}{rgb}{0.203921568627451,0.658823529411765,0.325490196078431}

\begin{axis}[
width=\linewidth,
height=0.4\linewidth,
legend cell align={left},
legend style={
  fill opacity=0.8,
  draw opacity=1,
  text opacity=1,
  at={(0.03,0.97)},
  anchor=north west,
  draw=white!80!black
},
tick align=outside,
tick pos=left,
x grid style={white!69.0196078431373!black},
xlabel={VMT size},
xmin=57.6, xmax=1606.4,
xtick style={color=black},
y grid style={white!50.1960784313725!black},
ylabel={Hit Rate},
ymajorgrids,
ymin=0.53563946392413, ymax=1.0045325018877,
ytick style={color=black},
xticklabels={128,256,512,1024,1536},
xtick={128,256,512,1024,1536},
ylabel style={font=\fontsize{5}{5}\selectfont},
xlabel style={font=\fontsize{5}{5}\selectfont},
tick label style={font=\fontsize{6}{6}\selectfont}
]
\path [draw=color0, fill=color0, opacity=0.2]
(axis cs:128,0.617447804264075)
--(axis cs:128,0.556952783831565)
--(axis cs:256,0.719939139142196)
--(axis cs:384,0.796318224832717)
--(axis cs:512,0.825520018620035)
--(axis cs:640,0.872738773474681)
--(axis cs:768,0.882343072834432)
--(axis cs:896,0.903277515877145)
--(axis cs:1024,0.899028593096706)
--(axis cs:1152,0.915064798621388)
--(axis cs:1280,0.922446120297055)
--(axis cs:1408,0.948136882713495)
--(axis cs:1536,0.932528572205809)
--(axis cs:1536,0.942948602219975)
--(axis cs:1536,0.942948602219975)
--(axis cs:1408,0.956889582212209)
--(axis cs:1280,0.939435787145407)
--(axis cs:1152,0.926357356522209)
--(axis cs:1024,0.912402274048688)
--(axis cs:896,0.918084047426559)
--(axis cs:768,0.898197852912185)
--(axis cs:640,0.882329741109397)
--(axis cs:512,0.841278723298048)
--(axis cs:384,0.828458226033279)
--(axis cs:256,0.752501389774034)
--(axis cs:128,0.617447804264075)
--cycle;

\path [draw=color1, fill=color1, opacity=0.2]
(axis cs:128,0.59798646130771)
--(axis cs:128,0.567978184072109)
--(axis cs:256,0.702789516765738)
--(axis cs:384,0.790765659079168)
--(axis cs:512,0.861041293569699)
--(axis cs:640,0.876744907759462)
--(axis cs:768,0.894333039174183)
--(axis cs:896,0.902411982395679)
--(axis cs:1024,0.910849658448237)
--(axis cs:1152,0.915063883131727)
--(axis cs:1280,0.931447461862283)
--(axis cs:1408,0.937943465967594)
--(axis cs:1536,0.945767251006878)
--(axis cs:1536,0.957863746845167)
--(axis cs:1536,0.957863746845167)
--(axis cs:1408,0.951206820501804)
--(axis cs:1280,0.941994832576743)
--(axis cs:1152,0.938793168656968)
--(axis cs:1024,0.934854713631122)
--(axis cs:896,0.922001385325214)
--(axis cs:768,0.907828834837242)
--(axis cs:640,0.899386866521208)
--(axis cs:512,0.873324430841468)
--(axis cs:384,0.817712993032809)
--(axis cs:256,0.76403752103896)
--(axis cs:128,0.59798646130771)
--cycle;

\path [draw=color2, fill=color2, opacity=0.2]
(axis cs:256,0.752535792726118)
--(axis cs:256,0.732597163290291)
--(axis cs:512,0.860652296217409)
--(axis cs:768,0.906158754469927)
--(axis cs:1024,0.921527931793997)
--(axis cs:1280,0.939927382117043)
--(axis cs:1536,0.96697750641503)
--(axis cs:1536,0.981183637415887)
--(axis cs:1536,0.981183637415887)
--(axis cs:1280,0.958498265404169)
--(axis cs:1024,0.940796081562138)
--(axis cs:768,0.930427281789759)
--(axis cs:512,0.874478695486549)
--(axis cs:256,0.752535792726118)
--cycle;

\path [draw=color3, fill=color3, opacity=0.2]
(axis cs:512,0.890958374264995)
--(axis cs:512,0.881841564972849)
--(axis cs:1024,0.945094332684931)
--(axis cs:1536,0.968212456533074)
--(axis cs:1536,0.983219181980269)
--(axis cs:1536,0.983219181980269)
--(axis cs:1024,0.971338846104968)
--(axis cs:512,0.890958374264995)
--cycle;

\addplot [semithick, color0, mark=*, mark size=3, mark options={solid,draw=white}, forget plot]
table {%
128 0.587546275521916
256 0.737052291141507
384 0.812763760556113
512 0.833308109199019
640 0.877570920275161
768 0.890336512088408
896 0.910604346959479
1024 0.905547374291018
1152 0.920755309326682
1280 0.93071363592675
1408 0.952635066003162
1536 0.938007439051319
};
\addplot [semithick, color1, mark=*, mark size=3, mark options={solid,draw=white}, forget plot]
table {%
128 0.583076599567788
256 0.734518690359405
384 0.804003078532348
512 0.867489495355117
640 0.888430585823482
768 0.901096397095589
896 0.912205499452056
1024 0.922955573034034
1152 0.927814415732235
1280 0.936652100176911
1408 0.944877916367219
1536 0.951817517521143
};
\addplot [semithick, color2, mark=*, mark size=3, mark options={solid,draw=white}, forget plot]
table {%
256 0.742250470901934
512 0.867620497351552
768 0.917958207765796
1024 0.931230416773863
1280 0.949408474837491
1536 0.974217313437209
};
\addplot [semithick, color3, mark=*, mark size=3, mark options={solid,draw=white}, forget plot]
table {%
512 0.886274560749061
1024 0.957665804872614
1536 0.97585195051331
};
\end{axis}

\end{tikzpicture}

%% file: figures/eval-results/hit-rate/hit-rate-trace-2.tex
\begin{tikzpicture}

\definecolor{color0}{rgb}{0.258823529411765,0.403921568627451,0.823529411764706}
\definecolor{color1}{rgb}{0.917647058823529,0.262745098039216,0.207843137254902}
\definecolor{color2}{rgb}{0.984313725490196,0.737254901960784,0.0156862745098039}
\definecolor{color3}{rgb}{0.203921568627451,0.658823529411765,0.325490196078431}

\begin{axis}[
width=\linewidth,
height=0.4\linewidth,
legend cell align={left},
legend style={
  fill opacity=0.8,
  draw opacity=1,
  text opacity=1,
  at={(0.03,0.97)},
  anchor=north west,
  draw=white!80!black
},
tick align=outside,
tick pos=left,
x grid style={white!69.0196078431373!black},
xlabel={VMT size},
xmin=57.6, xmax=1606.4,
xtick style={color=black},
y grid style={white!50.1960784313725!black},
ylabel={Hit Rate},
ymajorgrids,
ymin=0.653670247290028, ymax=1.00251103705287,
ytick style={color=black},
xticklabels={128,256,512,1024,1536},
xtick={128,256,512,1024,1536},
ylabel style={font=\fontsize{5}{5}\selectfont},
xlabel style={font=\fontsize{5}{5}\selectfont},
tick label style={font=\fontsize{6}{6}\selectfont}
]
\path [draw=color0, fill=color0, opacity=0.2]
(axis cs:128,0.719326046702533)
--(axis cs:128,0.669526646824702)
--(axis cs:256,0.757657835955275)
--(axis cs:384,0.824313431532755)
--(axis cs:512,0.8451826838066)
--(axis cs:640,0.878439946498582)
--(axis cs:768,0.897801974395669)
--(axis cs:896,0.889866487583423)
--(axis cs:1024,0.926084735396641)
--(axis cs:1152,0.930558343502333)
--(axis cs:1280,0.94104303101435)
--(axis cs:1536,0.951541052870708)
--(axis cs:1536,0.965630614534882)
--(axis cs:1536,0.965630614534882)
--(axis cs:1280,0.95317824338569)
--(axis cs:1152,0.945236433016162)
--(axis cs:1024,0.936832067147937)
--(axis cs:896,0.911158610752082)
--(axis cs:768,0.913376877622218)
--(axis cs:640,0.899987447332802)
--(axis cs:512,0.870181430508492)
--(axis cs:384,0.85515421373252)
--(axis cs:256,0.796588824065745)
--(axis cs:128,0.719326046702533)
--cycle;

\path [draw=color1, fill=color1, opacity=0.2]
(axis cs:128,0.758915682523852)
--(axis cs:128,0.720661165833621)
--(axis cs:256,0.778138628194753)
--(axis cs:384,0.838730288351494)
--(axis cs:512,0.861905582594382)
--(axis cs:640,0.894133424101284)
--(axis cs:768,0.915009132118239)
--(axis cs:896,0.928178485634768)
--(axis cs:1024,0.9319660741232)
--(axis cs:1152,0.941662361182261)
--(axis cs:1280,0.94022708693689)
--(axis cs:1408,0.952625274719855)
--(axis cs:1536,0.963534552173582)
--(axis cs:1536,0.971204426085284)
--(axis cs:1536,0.971204426085284)
--(axis cs:1408,0.966555596709743)
--(axis cs:1280,0.954821141926588)
--(axis cs:1152,0.956189148372945)
--(axis cs:1024,0.949270508898816)
--(axis cs:896,0.941062631158827)
--(axis cs:768,0.92878214677868)
--(axis cs:640,0.911822405720226)
--(axis cs:512,0.884944476298267)
--(axis cs:384,0.864055995962819)
--(axis cs:256,0.817036477326936)
--(axis cs:128,0.758915682523852)
--cycle;

\path [draw=color2, fill=color2, opacity=0.2]
(axis cs:256,0.848687570477573)
--(axis cs:256,0.825224251102035)
--(axis cs:512,0.878083848813083)
--(axis cs:768,0.913248292675584)
--(axis cs:1024,0.947892984754975)
--(axis cs:1280,0.962050681453409)
--(axis cs:1536,0.971529554242236)
--(axis cs:1536,0.980999775524921)
--(axis cs:1536,0.980999775524921)
--(axis cs:1280,0.973862368531318)
--(axis cs:1024,0.961058804125607)
--(axis cs:768,0.936480530830888)
--(axis cs:512,0.904572981168731)
--(axis cs:256,0.848687570477573)
--cycle;

\path [draw=color3, fill=color3, opacity=0.2]
(axis cs:512,0.933158618760748)
--(axis cs:512,0.923061577517675)
--(axis cs:1024,0.952463061988683)
--(axis cs:1536,0.979662930594384)
--(axis cs:1536,0.986654637518194)
--(axis cs:1536,0.986654637518194)
--(axis cs:1024,0.964472231408574)
--(axis cs:512,0.933158618760748)
--cycle;

\addplot [semithick, color0, mark=*, mark size=3, mark options={solid,draw=white}, forget plot]
table {%
128 0.694953249111099
256 0.777529168904707
384 0.840464470954251
512 0.85738024549959
640 0.888853823986511
768 0.90582343422283
896 0.90003626596969
1024 0.931043080665209
1152 0.937937844132797
1280 0.94740257659
1536 0.958683363525593
};
\addplot [semithick, color1, mark=*, mark size=3, mark options={solid,draw=white}, forget plot]
table {%
128 0.738587915733484
256 0.798356144580288
384 0.851912125722736
512 0.873116956748991
640 0.903084500568246
768 0.921807039448816
896 0.934647472617939
1024 0.940374376377825
1152 0.948915561989894
1280 0.947598767815303
1408 0.960056038588057
1536 0.967458712773117
};
\addplot [semithick, color2, mark=*, mark size=3, mark options={solid,draw=white}, forget plot]
table {%
256 0.837393906141835
512 0.892018329908261
768 0.926121987453236
1024 0.954884588654908
1280 0.968051445358437
1536 0.976288972457773
};
\addplot [semithick, color3, mark=*, mark size=3, mark options={solid,draw=white}, forget plot]
table {%
512 0.928221991024784
1024 0.95844194275416
1536 0.983115234707377
};
\end{axis}

\end{tikzpicture}

%% file: figures/eval-results/throughput/throughput-trace-1.tex
\pgfplotstableread{
x       y
125000 63940.3364341781
300000 156886.778560617
650000 354686.564381107
1100000 631933.418152971
1275000 760669.180945116
1800000 1191000.22914859
2050000 1401095.59925096
2550000 1861327.53517524
2800000 2105130.24224298
2875000 2177623.22638048
3025000 2312489.29987914
3625000 2856676.89749752
3875000 3087984.2864965
4125000 3332439.53299332
4350000 3528868.46900767
4800000 3932564.79888948
5900000 4969611.71758836
6200000 5249345.57968442
6250000 5294089.4561899
7100000 6089844.72346668
7300000 6294364.55606784
7350000 6355218.83620839
7775000 6743439.01089761
7925000 6864907.6464773
8300000 7186809.98868323
8400000 7316239.37986476
8625000 7482584.61084784
9075000 7886451.53260278
9450000 8238518.1280021
9475000 8259202.59983412
9875000 8561320.48804769
10075000 8732970.59323832
10375000 8981564.70362803
10450000 9027844.72247234
10550000 9077497.1926237
10675000 9205066.39355362
10750000 9242383.0321526
10825000 9295480.79062709
11100000 9495751.55056264
11150000 9498630.90962641
11550000 9804753.51497222
12025000 10085894.6328797
12100000 10082269.6908023
12500000 10244978.7378346
12750000 10363744.5700785
13225000 10583463.3231554
13300000 10622620.4308572
13325000 10638843.9742875
13375000 10709108.9272751
13700000 10872012.5573396
13825000 10897939.1970635
14050000 10965586.6622447
14250000 10983085.7808273
14400000 11041158.4374125
14500000 11071159.3206423
14850000 11267818.3339038
16175000 11512110.5825835
16575000 11704053.6904609
16700000 11674617.1075292
16750000 11680377.7052869
16850000 11620738.3726522
17525000 11415154.407708
17550000 11431751.5350937
18000000 11497590.298842
18150000 11428034.0028921
18325000 11339197.829984
18350000 11367822.8680433
18850000 11296433.2267127
18925000 11165910.2222708
19075000 11157850.4474446
19400000 11149784.0910113
19450000 11081896.4141527
19600000 11076004.9051392
19850000 11018278.9144496
19875000 11028813.7520524
20000000 10961426.0403143
20650000 10677508.1638131
20875000 10459889.0459369
20925000 10471559.014871
20975000 10463924.5575116
21075000 10447513.6378421
21525000 10177904.480521
21875000 10090673.9790079
22400000 9734522.05706697
22550000 9609512.83524824
22600000 9598852.98563469
22900000 9451082.05569759
23725000 8639229.03293552
23750000 8651540.3316489
24100000 8348572.48768616
24325000 8139669.85617989
24575000 8004382.57049304
24900000 7756948.26930391
25025000 7736650.84447055
25150000 7655825.33742586
25225000 7582671.84664727
25250000 7598292.94431488
25375000 7590058.94754703
25400000 7587149.29497553
26075000 7350562.06682547
26200000 7292839.09485629
26225000 7256972.37460283
26325000 7219229.53720002
26750000 6961179.65389164
27125000 6770135.2123677
27400000 6673226.90440499
27700000 6542499.53997174
27725000 6514565.37531449
28275000 6227793.66439758
28825000 6013853.9121436
29075000 5872977.04600211
29500000 5727520.48037876
30275000 5611967.62813494
30325000 5609926.4380926
30600000 5641012.60619858
30975000 5546533.94272305
31225000 5533498.82270154
31250000 5543521.25474958
31425000 5513433.74997023
31450000 5510033.08820606
31700000 5524032.22837353
31825000 5572515.23219272
32875000 5336012.43985514
33150000 5347349.230434
33325000 5312430.0405736
33600000 5269776.08999215
33800000 5204271.13766341
34375000 4985215.75489185
34625000 5001026.09762227
35425000 4925486.73115778
35575000 4909532.946259
36000000 4802445.18760496
36200000 4736237.10623814
36350000 4694330.88962921
36525000 4668680.47281091
36650000 4655089.76661011
36900000 4620001.25832995
37075000 4614430.6128488
37175000 4645588.8444165
37250000 4632602.02964255
37375000 4607542.88221085
37525000 4566579.02600486
37875000 4576592.59305083
37975000 4559277.84921976
38425000 4523307.60605293
38525000 4513053.80641232
39250000 4406457.04421869
39575000 4347260.8561675
39775000 4300991.61107309
39800000 4284690.65511367
40500000 4281347.76099993
40725000 4284852.21266929
40825000 4281312.6790637
40950000 4295443.98276623
41475000 4240111.22388636
41850000 4256651.89488765
42300000 4187037.5513781
42400000 4184322.9658895
42825000 4185968.09830459
43050000 4194939.83961194
43300000 4166333.43708943
43725000 4074533.25261952
43750000 4076713.27287628
44250000 3995010.78452707
44425000 3983739.54888348
44475000 3981097.63307853
45025000 3887369.66558659
45600000 3909549.48031976
45625000 3903038.70619584
46050000 3848224.16911248
46225000 3864777.92867719
46325000 3852744.12911038
46550000 3842815.36769528
46600000 3820956.23577016
47200000 3769178.6968009
47475000 3739187.64193902
47650000 3702975.73292287
47725000 3701936.31495698
47825000 3671339.8513187
47875000 3659813.39034862
48275000 3665339.59180325
48300000 3669491.42987247
48450000 3641832.62354862
49175000 3648688.80495518
49225000 3639283.87909681
49450000 3664706.93225491
49475000 3666052.54796233
49625000 3635423.73980372
49750000 3634327.60485211
49900000 3593360.25899201
50175000 3623879.80455735
50525000 3644258.99262903
51275000 3715998.73058097
51500000 3664803.74750424
51650000 3702204.21467886
51875000 3667497.32445596
52075000 3698321.13581286
53100000 3734025.95025901
53625000 3662747.23190848
53950000 3617375.23414295
54700000 3613252.09100276
55150000 3551284.15859294
55625000 3583949.68106785
55850000 3513141.39786891
56025000 3499502.22810752
56050000 3494773.4262492
56550000 3407777.69926059
56625000 3400429.79252003
56950000 3372798.27140864
57250000 3368156.03675803
57350000 3369724.35922269
57975000 3300955.31750977
58775000 3254819.89514487
59325000 3277907.35888267
59575000 3274406.51724199
60400000 3155095.81420348
60425000 3143721.386474
61075000 3090246.81976986
62575000 2947854.00493838
62650000 2973502.65446395
62725000 3045643.1768927
63325000 3174466.03879912
63525000 3158827.60150423
63950000 3254763.57117981
64575000 3238075.0464567
64825000 3231346.47967222
65500000 3353064.1915614
65525000 3354229.59957
65625000 3348600.650928
66600000 3764371.3047213
66775000 3850051.77254186
66925000 3791791.6228042
67075000 3874605.65893569
67525000 3840099.81042921
68225000 3778102.94704311
68450000 3785590.51709123
68975000 3800863.50022485
69650000 3770469.32978902
69850000 3750008.37723541
70225000 3725678.41938527
70975000 3734479.51766296
71300000 3816634.51811772
71400000 3788892.58101872
71750000 3728702.10057508
72325000 3738018.70279408
72800000 3733398.14604228
73925000 3940882.11278937
75525000 4266463.8691936
75700000 4250786.33895862
78600000 4345452.78972583
}\DataTableTraceOneThreeBlocs

\pgfplotstableread{
x       y
175000 90127.6730756692
225000 116623.441439761
275000 143330.707493002
475000 253834.557424182
850000 475215.17248238
1550000 942237.171393933
1575000 959555.710127062
2350000 1546829.17623694
2700000 1853749.71794082
3150000 2225010.98940981
3625000 2674388.29835162
3675000 2717027.30665371
4300000 3275382.33584667
4575000 3522978.0932433
4675000 3610140.07051452
4700000 3633409.16014562
5000000 3894727.79627073
5225000 4106370.29792176
5450000 4356267.95243437
5525000 4444444.56364653
5600000 4517588.83725232
5650000 4545779.59090531
5950000 4826995.97743761
5975000 4841714.31221276
6350000 5280986.26480043
6375000 5302497.31246205
6400000 5332261.39589073
6550000 5504474.13175569
6575000 5525718.86511614
6675000 5613021.18630267
6925000 5840938.78795679
6975000 5888900.22333909
7025000 5934646.7702366
7275000 6186941.59139047
7400000 6356226.18300222
7425000 6388763.93708567
7825000 6751026.30446972
8050000 6962755.5392178
8300000 7227485.88351665
8475000 7397925.10454976
8700000 7670199.29187157
8825000 7757996.57107837
8950000 7869064.12947673
9250000 8137612.12840028
9575000 8415237.19514165
9875000 8752590.66923056
10075000 8903205.77692305
10225000 9039944.64698285
10650000 9473435.28195372
10875000 9631673.140596
11175000 9868176.497658
11275000 9981505.92104537
11300000 10010454.4313139
11375000 10103714.5996182
11450000 10169307.6982062
11500000 10216863.9825041
11550000 10248263.6100903
11625000 10318814.9169808
11800000 10435397.0977794
12125000 10690357.8881915
12200000 10795695.9611592
12500000 10977329.8410038
12825000 11215843.932198
13375000 11606966.463845
13500000 11636261.6070306
13550000 11618170.4516119
13650000 11615317.3543908
13725000 11692928.8444516
13750000 11694678.1707464
13875000 11710853.3240295
13950000 11812798.3076322
14175000 12017068.0378488
14700000 12324033.9540178
15000000 12358121.8985222
15075000 12467327.5855294
15125000 12584905.6486402
15150000 12571843.6945257
15375000 12850255.6932479
15700000 13005902.3309936
15775000 13070360.7129805
16025000 13300165.9537506
16075000 13259576.0999593
16275000 13364687.013381
16375000 13392885.6736515
16575000 13515937.3913079
16750000 13588578.231519
17100000 13902097.1033017
17175000 13880738.3219216
17250000 13930534.4913549
17375000 13962589.5135157
17725000 14144123.2463937
17975000 14161040.8730086
18175000 13935577.6148839
18200000 13965985.2002802
18250000 13988491.8099259
18450000 14186251.0939049
18525000 14227883.61703
18900000 14332976.3713395
19050000 14253059.1468358
19200000 14188549.9039111
19450000 14210411.8546438
19525000 13931566.1324456
19800000 13968312.2384772
20000000 13796690.8320854
20525000 13451173.5223231
20875000 13077794.1893787
20975000 12985602.0274739
21225000 12660536.8406361
21525000 12620201.9748609
22025000 12606574.2101267
22275000 12625861.3886827
22475000 12853267.0060099
22700000 12591535.1062634
22750000 12732119.9619408
22925000 12468913.1764497
23125000 12392220.827361
23225000 12535310.6162086
23250000 12595463.8400124
23525000 12837767.6262094
23750000 13110323.6597423
23775000 12993874.7356856
24075000 13038403.7321642
24150000 13029477.1878586
24350000 12657263.6526514
24375000 12624822.7530131
25425000 11095195.4738594
25500000 10975978.953843
25650000 10613609.2044325
26000000 9929736.9279861
26025000 9920526.3999763
26050000 9902686.94673636
26225000 9824696.58637972
26450000 9593703.78231636
27025000 9276418.69615355
27500000 9106312.08978823
27600000 9160663.94572635
27900000 9119874.01055829
27925000 9082229.42210684
28050000 9094988.28749306
28200000 9066807.84799705
28225000 9031306.71371581
28650000 9019512.7523641
28800000 8865004.69576943
28850000 8822865.91288381
28875000 8811143.05184553
28925000 8847037.95650219
29275000 8578000.04567738
29450000 8461340.14493559
30200000 8108655.74307224
30825000 7678593.84767116
30850000 7618621.41928399
31075000 7349781.42896258
31175000 7431465.74774898
31225000 7449169.30771753
31625000 7171400.2594785
31650000 7193345.84262627
31700000 7137061.87616192
31800000 7181988.31801384
32025000 7244087.85736392
32575000 7113557.32433615
32650000 7062693.16760003
32750000 7077805.87043915
33100000 6956778.73575932
33225000 7019527.77483051
33500000 6947630.40853976
33600000 7022548.28287806
33725000 6936023.23583276
34425000 6832657.62643953
34450000 6843361.81863821
34625000 6890417.50738221
34725000 6849238.24321109
34775000 6831010.42215597
35100000 6978059.44717272
35175000 7042468.46588435
35700000 6788324.81336494
35825000 6732721.62957916
35900000 6685893.61957863
35950000 6761795.41802384
36200000 6765075.30114537
36450000 6543496.72683477
36550000 6543582.23057381
36700000 6496227.35608062
36725000 6462477.80007904
36750000 6455860.58831278
36925000 6473476.80758648
36950000 6453326.9355423
37050000 6365569.35220983
37250000 6427548.86960447
37425000 7056189.37825257
37500000 6946997.11233414
37700000 6763886.07508696
37950000 6493636.9853463
38000000 6469244.50725327
38550000 6310946.41528859
39025000 5907793.05097912
39325000 5785494.69993371
39575000 5662778.86176221
40275000 5582031.86860833
40300000 5513756.73392974
40525000 5449806.20695686
41075000 5400930.51883681
41200000 5370728.58112125
41250000 5351574.95301927
42325000 5116982.97807213
42775000 5177739.46626766
42925000 5100991.86865034
42950000 5112737.12612779
42975000 5128990.64653051
43325000 5169491.41918357
43775000 5060203.43988687
43800000 5036473.9606734
43950000 4982127.84699195
44275000 4969417.61720581
44500000 4903271.99417626
44675000 4882614.32705491
44775000 4892358.70362066
44875000 4870750.27743139
45200000 4875930.11112517
45350000 4805834.12212092
45375000 4797876.59031462
45650000 4807830.71910135
45925000 4672375.26218699
46175000 4584173.26824346
46325000 4561847.16262951
48450000 4407912.78409966
48525000 4370970.57295239
49250000 4479526.16452603
49350000 4572348.6484516
49900000 4644063.14476146
49950000 4700358.6550956
50325000 4712118.6288574
50725000 4684729.89465944
50775000 4713167.9110529
50800000 4752615.59907147
51125000 4734766.20564514
51300000 4999640.79204561
51725000 5158066.70770739
52775000 5075363.80442144
53675000 5105626.84594395
53825000 5081496.90778818
54350000 5092438.78088013
56375000 4730876.63208826
56950000 4911991.84576843
56975000 4879893.96643963
57425000 4828785.14904663
63525000 5774977.20473744
65275000 6061166.5390499
66650000 6586256.14755597
66800000 6837476.29829022
69175000 7092545.33169463
}\DataTableTraceOneFourBlocs

\pgfplotstableread{
x       y
150000 77002.8011245076
225000 116613.834431654
250000 129936.281943775
275000 143362.620053937
375000 198034.387415623
650000 354960.807183433
675000 369796.607109731
1000000 567045.100727133
1300000 760116.904654982
1325000 776626.160529182
1650000 1000158.20597965
2100000 1327318.59851066
2275000 1459915.29652909
2600000 1719434.39794663
2950000 2026223.32612952
3100000 2169724.42988374
3225000 2272193.35012979
3250000 2292649.18237299
3575000 2568753.79781578
3625000 2612402.14983008
3675000 2653915.04916064
3925000 2871744.72014997
4025000 2966495.73782573
4375000 3282376.78344168
4650000 3539323.65849268
4700000 3583027.24385834
4950000 3810881.59139983
5025000 3877692.14280463
5125000 3970390.10316715
5200000 4039847.9736927
5375000 4216347.49872705
5800000 4628339.90242239
5925000 4752651.69755912
6025000 4859865.98327329
6275000 5106151.90270473
6300000 5132308.69648998
6575000 5395150.44239063
6675000 5490274.65850596
6750000 5593822.92365264
6775000 5616542.62552941
7025000 5856773.33173632
7075000 5908645.26915107
7250000 6076644.00837453
7700000 6583319.15722893
7875000 6741636.11605931
7950000 6848141.93619047
8000000 6890097.08807335
8375000 7281786.00023134
8550000 7439897.21906407
9050000 7964032.23488255
9425000 8311474.22094995
9625000 8517613.018185
9775000 8770209.00193168
9850000 8859185.47247445
10025000 8969958.32583023
10075000 9002597.28407243
10100000 9004507.76941307
10150000 9040408.37389064
10400000 9297763.75269727
10700000 9587389.2299673
10750000 9632064.31051781
10775000 9650262.63351673
11300000 10134269.5024003
11325000 10156966.3925032
11475000 10260980.7052276
11500000 10284146.4923417
11550000 10285267.8637145
11625000 10350478.4835332
11875000 10603471.0793744
12500000 11142590.2068725
12550000 11198584.7069425
12800000 11529770.2982446
12975000 11641181.1779803
13625000 11914325.3426097
13850000 12021053.551807
14075000 12141659.5071375
14125000 12203362.8518427
14425000 12490928.5271759
14900000 12869395.8162791
14975000 12941412.4075732
15025000 13010261.6465916
15250000 13282651.9758011
15600000 13675241.7713508
15775000 13856347.9648093
15825000 13902649.431064
16200000 14228194.5603532
16500000 14435632.7254231
16700000 14543541.4192222
16900000 14733708.9260911
17175000 14985796.7588016
17500000 15203083.0272227
17775000 15296407.9706188
17925000 15363500.1701143
18050000 15404499.5448925
18100000 15565193.5985793
18125000 15536893.2177552
18300000 15731084.2136774
18600000 15851238.4368479
19150000 16119618.4834507
19900000 16093872.8507623
20700000 15898664.053205
20875000 15771281.1736699
21275000 15520479.8019115
21375000 15584276.5426118
21675000 15857250.7824386
21825000 15776547.0147791
22225000 15740128.8302229
22350000 15501897.6818441
22575000 15600960.4732397
22625000 15498803.7980805
22650000 15552048.5423787
22850000 15441255.2501351
23025000 15179690.5761699
23350000 15620226.4216317
23450000 15719449.2202693
23500000 15797183.026646
23575000 15742217.7076117
23800000 15883950.2055054
24200000 15698449.2594485
24425000 15431510.5544552
24525000 15438219.8660796
24775000 15031984.8629692
24850000 15086437.1468183
25000000 14817946.6701637
25250000 14606990.6172906
25675000 13988596.0921851
26225000 14003705.6377355
26500000 13758091.1457443
26825000 13959749.0015547
26950000 13788185.1030022
27050000 13681709.9474819
27125000 13633883.4116014
27200000 13554260.6899228
27325000 13805543.7685686
27350000 13754863.1717799
27550000 13686102.5696744
28875000 13566853.9844574
29875000 11652412.8241952
29900000 11787169.4354924
30300000 11317650.3504228
31425000 10551311.9185193
31575000 10590809.5998118
32050000 10493125.6286853
32075000 10507672.1849857
32250000 10657101.4526014
32275000 10588579.855491
32700000 10525357.9241017
32750000 10464164.350988
32875000 10295639.600938
33175000 10253389.9835245
33575000 10131888.8593761
33975000 9626183.87575262
34350000 9549856.54028738
34375000 9629271.11792893
34400000 9573182.46855769
34525000 9475140.25515674
34575000 9506913.79498236
35225000 9231930.18874925
35250000 9227382.53798946
35300000 9158345.38445733
35375000 9017948.02193088
35475000 9043892.3771422
35750000 8942227.92507112
35775000 8872895.02457949
35900000 8739929.90905862
36225000 8601726.33787721
36275000 8500897.57581428
36525000 8396350.54997374
36875000 8383260.23664485
37050000 8238743.6063096
37175000 8226683.86531426
37250000 8212303.45272285
37425000 8489323.64551991
37550000 8341953.52487394
38075000 8054086.61599778
38450000 7773424.68256776
38650000 7896982.8723812
38975000 7651148.25057684
39000000 7642792.40252631
39050000 7586182.30410705
39200000 7443326.63001772
39875000 7028637.41011782
39925000 7082924.26962498
40100000 7137098.49029492
40175000 7099966.32302385
40200000 7089550.31669005
40575000 6997567.4135348
40925000 6872879.04450991
41150000 6852715.46508707
41400000 6858952.44083899
41475000 6794318.58866404
41725000 6589855.56587123
42025000 6651710.33365623
42400000 6565498.47536551
42450000 6497930.41974129
42550000 6427094.67514462
42625000 6374521.96786865
42675000 6367605.69768229
42750000 6322499.32562604
42875000 6215446.1661652
42975000 6177910.53390222
43150000 6189595.73497786
43275000 6241962.75863467
43475000 6201507.53235258
43600000 6115077.87380929
43775000 6191294.7537634
44250000 5980829.8464954
44325000 6052542.71216896
44750000 5870232.4412669
44900000 5815712.15036156
44975000 5810806.24661853
45000000 5796705.47054104
45175000 5931403.92789603
45625000 5811460.39222841
45825000 5720192.56574367
46150000 5764558.75369001
46225000 5666584.32687718
46600000 5607459.1790638
46750000 5784530.85013405
46775000 5769760.63738632
46900000 5642500.34183216
47275000 5448803.32456759
47750000 5256894.80465487
47900000 5262026.01007981
47925000 5282868.63733315
48000000 5236628.83250014
48100000 5355993.69894936
48175000 5311197.52107914
48300000 5259001.85407657
48350000 5260497.74517163
48500000 5245188.01285833
49075000 5588033.62976862
49350000 5668663.8119061
50125000 5552836.60335241
50175000 5584166.28542138
50200000 5571241.45629547
50550000 5639878.6252947
50725000 5696245.72583477
51725000 6118632.15588573
54175000 5975355.76572861
55675000 5762275.01242663
56225000 5619756.55213224
56375000 5585055.4432669
56400000 5548583.76790873
58350000 5763800.49739812
63225000 6199655.92577073
64400000 6891411.89152347
66000000 6971791.42531546
66175000 7135089.63266222
68850000 7642903.71544036
}\DataTableTraceOneFiveBlocs

\begin{tikzpicture}
  \pgfplotsset{every tick label/.append style={font=\scriptsize}}
    \definecolor{darkgray176}{RGB}{176,176,176}
    \definecolor{lightgray204}{RGB}{204,204,204}
    \begin{axis}[
        x grid style={lightgray204},
        width=\linewidth,
        height=0.8\linewidth,
        legend style={
            font=\tiny, 
            row sep=-3pt, 
            column sep=1pt,
            fill opacity=0.8,
            draw opacity=1,
            text opacity=1,
            draw=lightgray204,
            at={(1.28,1.1)}, 
            anchor=north east,
            font=\scriptsize, 
            /tikz/every even column/.append style={column sep=5mm} 
        },
        tick align=outside,
        xlabel={Input Rate (Mpps)},
        ylabel={Throughput (Mpps)},
        xticklabel style={align=center, text width=1.5cm},
        xmajorgrids,
        ymajorgrids,
        ymajorticks=true,
        xtick style={color=black},
        xmax=80,xmin=0,
        y grid style={darkgray176},
        ytick style={color=black},
        label style={font=\footnotesize},
        tick label style={font=\scriptsize},
        ylabel style={font=\fontsize{5}{5}\selectfont},
        xlabel style={font=\fontsize{5}{5}\selectfont},
        tick label style={font=\fontsize{6}{6}\selectfont}
    ]
    \addplot [color=color0, mark=*, mark size=1.0] table [x expr=\thisrow{x}/1000000, y expr=\thisrow{y}/1000000] {\DataTableTraceOneThreeBlocs};
    \addlegendentry{\#PMU=3}

    \addplot [color=color1, mark=square*, mark size=1.0] table [x expr=\thisrow{x}/1000000, y expr=\thisrow{y}/1000000] {\DataTableTraceOneFourBlocs};
    \addlegendentry{\#PMU=4}

    \addplot [color=color2, mark=triangle*, mark size=1.0] table [x expr=\thisrow{x}/1000000, y expr=\thisrow{y}/1000000] {\DataTableTraceOneFiveBlocs};
    \addlegendentry{\#PMU=5}
    \end{axis}
\end{tikzpicture}

%% file: figures/eval-results/throughput/throughput-trace-2.tex
\pgfplotstableread{
x       y
350000 0.00181888788248113
375000 0.00235883745926224
575000 0.00730331796007602
1900000 0.0910531332321528
2025000 0.0919831545333449
2200000 0.0999681109957054
2550000 0.111522751764545
2675000 0.115011554555283
2800000 0.110187581612686
3550000 0.123261658709107
3650000 0.135631053640985
4775000 0.107847582147259
5225000 0.124301324845394
5675000 0.144347286326928
5800000 0.149962296178047
5950000 0.136353499167103
6025000 0.124970951167598
6525000 0.143676814078269
6550000 0.144703995545451
7250000 0.150716208338149
7400000 0.135769453183942
8250000 0.135121448777802
8275000 0.138533448777802
8300000 0.142130041370394
8375000 0.144240012066365
8575000 0.150671744262484
9450000 0.153412292235125
9475000 0.155914870632338
9500000 0.156283879641347
9625000 0.159993110831377
10025000 0.167304630286023
11950000 0.187044954345172
13275000 0.222259624557727
13325000 0.218005333750495
14300000 0.215230193552848
15225000 0.237795169361145
15550000 0.232612027832393
15925000 0.250064102391042
16425000 0.236016023443223
16875000 0.249852733145191
17500000 0.262807843640978
17675000 0.270723188405797
18775000 0.266311289471566
18800000 0.270285184208408
18825000 0.272538517541741
18950000 0.2787592350493
20100000 0.274261563041323
20575000 0.275584796672799
20700000 0.278267985128787
21350000 0.283839293020705
21375000 0.285883600713012
21450000 0.282245036610448
22000000 0.295109411153552
22175000 0.280437398200241
22850000 0.286281672041065
22925000 0.286683776853899
23500000 0.283669120255183
23725000 0.281367969632521
24400000 0.289432882391784
25125000 0.279159824138351
25500000 0.279843324111643
25825000 0.287660326373042
26000000 0.273330528305681
26750000 0.26625159752322
27100000 0.283153209690894
27650000 0.285546963938023
27750000 0.283400989912049
27775000 0.287383535366594
28175000 0.281537991685095
28225000 0.281238623264042
28350000 0.281261495739348
29875000 0.278070258434376
30400000 0.277450875408733
32275000 0.275886746897409
32325000 0.275440276309174
32725000 0.27539322157581
32950000 0.281213714294649
33650000 0.2853416352014
33850000 0.276950124607419
34775000 0.285238708624709
34850000 0.287177263951735
34925000 0.281330101558572
35075000 0.285782532531108
35775000 0.280012103347633
36175000 0.289338571428571
36675000 0.279991872917588
38275000 0.280834435786436
38400000 0.279049050505051
38450000 0.277690828282828
38625000 0.281752288600289
38875000 0.288081564517354
39250000 0.284774204240204
40425000 0.274614091163738
43300000 0.285300170940171
43400000 0.279733555555556
43700000 0.285272351204351
43925000 0.282046116550117
44100000 0.276915108225108
44850000 0.276332063492064
45300000 0.287474043290043
45975000 0.263735688311688
46575000 0.27953902402173
46825000 0.289859465423465
47100000 0.280804820512821
47775000 0.265776444444444
48525000 0.273192155844156
48550000 0.265030251082251
48900000 0.275436825396825
49525000 0.282609142857143
49900000 0.278261142857143
50050000 0.28007980952381
50800000 0.296208
50975000 0.279972380952381
51000000 0.275305714285714
51025000 0.274985714285714
51950000 0.277595047619048
52750000 0.270997333333333
52800000 0.261104
53100000 0.288762666666667
53300000 0.284717333333333
54000000 0.284646666666667
54175000 0.292929904761905
56675000 0.283265523809524
56800000 0.273566857142857
57925000 0.28136673015873
58475000 0.27087580952381
58700000 0.283901333333333
58850000 0.294213333333333
58975000 0.296693333333333
59400000 0.27895204040404
59925000 0.275
61350000 0.256373333333333
62325000 0.30768
62525000 0.304697142857143
63300000 0.296986666666667
63325000 0.292986666666667
63375000 0.28464
64725000 0.2746
66925000 0.250453333333333
67300000 0.25584
68900000 0.312266666666667
68975000 0.33088
69850000 0.27136
70150000 0.31536
70750000 0.314106666666667
70975000 0.25976
76925000 0.268053333333333
77975000 0.28128
78300000 0.29968
78625000 0.28912
}\DataTableTraceTwoThreeBlocs

\pgfplotstableread{
x       y
225000 0.00033981809535317
425000 0.00326962362280046
625000 0.00976502261840084
1200000 0.0371916450017416
1550000 0.051793865008057
2125000 0.079369703760292
2450000 0.0800058874458874
2600000 0.0817932813852814
2975000 0.0858263416583416
3425000 0.079800673992674
3475000 0.0882968394742512
3700000 0.11791793837535
4800000 0.0796926007326007
5000000 0.0793140634920635
5125000 0.0876706406926407
5200000 0.0949857522477523
6900000 0.0975492773892774
7225000 0.0990040404040404
7450000 0.110061414141414
7825000 0.101879558663559
8150000 0.0896198499278499
9200000 0.130149206349206
9250000 0.129694095238095
9400000 0.143441858585859
9425000 0.142401858585859
9625000 0.114287930735931
10250000 0.15869038961039
10300000 0.151507157287157
10775000 0.154837915417915
11350000 0.145226290598291
11475000 0.135881811521812
11650000 0.137453811965812
12425000 0.126675601825076
13300000 0.186727111111111
14000000 0.22820303030303
14325000 0.233239555555556
14525000 0.233799619047619
14600000 0.22953158974359
15525000 0.207269079365079
15750000 0.208560426240426
16350000 0.228101171717172
16700000 0.238026666666667
17075000 0.210547428571429
17650000 0.22146111976912
17725000 0.213275341991342
17775000 0.199532611832612
17800000 0.201702793650794
17925000 0.200689633477633
18425000 0.208045858585859
19075000 0.241593934065934
20075000 0.267513766233766
20275000 0.26835301010101
20375000 0.26731898989899
20750000 0.267945714285714
21000000 0.276080507936508
21200000 0.259201396825397
22500000 0.240579047619048
23175000 0.278771238095238
23300000 0.258156190476191
23900000 0.235205714285714
24500000 0.272908571428571
24825000 0.298176317460317
24900000 0.288938793650794
25350000 0.298104380952381
25475000 0.295593333333333
25550000 0.283433333333333
26000000 0.28545995959596
26200000 0.275296277056277
26675000 0.272044126984127
27275000 0.295553523809524
27475000 0.265891601731602
27725000 0.298541333333333
27925000 0.298456888888889
28050000 0.280892698412698
28650000 0.293354666666667
29025000 0.288719492063492
29100000 0.269155301587302
29150000 0.277670857142857
29475000 0.288607238095238
29500000 0.289452952380952
29525000 0.290028952380952
30025000 0.276076380952381
30500000 0.294813333333333
30700000 0.293335111111111
30875000 0.261124444444444
31075000 0.255586095238095
31675000 0.251361904761905
31700000 0.257453333333333
32175000 0.283855619047619
32250000 0.269876952380952
32500000 0.282069333333333
32525000 0.285383619047619
32750000 0.293614476190476
32950000 0.280890666666667
33175000 0.251106666666667
33300000 0.256994666666667
33500000 0.277449333333333
33575000 0.302034222222222
33675000 0.301872317460317
34300000 0.278974476190476
34350000 0.26209980952381
35000000 0.269
35025000 0.269133333333333
36000000 0.281071746031746
36350000 0.273647792207792
37775000 0.272669968253968
37800000 0.273003301587302
38100000 0.291504
38750000 0.272826666666667
39200000 0.278501333333333
39675000 0.271149333333333
39725000 0.287813333333333
39925000 0.312058857142857
40525000 0.296010285714286
40700000 0.276852952380952
41000000 0.309893333333333
41200000 0.303973333333333
41650000 0.277893333333333
42400000 0.279826666666667
43025000 0.287941333333333
43075000 0.285488
43725000 0.304913333333333
44150000 0.291331428571429
44675000 0.275378666666667
44775000 0.293408
44800000 0.296528
45450000 0.26112
45825000 0.254730666666667
46425000 0.260906666666667
47000000 0.307586666666667
47475000 0.327893333333333
47500000 0.314613333333333
48850000 0.27016
49175000 0.301893333333333
49275000 0.327605333333333
50100000 0.261973333333333
50175000 0.299173333333333
50800000 0.27016
50925000 0.28424
53025000 0.31024
55525000 0.28912
56450000 0.27344
56975000 0.2968
57100000 0.30096
57575000 0.29648
58350000 0.26752
63825000 0.27776
65275000 0.25432
65525000 0.23928
67200000 0.289226666666667
}\DataTableTraceTwoFourBlocs

\pgfplotstableread{
x       y
650000 0.00965537542481945
900000 0.0198644137666758
1250000 0.0349323373107097
1500000 0.0501315462576639
1975000 0.0757837354177508
2400000 0.0688561038961039
2850000 0.0770782857142857
2875000 0.0671382857142857
2925000 0.0843306666666667
3525000 0.0751228804137039
3600000 0.0938685507433743
3975000 0.081698897990898
4225000 0.0780309956709957
4250000 0.075967619047619
4650000 0.06818778998779
4900000 0.0881143492063492
5550000 0.0909401904761905
6225000 0.112002961038961
6400000 0.0874185006105006
6425000 0.0911945006105006
6525000 0.0936558974358974
7175000 0.0971971717171717
7525000 0.0798138239538239
7550000 0.0780204906204906
7975000 0.100680909090909
8125000 0.0797746031746032
8300000 0.0773645650311161
9450000 0.120596467532468
9525000 0.106631861471861
9775000 0.12523619047619
10125000 0.0980674343434343
10225000 0.118291751803752
10400000 0.0952933275613276
10875000 0.0937223807303807
11075000 0.118497076923077
11125000 0.118320188034188
11175000 0.127543902319902
11800000 0.0782907936507937
11975000 0.121812738816739
12500000 0.108537788445473
12625000 0.122374126984127
13600000 0.16197675036075
14000000 0.184859134199134
14125000 0.175814256854257
14300000 0.199746222222222
14325000 0.207362222222222
14900000 0.163633333333333
15300000 0.172157777777778
15500000 0.172618095238095
16300000 0.162784588744589
16675000 0.192201333333333
17375000 0.192528935508936
18550000 0.176790505050505
18600000 0.180012727272727
19475000 0.233050857142857
19875000 0.218428455988456
20000000 0.222406095238095
20500000 0.248143111111111
20550000 0.234934857142857
20700000 0.225703428571429
20875000 0.247147746031746
20900000 0.257554412698413
20975000 0.243112126984127
21100000 0.248380952380952
21225000 0.243054776334776
21425000 0.237196173160173
21650000 0.255503555555556
21750000 0.260466871794872
22175000 0.235329523809524
22675000 0.265959428571429
22850000 0.246787809523809
23075000 0.269801333333333
23100000 0.270821333333333
23425000 0.235809924297924
23575000 0.229368971916972
23775000 0.233359428571429
23800000 0.233647428571429
24025000 0.23397980952381
24250000 0.260720929292929
24475000 0.290982222222222
24625000 0.280975238095238
24750000 0.278325904761905
25050000 0.241067353535354
25100000 0.253989258297258
25450000 0.280108380952381
25675000 0.266238222222222
25800000 0.267632
26350000 0.286624
27125000 0.274238222222222
27875000 0.265250158730159
28350000 0.270360571428571
28950000 0.294127111111111
28975000 0.293047111111111
29250000 0.273388952380952
29400000 0.267595047619048
29900000 0.285666666666667
30975000 0.300904
31125000 0.276744
31200000 0.280605818181818
31525000 0.275669396825397
31575000 0.275518285714286
32125000 0.294249523809524
32150000 0.298249523809524
32650000 0.28257219047619
32700000 0.281270857142857
33025000 0.251514285714286
33225000 0.270850666666667
33400000 0.279698666666667
33800000 0.294916317460317
34100000 0.272277333333333
35100000 0.271194666666667
36150000 0.269187301587302
36275000 0.260386493506494
36375000 0.274415826839827
37225000 0.280201523809524
38400000 0.297143619047619
38575000 0.31328380952381
38825000 0.267453333333333
39225000 0.261229333333333
39425000 0.275297142857143
39650000 0.235784
41750000 0.268130666666667
41975000 0.286026666666667
42175000 0.284296
42800000 0.283090666666667
42825000 0.286184
43325000 0.292272
43550000 0.291354666666667
44400000 0.279013333333333
44500000 0.280066666666667
44850000 0.280792
46225000 0.27232
46250000 0.271424
46575000 0.282853333333333
48025000 0.288133333333333
48775000 0.28588
48975000 0.31476
49500000 0.281450666666667
50375000 0.25656
50425000 0.270853333333333
50600000 0.281893333333333
51400000 0.26928
53825000 0.27848
54175000 0.30688
55900000 0.26008
56975000 0.28528
58350000 0.27288
66000000 0.25608
66450000 0.29952
69025000 0.27368

}\DataTableTraceTwoFiveBlocs

\begin{tikzpicture}
    \pgfplotsset{every tick label/.append style={font=\scriptsize}}
    \definecolor{darkgray176}{RGB}{176,176,176}
    \definecolor{lightgray204}{RGB}{204,204,204}
    \definecolor{lightcoral246112136}{RGB}{246,112,136}
    \definecolor{mediumseagreen51176122}{RGB}{51,176,122}
    \definecolor{peru17315649}{RGB}{173,156,49}
    \begin{axis}[
        width=\linewidth,
        height=0.8\linewidth,
        x grid style={lightgray204},
        legend style={
            fill opacity=0.8,
            draw opacity=1,
            text opacity=1,
            draw=lightgray204,
            at={(0.98,0.98)}, 
            anchor=north east,
            font=\scriptsize, 
            /tikz/every even column/.append style={column sep=5mm} 
        },
        tick align=outside,
        xlabel={Input Rate (Mpps)},
        ylabel={Memory BW (GB/s)},
        xticklabel style={align=center, text width=1.5cm},
        xmajorgrids,
        xmax=80,xmin=0,
        ymax=0.8,ymin=0,
        ymajorgrids,
        ymajorticks=true,
        xtick style={color=black},
        y grid style={darkgray176},
        ytick style={color=black},
        label style={font=\footnotesize},
        tick label style={font=\scriptsize},
        ylabel style={font=\fontsize{5}{5}\selectfont},
        xlabel style={font=\fontsize{5}{5}\selectfont},
        tick label style={font=\fontsize{6}{6}\selectfont}
    ]
    \addplot [color=color0, mark=*, mark size=1.0] table 
    [x expr=\thisrow{x}/1000000, y expr=\thisrow{y} * 2] {\DataTableTraceTwoThreeBlocs};

    \addplot [color=color1, mark=square*, mark size=1.0] table [x expr=\thisrow{x}/1000000, y expr=\thisrow{y} * 2] {\DataTableTraceTwoFourBlocs};

    \addplot [color=color2, mark=triangle*, mark size=1.0] table [x expr=\thisrow{x}/1000000, y expr=\thisrow{y} * 2] {\DataTableTraceTwoFiveBlocs};
    \end{axis}
\end{tikzpicture}

%% file: figures/eval-results/adapt-vs-static/adapt-static-t1.tex
\begin{tikzpicture}

\definecolor{color0}{rgb}{0.258823529411765,0.403921568627451,0.823529411764706}
\definecolor{color1}{rgb}{0.917647058823529,0.262745098039216,0.207843137254902}
\definecolor{color2}{rgb}{0.984313725490196,0.737254901960784,0.0156862745098039}
\definecolor{color3}{rgb}{0.203921568627451,0.658823529411765,0.325490196078431}

\begin{groupplot}[group style={group size=1 by 3}]
\nextgroupplot[
width=\linewidth,
height=0.7\linewidth,
legend cell align={left},
legend style={
  fill opacity=0.8,
  draw opacity=1,
  text opacity=1,
  at={(0.97,0.03)},
  anchor=south east,
  draw=white!80!black
},
scaled x ticks=manual:{}{\pgfmathparse{#1}},
tick align=outside,
tick pos=left,
x grid style={white!69.0196078431373!black},
xmajorgrids,
xmin=-25, xmax=525,
xminorgrids,
xtick style={color=black},
xticklabels={},
y grid style={white!69.0196078431373!black},
ylabel={Hit Rate},
ymajorgrids,
ymin=0.0814735469448584, ymax=1.03905551415797,
yminorgrids,
ytick style={color=black},
ylabel style={font=\fontsize{5}{5}\selectfont},
xlabel style={font=\fontsize{5}{5}\selectfont},
tick label style={font=\fontsize{6}{6}\selectfont}
]
\addplot [semithick, color0, mark=*, mark size=0.5, mark options={solid}]
table {%
1 0.125
6 0.814814814814815
11 0.923076923076923
16 0.946153846153846
21 0.924050632911392
26 0.961165048543689
31 0.986238532110092
36 0.96958174904943
41 0.96742671009772
46 0.978260869565217
51 0.972527472527472
56 0.98
61 0.977011494252874
66 0.980237154150198
71 0.990689013035382
76 0.973867595818815
81 0.983812949640288
86 0.987823439878234
91 0.986466165413534
96 0.98465829846583
101 0.980337078651685
106 0.978523489932886
111 0.99
116 0.99055330634278
121 0.991428571428571
127 0.981279251170047
132 0.988636363636364
137 0.988993710691824
142 0.98169014084507
147 0.989296636085627
152 0.991735537190083
157 0.985974754558205
162 0.986706056129985
167 0.992774566473988
172 0.979259259259259
177 0.984567901234568
182 0.986404833836858
187 0.991379310344828
192 0.986425339366516
197 0.982808022922636
202 0.979197622585438
207 0.987915407854985
212 0.995529061102832
217 0.988522238163558
222 0.992412746585736
227 0.9860529986053
232 0.981981981981982
237 0.990801576872536
242 0.974242424242424
247 0.994310099573257
253 0.987293519695044
258 0.986754966887417
263 0.991379310344828
268 0.995092024539877
273 0.99009900990099
278 0.988341968911917
283 0.990527740189445
288 0.986899563318777
293 0.991880920162382
298 0.992548435171386
303 0.989473684210526
308 0.991724137931034
313 0.994535519125683
318 0.985123966942149
323 0.989329268292683
328 0.98169014084507
333 0.989345509893455
338 0.986706056129985
343 0.98079561042524
348 0.993270524899058
353 0.990338164251208
358 0.992260061919505
363 0.984615384615385
368 0.985632183908046
373 0.982404692082112
379 0.988522238163558
384 0.982093663911846
389 0.992138364779874
394 0.991202346041056
399 0.989079563182527
404 0.985623003194888
409 0.982945736434108
414 0.990811638591118
419 0.986024844720497
424 0.985380116959064
429 0.990415335463259
434 0.988745980707396
439 0.98
444 0.988235294117647
449 0.981001727115717
454 0.978886756238004
459 0.980988593155894
464 0.975700934579439
469 0.972875226039783
474 0.983812949640288
479 0.983546617915905
484 0.989473684210526
489 0.985559566787004
494 0.992331288343558
500 0.990014265335235
};
\addplot [semithick, color1, mark=*, mark size=0.5, mark options={solid}]
table {%
1 0.125
6 0.814814814814815
11 0.923076923076923
16 0.946153846153846
21 0.930379746835443
26 0.961165048543689
31 0.986238532110092
36 0.954372623574144
41 0.944625407166124
46 0.923913043478261
51 0.793956043956044
56 0.666666666666667
61 0.880165289256198
66 0.803174603174603
71 0.944275582573455
76 0.970383275261324
81 0.976618705035971
86 0.980213089802131
91 0.972932330827068
96 0.97071129707113
101 0.952247191011236
106 0.959731543624161
111 0.9575
116 0.947368421052632
121 0.937142857142857
127 0.940717628705148
132 0.939935064935065
137 0.937106918238994
142 0.956338028169014
147 0.94954128440367
152 0.964187327823691
157 0.948106591865358
162 0.954209748892171
167 0.947976878612717
172 0.948148148148148
177 0.938271604938272
182 0.944108761329305
187 0.945402298850575
192 0.950226244343891
197 0.932664756446991
202 0.931649331352155
207 0.930513595166163
212 0.956780923994039
217 0.959827833572453
222 0.963581183611533
227 0.947001394700139
232 0.951093951093951
237 0.948751642575558
242 0.943939393939394
247 0.951635846372688
253 0.95425667090216
258 0.908902691511387
263 0.945812807881773
268 0.98849472674976
273 0.981553398058252
278 0.984652665589661
283 0.984232365145228
288 0.979033728350046
293 0.985306828003457
298 0.976811594202898
303 0.973496432212028
308 0.990152193375112
313 0.983420593368237
318 0.97914597815293
323 0.979087452471483
328 0.969135802469136
333 0.983855650522317
338 0.975609756097561
343 0.987057808455565
348 0.990376202974628
353 0.983281086729362
358 0.984472049689441
363 0.97029702970297
368 0.981132075471698
373 0.973790322580645
379 0.978199052132701
384 0.965225563909774
389 0.97410358565737
394 0.975728155339806
399 0.982273201251304
404 0.974630021141649
409 0.977933801404213
414 0.96969696969697
419 0.974257425742574
424 0.972677595628415
429 0.982688391038696
434 0.985200845665962
439 0.975215517241379
444 0.977917981072555
449 0.969727547931382
454 0.97055359246172
459 0.980414746543779
464 0.968936678614098
469 0.965669988925803
474 0.972406181015452
479 0.974808324205915
484 0.96969696969697
489 0.980681818181818
494 0.97931654676259
500 0.983249581239531
};

\nextgroupplot[
width=\linewidth,
height=0.7\linewidth,
legend cell align={left},
legend style={
  fill opacity=0.8,
  draw opacity=1,
  text opacity=1,
  at={(0.03,0.97)},
  anchor=north west,
  draw=white!80!black
},
scaled x ticks=manual:{}{\pgfmathparse{#1}},
tick align=outside,
tick pos=left,
x grid style={white!69.0196078431373!black},
xmajorgrids,
xmin=-25, xmax=525,
xminorgrids,
xtick style={color=black},
xticklabels={},
y grid style={white!69.0196078431373!black},
ylabel={Throughput (pps)},
ymajorgrids,
ymin=-0, ymax=77057400,
ytick={0,10000000,20000000,30000000,40000000,50000000,60000000,70000000},
yticklabels={0,1,2,3,4,5,6,7},
yminorgrids,
ytick style={color=black},
ylabel style={font=\fontsize{5}{5}\selectfont},
xlabel style={font=\fontsize{5}{5}\selectfont},
tick label style={font=\fontsize{6}{6}\selectfont}
]
\addplot [semithick, color0, mark=square*, mark size=0.5, mark options={solid}]
table {%
0 0
100 37839000
200 69126000
300 73388000
400 68285000
500 60079000
};
\addplot [semithick, color1, mark=square*, mark size=0.5, mark options={solid}]
table {%
0 0
100 37013000
200 69098000
300 73065000
400 68286000
500 60080000
};

\nextgroupplot[
width=\linewidth,
height=0.7\linewidth,
legend cell align={left},
legend style={
  fill opacity=0.8,
  draw opacity=1,
  text opacity=1,
  at={(0.97,0.03)},
  anchor=south east,
  draw=white!80!black
},
tick align=outside,
tick pos=left,
x grid style={white!69.0196078431373!black},
xlabel={Time (s)},
xmajorgrids,
xmin=-25, xmax=525,
xminorgrids,
xtick style={color=black},
y grid style={white!69.0196078431373!black},
ylabel={Active \#PMTs},
ymajorgrids,
ymin=0.7, ymax=7.3,
yminorgrids,
xtick={0,100,200,300,400,500},
xticklabels={0,1,2,3,4,5},
ytick style={color=black},
ylabel style={font=\fontsize{5}{5}\selectfont},
xlabel style={font=\fontsize{5}{5}\selectfont},
tick label style={font=\fontsize{6}{6}\selectfont}
]
\addplot [semithick, color0, dashed, mark=x, mark size=0.5, mark options={solid}]
table {%
1 6
6 6
11 6
16 6
21 6
26 6
31 6
36 6
41 6
46 6
51 6
56 6
61 6
66 6
71 6
76 6
81 6
86 6
91 6
96 6
101 6
106 6
111 6
116 6
121 6
127 6
132 6
137 6
142 6
147 6
152 6
157 6
162 6
167 6
172 6
177 6
182 6
187 6
192 6
197 6
202 6
207 6
212 6
217 6
222 6
227 6
232 6
237 6
242 6
247 6
253 6
258 6
263 6
268 6
273 6
278 6
283 6
288 6
293 6
298 6
303 6
308 6
313 6
318 6
323 6
328 6
333 6
338 6
343 6
348 6
353 6
358 6
363 6
368 6
373 6
379 6
384 6
389 6
394 6
399 6
404 6
409 6
414 6
419 6
424 6
429 6
434 6
439 6
444 6
449 6
454 6
459 6
464 6
469 6
474 6
479 6
484 6
489 6
494 6
500 6
};
\addplot [semithick, color1, dashed, mark=x, mark size=0.5, mark options={solid}]
table {%
1 1
6 1
11 1
16 1
21 1
26 1
31 1
36 1
41 1
46 1
51 2
56 3
61 3
66 4
71 5
76 5
81 6
86 6
91 6
96 6
101 6
106 6
111 6
116 6
121 6
127 6
132 6
137 6
142 6
147 6
152 6
157 6
162 6
167 6
172 6
177 6
182 6
187 6
192 6
197 6
202 6
207 6
212 6
217 6
222 6
227 6
232 6
237 6
242 6
247 6
253 6
258 7
263 6
268 7
273 6
278 7
283 6
288 6
293 6
298 6
303 6
308 6
313 6
318 6
323 6
328 6
333 6
338 6
343 6
348 6
353 6
358 6
363 6
368 6
373 6
379 6
384 6
389 6
394 6
399 6
404 6
409 6
414 6
419 6
424 6
429 6
434 6
439 6
444 6
449 5
454 5
459 5
464 5
469 5
474 5
479 5
484 5
489 5
494 6
500 6
};
\end{groupplot}
\end{tikzpicture}

%% file: figures/eval-results/adapt-vs-static/adapt-static-t2.tex
\begin{tikzpicture}

\definecolor{color0}{rgb}{0.258823529411765,0.403921568627451,0.823529411764706}
\definecolor{color1}{rgb}{0.917647058823529,0.262745098039216,0.207843137254902}
\definecolor{color2}{rgb}{0.984313725490196,0.737254901960784,0.0156862745098039}
\definecolor{color3}{rgb}{0.203921568627451,0.658823529411765,0.325490196078431}

\begin{groupplot}[group style={group size=1 by 3}]
\nextgroupplot[
width=\linewidth,
height=0.7\linewidth,
legend cell align={left},
legend style={
  fill opacity=0.8,
  draw opacity=1,
  text opacity=1,
  at={(0.97,0.03)},
  anchor=south east,
  draw=white!80!black
},
scaled x ticks=manual:{}{\pgfmathparse{#1}},
tick align=outside,
tick pos=left,
x grid style={white!69.0196078431373!black},
xmajorgrids,
xmin=-25, xmax=525,
xminorgrids,
xtick style={color=black},
xticklabels={},
y grid style={white!69.0196078431373!black},
ylabel={Hit Rate},
ymajorgrids,
ymin=0.300030321406913, ymax=1.03269658378815,
yminorgrids,
ytick style={color=black},
ylabel style={font=\fontsize{5}{5}\selectfont},
xlabel style={font=\fontsize{5}{5}\selectfont},
tick label style={font=\fontsize{6}{6}\selectfont}
]
\addplot [semithick, color0, mark=*, mark size=0.5, mark options={solid}]
table {%
1 0.333333333333333
6 0.692307692307692
11 0.870967741935484
16 0.790697674418605
21 0.923076923076923
26 0.969512195121951
31 0.984293193717278
36 0.93717277486911
41 0.966666666666667
46 0.975409836065574
51 0.9812734082397
56 0.969387755102041
61 0.99079754601227
66 0.978494623655914
71 0.988700564971751
76 0.99273607748184
81 0.988066825775656
86 0.990804597701149
91 0.984269662921348
96 0.99452804377565
101 0.993351063829787
106 0.99501246882793
111 0.994825355756792
116 0.996231155778894
121 0.99375
127 0.991606714628297
132 0.993150684931507
137 0.995584988962472
142 0.996376811594203
147 0.991256830601093
152 0.990686845168801
157 0.99642431466031
162 0.991442542787286
167 0.982323232323232
172 0.987991266375546
177 0.994138335287222
182 0.990055248618785
187 0.99566630552546
192 0.990055248618785
197 0.99025974025974
202 0.996879875195008
207 0.983988355167394
212 0.993174061433447
217 0.993421052631579
222 0.984709480122324
227 0.992263056092843
232 0.991150442477876
237 0.99163179916318
242 0.989550679205852
247 0.994272623138602
253 0.985945945945946
258 0.993865030674847
263 0.992558139534884
268 0.991596638655462
273 0.997995991983968
278 0.989785495403473
283 0.989594172736733
288 0.990654205607477
293 0.991649269311065
298 0.993670886075949
303 0.993670886075949
308 0.991219512195122
313 0.989
318 0.995327102803738
323 0.995425434583715
328 0.996694214876033
333 0.995483288166215
338 0.991071428571429
343 0.995029821073559
348 0.986748216106014
353 0.99232245681382
358 0.994954591321897
363 0.988659793814433
368 0.990740740740741
373 0.989775051124744
379 0.992290748898678
384 0.989339019189765
389 0.988577362409138
394 0.995069033530572
399 0.994360902255639
404 0.995636998254799
409 0.986200551977921
414 0.994100294985251
419 0.991071428571429
424 0.992373689227836
429 0.988876529477197
434 0.987179487179487
439 0.992042440318302
444 0.994505494505494
449 0.994805194805195
454 0.993227990970655
459 0.994736842105263
464 0.995570321151716
469 0.990280777537797
474 0.991774383078731
479 0.990877993158495
484 0.992647058823529
489 0.990958408679928
494 0.997247706422018
500 0.99335232668566
};
\addplot [semithick, color1, mark=*, mark size=0.5, mark options={solid}]
table {%
1 0.333333333333333
6 0.730769230769231
11 0.838709677419355
16 0.790697674418605
21 0.923076923076923
26 0.963414634146341
31 0.984293193717278
36 0.931937172774869
41 0.966666666666667
46 0.967213114754098
51 0.958801498127341
56 0.91156462585034
61 0.938650306748466
66 0.89247311827957
71 0.858757062146893
76 0.726392251815981
81 0.904534606205251
86 0.931034482758621
91 0.966292134831461
96 0.986320109439124
101 0.988031914893617
106 0.99501246882793
111 0.96895213454075
116 0.991206030150754
121 0.992205767731878
127 0.993705743509048
132 0.995530726256983
137 0.995680345572354
142 0.997109826589595
147 0.99292324442025
152 0.992911990549321
157 0.999393571861734
162 0.993397358943577
167 0.988456865127582
172 0.99559277214632
177 0.994823529411765
182 0.995882891125343
187 0.996019460415745
192 0.990893321769298
197 0.994402985074627
202 0.998289136013687
207 0.984323432343234
212 0.996916752312436
217 0.994041708043694
222 0.981800766283525
227 0.991935483870968
232 0.991235758106924
237 0.992424242424242
242 0.9826435246996
247 0.996827914353688
253 0.979913916786227
258 0.984615384615385
263 0.990995762711864
268 0.995472837022133
273 0.996569468267582
278 0.989977728285078
283 0.988727858293076
288 0.992781787895614
293 0.991549295774648
298 0.990046838407494
303 0.992289442467378
308 0.992920353982301
313 0.991694352159468
318 0.993847874720358
323 0.997421351211965
328 0.99386816555953
333 0.99840848806366
338 0.991423670668954
343 0.998260869565217
348 0.986533957845433
353 0.991374353076481
358 0.994639666468136
363 0.977761836441894
368 0.986890021849964
373 0.984581497797357
379 0.989344262295082
384 0.992743105950653
389 0.984593837535014
394 0.997912317327766
399 0.994459833795014
404 0.997433264887064
409 0.988717339667458
414 0.997563946406821
419 0.986348122866894
424 0.998289623717218
429 0.986284289276808
434 0.993630573248408
439 0.991116751269036
444 0.994743758212878
449 0.993073047858942
454 0.986308583464982
459 0.993411764705882
464 0.998429319371728
469 0.994214079074253
474 0.99393605292172
479 0.991692627206646
484 0.994869402985075
489 0.989081885856079
494 0.999043977055449
500 0.995444191343964
};

\nextgroupplot[
width=\linewidth,
height=0.7\linewidth,
legend cell align={left},
legend style={
  fill opacity=0.8,
  draw opacity=1,
  text opacity=1,
  at={(0.03,0.97)},
  anchor=north west,
  draw=white!80!black
},
scaled x ticks=manual:{}{\pgfmathparse{#1}},
tick align=outside,
tick pos=left,
x grid style={white!69.0196078431373!black},
xmajorgrids,
xmin=-25, xmax=525,
xminorgrids,
xtick style={color=black},
xticklabels={},
y grid style={white!69.0196078431373!black},
ylabel={Throughput (pps)},
ymajorgrids,
ymin=-5114950, ymax=107413950,
yminorgrids,
ytick={0,20000000,40000000,60000000,80000000,100000000},
yticklabels={0,.2,.4,.6,.8,1.},
ytick style={color=black},
ylabel style={font=\fontsize{5}{5}\selectfont},
xlabel style={font=\fontsize{5}{5}\selectfont},
tick label style={font=\fontsize{6}{6}\selectfont}
]
\addplot [semithick, color0, mark=square*, mark size=0.5, mark options={solid}]
table {%
0 0
100 28346000
200 82317000
300 85308000
400 102295000
500 93510000
};
\addplot [semithick, color1, mark=square*, mark size=0.5, mark options={solid}]
table {%
0 0
100 28346000
200 82317000
300 85305000
400 102299000
500 93509000
};

\nextgroupplot[
width=\linewidth,
height=0.7\linewidth,
legend cell align={left},
legend style={
  fill opacity=0.8,
  draw opacity=1,
  text opacity=1,
  at={(0.97,0.03)},
  anchor=south east,
  draw=white!80!black
},
tick align=outside,
tick pos=left,
x grid style={white!69.0196078431373!black},
xlabel={Time (s)},
xmajorgrids,
xmin=-25, xmax=525,
xminorgrids,
xtick style={color=black},
y grid style={white!69.0196078431373!black},
ylabel={Active \#PMTs},
ymajorgrids,
ymin=0.7, ymax=7.3,
yminorgrids,
xtick={0,100,200,300,400,500},
xticklabels={0,1,2,3,4,5},
ytick style={color=black},
ylabel style={font=\fontsize{5}{5}\selectfont},
xlabel style={font=\fontsize{5}{5}\selectfont},
tick label style={font=\fontsize{6}{6}\selectfont}
]
\addplot [semithick, color0, dashed, mark=x, mark size=0.5, mark options={solid}]
table {%
1 5
6 5
11 5
16 5
21 5
26 5
31 5
36 5
41 5
46 5
51 5
56 5
61 5
66 5
71 5
76 5
81 5
86 5
91 5
96 5
101 5
106 5
111 5
116 5
121 5
127 5
132 5
137 5
142 5
147 5
152 5
157 5
162 5
167 5
172 5
177 5
182 5
187 5
192 5
197 5
202 5
207 5
212 5
217 5
222 5
227 5
232 5
237 5
242 5
247 5
253 5
258 5
263 5
268 5
273 5
278 5
283 5
288 5
293 5
298 5
303 5
308 5
313 5
318 5
323 5
328 5
333 5
338 5
343 5
348 5
353 5
358 5
363 5
368 5
373 5
379 5
384 5
389 5
394 5
399 5
404 5
409 5
414 5
419 5
424 5
429 5
434 5
439 5
444 5
449 5
454 5
459 5
464 5
469 5
474 5
479 5
484 5
489 5
494 5
500 5
};
\addplot [semithick, color1, dashed, mark=x, mark size=0.5, mark options={solid}]
table {%
1 1
6 1
11 1
16 1
21 1
26 1
31 1
36 1
41 1
46 1
51 1
56 1
61 1
66 1
71 2
76 3
81 3
86 4
91 4
96 5
101 6
106 6
111 7
116 6
121 7
127 6
132 7
137 7
142 7
147 7
152 7
157 7
162 7
167 6
172 7
177 7
182 7
187 7
192 7
197 6
202 6
207 6
212 6
217 6
222 6
227 5
232 6
237 7
242 7
247 7
253 7
258 7
263 7
268 7
273 7
278 7
283 7
288 7
293 7
298 7
303 7
308 7
313 7
318 7
323 7
328 7
333 7
338 7
343 7
348 7
353 7
358 7
363 7
368 7
373 7
379 7
384 7
389 7
394 7
399 7
404 7
409 7
414 7
419 7
424 7
429 7
434 6
439 6
444 6
449 6
454 7
459 7
464 7
469 7
474 7
479 7
484 7
489 7
494 7
500 7
};
\end{groupplot}

\end{tikzpicture}

%% file: figures/eval-results/adapt-vs-static/adapt-static-t3.tex
\begin{tikzpicture}

\definecolor{color0}{rgb}{0.258823529411765,0.403921568627451,0.823529411764706}
\definecolor{color1}{rgb}{0.917647058823529,0.262745098039216,0.207843137254902}
\definecolor{color2}{rgb}{0.984313725490196,0.737254901960784,0.0156862745098039}
\definecolor{color3}{rgb}{0.203921568627451,0.658823529411765,0.325490196078431}
\begin{groupplot}[group style={group size=1 by 3}]
\nextgroupplot[
width=\linewidth,
height=0.7\linewidth,
legend cell align={left},
legend style={
  fill opacity=0.8,
  draw opacity=1,
  text opacity=1,
  at={(0.97,0.03)},
  anchor=south east,
  draw=white!80!black
},
scaled x ticks=manual:{}{\pgfmathparse{#1}},
tick align=outside,
tick pos=left,
x grid style={white!69.0196078431373!black},
xmajorgrids,
xmin=-25, xmax=525,
xminorgrids,
xtick style={color=black},
xticklabels={},
y grid style={white!69.0196078431373!black},
ylabel={Hit Rate},
ymajorgrids,
ymin=-0.049812382739212, ymax=1.04606003752345,
yminorgrids,
ytick style={color=black},
ylabel style={font=\fontsize{5}{5}\selectfont},
xlabel style={font=\fontsize{5}{5}\selectfont},
tick label style={font=\fontsize{6}{6}\selectfont}
]
\addplot [semithick, color0, mark=*, mark size=0.5, mark options={solid}]
table {%
1 0
6 0.80952380952381
11 0.846153846153846
16 0.882352941176471
21 0.93984962406015
26 0.969230769230769
31 0.981818181818182
36 0.968325791855204
41 0.95850622406639
46 0.974358974358974
51 0.981191222570533
56 0.965616045845272
61 0.988235294117647
66 0.981182795698925
71 0.986522911051213
76 0.990291262135922
81 0.983758700696056
86 0.991452991452992
91 0.987249544626594
96 0.989754098360656
101 0.986754966887417
106 0.992307692307692
111 0.996138996138996
116 0.996086105675147
121 0.991334488734835
127 0.986817325800377
132 0.982847341337907
137 0.991554054054054
142 0.992619926199262
147 0.981196581196581
152 0.985655737704918
157 0.99624765478424
162 0.980544747081712
167 0.973118279569892
172 0.98019801980198
177 0.993492407809111
182 0.986138613861386
187 0.99554565701559
192 0.981171548117155
197 0.979959919839679
202 0.995833333333333
207 0.970786516853933
212 0.984905660377358
217 0.989384288747346
222 0.983606557377049
227 0.988344988344988
232 0.984409799554566
237 0.984198645598194
242 0.97737556561086
247 0.984478935698448
253 0.984158415841584
258 0.983471074380165
263 0.983985765124555
268 0.980530973451327
273 0.990234375
278 0.981818181818182
283 0.98974358974359
288 0.977401129943503
293 0.983221476510067
298 0.98936170212766
303 0.986254295532646
308 0.980968858131488
313 0.980551053484603
318 0.983164983164983
323 0.987676056338028
328 0.992792792792793
333 0.988304093567252
338 0.977917981072555
343 0.987860394537178
348 0.989230769230769
353 0.982905982905983
358 0.984962406015038
363 0.97887323943662
368 0.983633387888707
373 0.975778546712803
379 0.991596638655462
384 0.984536082474227
389 0.981574539363484
394 0.992844364937388
399 0.980657640232108
404 0.989406779661017
409 0.979919678714859
414 0.981891348088531
419 0.983455882352941
424 0.985446985446986
429 0.974409448818898
434 0.988304093567252
439 0.979452054794521
444 0.984304932735426
449 0.987730061349693
454 0.992047713717694
459 0.985386221294363
464 0.991718426501035
469 0.982490272373541
474 0.981718464351006
479 0.986193293885602
484 0.982035928143712
489 0.984644913627639
494 0.986538461538462
500 0.98932384341637
};
\addplot [semithick, color1, mark=*, mark size=0.5, mark options={solid}]
table {%
1 0
6 0.80952380952381
11 0.846153846153846
16 0.892156862745098
21 0.93984962406015
26 0.969230769230769
31 0.981818181818182
36 0.963800904977376
41 0.95850622406639
46 0.967032967032967
51 0.934169278996865
56 0.896848137535817
61 0.923529411764706
66 0.881720430107527
71 0.913746630727763
76 0.9
81 0.870229007633588
86 0.961538461538462
91 0.983128834355828
96 0.974182444061962
101 0.987084870848708
106 0.985576923076923
111 0.990353697749196
116 0.995008319467554
121 0.985207100591716
127 0.970636215334421
132 0.976296296296296
137 0.983918128654971
142 0.9712
147 0.971139971139971
152 0.979166666666667
157 0.995253164556962
162 0.978688524590164
167 0.957317073170732
172 0.97319932998325
177 0.987179487179487
182 0.978260869565217
187 0.993150684931507
192 0.980952380952381
197 0.98191933240612
202 0.994342291371994
207 0.971036585365854
212 0.982014388489209
217 0.98079561042524
222 0.974967061923584
227 0.984172661870504
232 0.986928104575163
237 0.986861313868613
242 0.962848297213622
247 0.983231707317073
253 0.982804232804233
258 0.981944444444444
263 0.977829638273046
268 0.981566820276498
273 0.983565107458913
278 0.976689976689977
283 0.986401673640167
288 0.978149100257069
293 0.984530386740331
298 0.984242424242424
303 0.986348122866894
308 0.977299880525687
313 0.970484061393152
318 0.973611111111111
323 0.982404692082112
328 0.982758620689655
333 0.978125
338 0.971279373368146
343 0.979952830188679
348 0.985981308411215
353 0.983957219251337
358 0.97841726618705
363 0.981843575418994
368 0.974968710888611
373 0.979513444302177
379 0.978589420654912
384 0.980237154150198
389 0.972010178117048
394 0.98828125
399 0.982758620689655
404 0.973597359735974
409 0.975920679886686
414 0.979591836734694
419 0.977900552486188
424 0.971810089020772
429 0.971264367816092
434 0.967816091954023
439 0.981067125645439
444 0.977853492333901
449 0.983660130718954
454 0.988617886178862
459 0.982547993019197
464 0.985530546623794
469 0.978090766823161
474 0.976083707025411
479 0.989312977099237
484 0.974358974358974
489 0.976744186046512
494 0.980122324159022
500 0.986731001206273
};

\nextgroupplot[
width=\linewidth,
height=0.7\linewidth,
legend cell align={left},
legend style={
  fill opacity=0.8,
  draw opacity=1,
  text opacity=1,
  at={(0.03,0.97)},
  anchor=north west,
  draw=white!80!black
},
scaled x ticks=manual:{}{\pgfmathparse{#1}},
tick align=outside,
tick pos=left,
x grid style={white!69.0196078431373!black},
xmajorgrids,
xmin=-25, xmax=525,
xminorgrids,
xtick style={color=black},
xticklabels={},
y grid style={white!69.0196078431373!black},
ylabel={Throughput (pps)},
ymajorgrids,
ymin=-2905500, ymax=61015500,
yminorgrids,
ytick={0,10000000,20000000,30000000,40000000,50000000,60000000},
yticklabels={0,1,2,3,4,5,6},
ytick style={color=black},
ylabel style={font=\fontsize{5}{5}\selectfont},
xlabel style={font=\fontsize{5}{5}\selectfont},
tick label style={font=\fontsize{6}{6}\selectfont}
]
\addplot [semithick, color0, mark=square*, mark size=0.5, mark options={solid}]
table {%
0 0
100 27696000
200 51916000
300 49961000
400 58110000
500 50043000
};
\addplot [semithick, color1, mark=square*, mark size=0.5, mark options={solid}]
table {%
0 0
100 27696000
200 51916000
300 49962000
400 58109000
500 50043000
};

\nextgroupplot[
width=\linewidth,
height=0.7\linewidth,
legend cell align={left},
legend style={
  fill opacity=0.8,
  draw opacity=1,
  text opacity=1,
  at={(0.97,0.03)},
  anchor=south east,
  draw=white!80!black
},
tick align=outside,
tick pos=left,
x grid style={white!69.0196078431373!black},
xlabel={Time (s)},
xmajorgrids,
xmin=-25, xmax=525,
xminorgrids,
xtick style={color=black},
y grid style={white!69.0196078431373!black},
ylabel={Active \#PMTs},
ymajorgrids,
ymin=0.75, ymax=6.25,
yminorgrids,
xtick={0,100,200,300,400,500},
xticklabels={0,1,2,3,4,5},
ytick style={color=black},
ylabel style={font=\fontsize{5}{5}\selectfont},
xlabel style={font=\fontsize{5}{5}\selectfont},
tick label style={font=\fontsize{6}{6}\selectfont}
]
\addplot [semithick, color0, dashed, mark=x, mark size=0.5, mark options={solid}]
table {%
1 5
6 5
11 5
16 5
21 5
26 5
31 5
36 5
41 5
46 5
51 5
56 5
61 5
66 5
71 5
76 5
81 5
86 5
91 5
96 5
101 5
106 5
111 5
116 5
121 5
127 5
132 5
137 5
142 5
147 5
152 5
157 5
162 5
167 5
172 5
177 5
182 5
187 5
192 5
197 5
202 5
207 5
212 5
217 5
222 5
227 5
232 5
237 5
242 5
247 5
253 5
258 5
263 5
268 5
273 5
278 5
283 5
288 5
293 5
298 5
303 5
308 5
313 5
318 5
323 5
328 5
333 5
338 5
343 5
348 5
353 5
358 5
363 5
368 5
373 5
379 5
384 5
389 5
394 5
399 5
404 5
409 5
414 5
419 5
424 5
429 5
434 5
439 5
444 5
449 5
454 5
459 5
464 5
469 5
474 5
479 5
484 5
489 5
494 5
500 5
};
\addplot [semithick, color1, dashed, mark=x, mark size=0.5, mark options={solid}]
table {%
1 1
6 1
11 1
16 1
21 1
26 1
31 1
36 1
41 1
46 1
51 1
56 1
61 1
66 1
71 2
76 3
81 4
86 3
91 4
96 5
101 5
106 5
111 5
116 5
121 5
127 5
132 5
137 5
142 5
147 5
152 5
157 5
162 5
167 5
172 4
177 4
182 5
187 4
192 4
197 5
202 4
207 5
212 4
217 5
222 4
227 4
232 4
237 4
242 4
247 4
253 5
258 5
263 5
268 5
273 5
278 5
283 5
288 5
293 5
298 5
303 5
308 5
313 5
318 6
323 5
328 5
333 5
338 5
343 6
348 6
353 5
358 5
363 5
368 6
373 6
379 5
384 5
389 5
394 5
399 5
404 5
409 5
414 5
419 5
424 5
429 4
434 5
439 4
444 4
449 4
454 4
459 4
464 4
469 5
474 5
479 5
484 5
489 5
494 5
500 5
};
\end{groupplot}

\end{tikzpicture}

%% file: figures/eval-results/adapt-vs-static/adapt-static-t4.tex
\begin{tikzpicture}

\definecolor{color0}{rgb}{0.258823529411765,0.403921568627451,0.823529411764706}
\definecolor{color1}{rgb}{0.917647058823529,0.262745098039216,0.207843137254902}
\definecolor{color2}{rgb}{0.984313725490196,0.737254901960784,0.0156862745098039}
\definecolor{color3}{rgb}{0.203921568627451,0.658823529411765,0.325490196078431}
\begin{groupplot}[group style={group size=1 by 3}]
\nextgroupplot[
width=\linewidth,
height=0.7\linewidth,
legend cell align={left},
legend style={
  fill opacity=0.8,
  draw opacity=1,
  text opacity=1,
  at={(0.97,0.03)},
  anchor=south east,
  draw=white!80!black
},
scaled x ticks=manual:{}{\pgfmathparse{#1}},
tick align=outside,
tick pos=left,
x grid style={white!69.0196078431373!black},
xmajorgrids,
xmin=-25, xmax=525,
xminorgrids,
xtick style={color=black},
xticklabels={},
y grid style={white!69.0196078431373!black},
ylabel={Hit Rate},
ymajorgrids,
ymin=-0.05, ymax=1.05,
yminorgrids,
ytick style={color=black},
ylabel style={font=\fontsize{5}{5}\selectfont},
xlabel style={font=\fontsize{5}{5}\selectfont},
tick label style={font=\fontsize{6}{6}\selectfont}
]
\addplot [semithick, color0, mark=*, mark size=0.5, mark options={solid}]
table {%
1 0
6 0.86046511627907
11 0.935483870967742
16 0.946666666666667
21 0.9375
26 0.969072164948454
31 0.972727272727273
36 0.962121212121212
41 0.932432432432432
46 0.993103448275862
51 0.990783410138249
56 0.981981981981982
61 0.995024875621891
66 0.976
71 0.989864864864865
76 0.993377483443709
81 0.99009900990099
86 0.98951048951049
91 0.984084880636605
96 0.985549132947977
101 0.979651162790698
106 0.975069252077562
111 0.959785522788204
116 0.963076923076923
121 0.958100558659218
127 0.96927374301676
132 0.954022988505747
137 0.955678670360111
142 0.964646464646465
147 0.957142857142857
152 0.948863636363636
157 0.955678670360111
162 0.961748633879781
167 0.936
172 0.958702064896755
177 0.950769230769231
182 0.962264150943396
187 0.966850828729282
192 0.959239130434783
197 0.96078431372549
202 0.978609625668449
207 0.968992248062015
212 0.965968586387434
217 0.971576227390181
222 0.96319018404908
227 0.980582524271845
232 0.966257668711656
237 0.959375
242 0.955489614243324
247 0.942122186495177
253 0.945945945945946
258 0.955479452054795
263 0.969040247678019
268 0.957928802588997
273 0.965625
278 0.951127819548872
283 0.959322033898305
288 0.954128440366972
293 0.975
298 0.97
303 0.924012158054711
308 0.930795847750865
313 0.941747572815534
318 0.959183673469388
323 0.954063604240283
328 0.969283276450512
333 0.957654723127036
338 0.941379310344828
343 0.967741935483871
348 0.932203389830508
353 0.953271028037383
358 0.966887417218543
363 0.941860465116279
368 0.970845481049563
373 0.948660714285714
379 0.949152542372881
384 0.971291866028708
389 0.92764857881137
394 0.953038674033149
399 0.968838526912181
404 0.964285714285714
409 0.942622950819672
414 0.962059620596206
419 0.945558739255014
424 0.977707006369427
429 0.957446808510638
434 0.959390862944162
439 0.962352941176471
444 0.977464788732394
449 0.962871287128713
454 0.9609375
459 0.969613259668508
464 0.96551724137931
469 0.928571428571429
474 0.951219512195122
479 0.947530864197531
484 0.957055214723926
489 0.950769230769231
494 0.957446808510638
500 0.920454545454545
};
\addplot [semithick, color1, mark=*, mark size=0.5, mark options={solid}]
table {%
1 0
6 0.86046511627907
11 0.935483870967742
16 0.946666666666667
21 0.9375
26 0.969072164948454
31 0.972727272727273
36 0.962121212121212
41 0.932432432432432
46 0.993103448275862
51 0.990783410138249
56 0.981981981981982
61 0.995024875621891
66 0.968
71 0.969594594594595
76 0.963576158940397
81 0.957095709570957
86 0.933566433566434
91 0.941644562334218
96 0.913294797687861
101 0.936046511627907
106 0.944598337950138
111 0.915254237288136
116 0.937321937321937
121 0.924812030075188
127 0.942982456140351
132 0.938864628820961
137 0.947162426614481
142 0.949820788530466
147 0.954063604240283
152 0.95446584938704
157 0.96085409252669
162 0.961424332344214
167 0.937413073713491
172 0.953051643192488
177 0.958402662229617
182 0.958100558659218
187 0.968980797636632
192 0.950797872340426
197 0.972121212121212
202 0.969211822660098
207 0.976442873969376
212 0.981572481572482
217 0.994110718492344
222 0.968888888888889
227 0.996904024767802
232 0.985096870342772
237 0.993730407523511
242 0.980169971671388
247 0.972434915773354
253 0.982210927573062
258 0.981099656357388
263 0.976780185758514
268 0.984154929577465
273 0.998366013071895
278 0.982421875
283 0.982394366197183
288 0.983508245877061
293 0.989441930618401
298 0.994854202401372
303 0.9904
308 0.982594936708861
313 0.985007496251874
318 0.98943661971831
323 0.992805755395684
328 0.985507246376812
333 0.984544049459042
338 0.990322580645161
343 0.998106060606061
348 0.956521739130435
353 0.98951048951049
358 1
363 0.971786833855799
368 0.992435703479576
373 0.980810234541578
379 0.991218441273326
384 0.990936555891239
389 0.976116303219107
394 0.993769470404984
399 0.993827160493827
404 0.98314606741573
409 0.9599609375
414 0.985977212971078
419 0.961767204757859
424 1
429 0.985422740524781
434 0.98489666136725
439 0.990513833992095
444 0.996044303797468
449 0.987987987987988
454 0.99268130405855
459 0.994121969140338
464 0.993256911665543
469 0.982348426707598
474 0.991961414790997
479 0.982940698619009
484 0.962641181581234
489 0.988954970263382
494 1
500 0.973129992737836
};

\nextgroupplot[
width=\linewidth,
height=0.7\linewidth,
legend cell align={left},
legend style={
  fill opacity=0.8,
  draw opacity=1,
  text opacity=1,
  at={(0.03,0.97)},
  anchor=north west,
  draw=white!80!black
},
scaled x ticks=manual:{}{\pgfmathparse{#1}},
tick align=outside,
tick pos=left,
x grid style={white!69.0196078431373!black},
xmajorgrids,
xmin=-25, xmax=525,
xminorgrids,
xtick style={color=black},
xticklabels={},
y grid style={white!69.0196078431373!black},
ylabel={Throughput (pps)},
ymajorgrids,
ymin=-1792800, ymax=37648800,
yminorgrids,
ytick={0,10000000,20000000,30000000},
yticklabels={0,1,2,3},
ytick style={color=black},
ylabel style={font=\fontsize{5}{5}\selectfont},
xlabel style={font=\fontsize{5}{5}\selectfont},
tick label style={font=\fontsize{6}{6}\selectfont}
]
\addplot [semithick, color0, mark=square*, mark size=0.5, mark options={solid}]
table {%
0 0
100 19343000
200 35489000
300 33226000
400 33373000
500 35856000
};
\addplot [semithick, color1, mark=square*, mark size=0.5, mark options={solid}]
table {%
0 0
100 19341000
200 35491000
300 33225000
400 33374000
500 35856000
};

\nextgroupplot[
width=\linewidth,
height=0.7\linewidth,
legend cell align={left},
legend style={
  fill opacity=0.8,
  draw opacity=1,
  text opacity=1,
  at={(0.03,0.03)},
  anchor=south west,
  draw=white!80!black
},
tick align=outside,
tick pos=left,
x grid style={white!69.0196078431373!black},
xlabel={Time (s)},
xmajorgrids,
xmin=-25, xmax=525,
xminorgrids,
xtick style={color=black},
y grid style={white!69.0196078431373!black},
ylabel={Active \#PMTs},
ymajorgrids,
ymin=0.85, ymax=4.15,
yminorgrids,
xtick={0,100,200,300,400,500},
xticklabels={0,1,2,3,4,5},
ytick style={color=black},
xtick={0,100,200,300,400,500},
xticklabels={0,1,2,3,4,5},
ylabel style={font=\fontsize{5}{5}\selectfont},
xlabel style={font=\fontsize{5}{5}\selectfont},
tick label style={font=\fontsize{6}{6}\selectfont}
]
\addplot [semithick, color0, dashed, mark=x, mark size=0.5, mark options={solid}]
table {%
1 3
6 3
11 3
16 3
21 3
26 3
31 3
36 3
41 3
46 3
51 3
56 3
61 3
66 3
71 3
76 3
81 3
86 3
91 3
96 3
101 3
106 3
111 3
116 3
121 3
127 3
132 3
137 3
142 3
147 3
152 3
157 3
162 3
167 3
172 3
177 3
182 3
187 3
192 3
197 3
202 3
207 3
212 3
217 3
222 3
227 3
232 3
237 3
242 3
247 3
253 3
258 3
263 3
268 3
273 3
278 3
283 3
288 3
293 3
298 3
303 3
308 3
313 3
318 3
323 3
328 3
333 3
338 3
343 3
348 3
353 3
358 3
363 3
368 3
373 3
379 3
384 3
389 3
394 3
399 3
404 3
409 3
414 3
419 3
424 3
429 3
434 3
439 3
444 3
449 3
454 3
459 3
464 3
469 3
474 3
479 3
484 3
489 3
494 3
500 3
};
\addplot [semithick, color1, dashed, mark=x, mark size=0.5, mark options={solid}]
table {%
1 1
6 1
11 1
16 1
21 1
26 1
31 1
36 1
41 1
46 1
51 1
56 1
61 1
66 1
71 1
76 1
81 1
86 1
91 1
96 2
101 2
106 1
111 2
116 1
121 1
127 2
132 1
137 1
142 2
147 1
152 1
157 1
162 2
167 1
172 1
177 1
182 2
187 1
192 2
197 2
202 3
207 3
212 2
217 2
222 3
227 1
232 1
237 1
242 1
247 1
253 1
258 1
263 1
268 1
273 1
278 1
283 1
288 1
293 1
298 1
303 1
308 1
313 1
318 1
323 1
328 1
333 1
338 1
343 1
348 1
353 1
358 1
363 1
368 1
373 2
379 3
384 2
389 3
394 2
399 2
404 1
409 2
414 3
419 2
424 1
429 1
434 1
439 2
444 4
449 2
454 3
459 2
464 2
469 3
474 1
479 1
484 1
489 1
494 1
500 1
};
\end{groupplot}

\end{tikzpicture}

%% file: nsections/rwork.tex
\section{Related work}

\noindent\textbf{Memory Management.}
Hogan et al. introduced P4All, an extension of the P4 language with elastic data structures that adjust size dynamically based on switch resources, optimizing routing, monitoring, and caching applications. P4All improves modularity and reduces compile-time complexities using symbolic primitives and objective functions for resource allocation \cite{Hogan.p4all}. Zhu et al. proposed NetVRM, a virtual register memory abstraction enabling dynamic memory sharing among concurrent applications on programmable networks, improving memory allocation efficiency and performance through a utility-based allocation policy \cite{NetVRM.Zhu}. Das et al. developed ActiveRMT, supporting dynamic memory allocation and reallocation, optimizing network performance with efficient memory synchronization and state management, facilitating in-network cache services and memory-intensive applications~\cite{Das.MemMGMTActiveRMT}.


\noindent\textbf{Virtualization in Programmable Data Planes.}
Zheng et al. introduced P4Visor, a virtualization abstraction allowing concurrent execution of multiple P4 programs by embedding testing primitives and optimizing compiler algorithms, reducing resource overhead \cite{Zheng.P4visor}. Zhang et al. proposed HyperV, a high-performance hypervisor virtualizing programmable data planes, supporting multiple networking contexts and enabling hot-swappable snapshots, improving performance and flexibility \cite{Zhang.hyperv}. Hancock et al. developed HyPer4, a portable virtualization solution for dynamically configuring programmable data planes, enabling network slicing and multi-tenancy without disrupting active programs~\cite{Hancock.Hyper4}.


%% file: nsections/limitations.tex
\section{Discussion and Limitations}

\noindent\textbf{In-Memory Processing.}
Our design relies heavily on high-bandwidth external memory to handle cache-missed requests. However, achieving even higher bandwidths remains a significant challenge due to technological constraints. One emerging solution is \gls{pim}, which aims to offload part of the computation to the memory itself. By integrating processing capabilities within memory modules, specifically for match operations, \gls{pim} can substantially reduce data movement overhead. This leads to lower latency, improved throughput, and reduced energy consumption.

\noindent\textbf{Security Considerations.}
In Synapse, data plane cache blocks are exposed to network traffic, making them susceptible to performance manipulation through traffic flooding. Malicious actors can generate high rates of anomalous packets, causing a high cache miss rate and degrading performance. Therefore, robust packet authentication is crucial to ensure that only authenticated packets enter the pipeline, preventing unauthorized access and mitigating denial-of-service attacks.

\noindent\textbf{Hardware Implementation Limitations.}
Our implementation of FPGA-based \glspl{pmu} serves as a prototype, providing flexibility and rapid prototyping capabilities. However, it poses significant scalability challenges. Utilizing \gls{fpga} LUTs to design \glspl{pmu} is not scalable. Scaling to a high number of \glspl{pmu} consumes most of the \gls{fpga} resources, particularly LUTs, making it infeasible to manage with high frequencies (\~250MHz). The bottleneck shifts to the synthesis/implementation process, specifically placement and routing, which becomes increasingly difficult. Therefore, for a production-ready product, integrating the \glspl{pmu} within an ASIC is essential. The programmable logic should primarily serve for the interconnection network and other architectural logic. This approach aligns with the design practices in modern programmable switches and Smart \glspl{nic}, where the critical, performance-intensive components are implemented as ASICs to ensure scalability and efficiency.

\noindent\textbf{Static Operations Scheduling}. Our design relies on static operation scheduling for the~\gls{elu} as it is implemented using HLS, which limits the exploitation of runtime memory bank availability. As a result, it leads to less efficient use of the available memory bandwidth. A more power-efficient future design could incorporate dynamic scheduling for the \gls{elu}, which can be realized using open-source processors such as RISC-V~\cite{xilinx_riscv_2024}, providing the flexibility to dynamically manage memory access patterns and improve bandwidth utilization.


%% file: nsections/conclusion.tex
\section{Conclusion}

Synapse enhances the management of match tables in programmable networks, providing elasticity and efficient resource allocation through the \gls{vmt} framework. By leveraging a hybrid memory system and the Runtime OPT, Synapse enables flexible and scalable match table allocation, ensuring efficient resource utilization and improved network performance. The prototype on \gls{fpga} and evaluation demonstrate the feasibility and effectiveness of the design, while also highlighting the limitations of \gls{fpga}-based \glspl{pmu} in terms of scalability. For a production-ready solution, integrating \gls{pmu} clusters as ASICs is recommended to ensure optimal performance and resource efficiency. Overall, we aimed to provide additional flexibility to  programmable data planes, enhancing runtime adaptation in hardware through improved memory management and virtualization of the \glspl{pmt}.

%% file: nsections/appendix.tex
\section{Reproducibility}

We plan to release the artifacts related to the design and evaluation of Synapse through a publicly accessible repository. In particular, we plan to release the following artifacts:

\begin{itemize}
    \item The HLS source code implementing the Synapse components (\gls{vmu}, \gls{pmu}, \gls{elu});
    \item The source code of the OPT module;
    \item The Synapse simulator source code;
    \item Instructions and datasets to reproduce the results presented in the paper.
\end{itemize}

\section{Runtime OPT -- Scalability and Execution Time}

\begin{figure}[h]
    \centering
    \input{figures/eval-results/ras-exec-time}
    \caption{[U50]~Runtime OPT execution time vs the CFG size}
    \label{fig:runtime-opt-exec-time}
\end{figure}
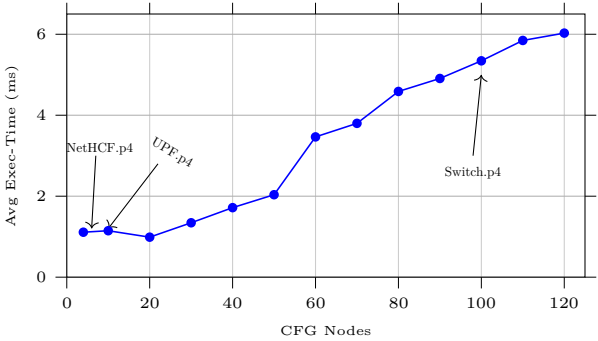

To further evaluate the scalability of Synapse, we measured the execution time of the runtime optimization process across various  \gls{cfg} sizes. Figure~\ref{fig:runtime-opt-exec-time} presents these results, focusing on three key CFGs from the NetHCF.p4, UPF.p4, and Switch.p4 implementations. The data shows that, for reasonably sized CFGs, the optimization can be solved in just a few milliseconds, highlighting the practicality of Synapse in real-world applications. The results were obtained using the Gurobi solver, ensuring efficient resolution of the optimization problem.

\section{VMT's Capacity Estimation}
\label{sec:vmt-capacity}

For each \gls{vmt}, the CPU collects statistics over a time window of 10 microseconds. These statistics include the number of \glspl{phv} that arrived and were processed, as well as the number of \gls{pmu} allocated to the \gls{vmt}.

\begin{figure}[t]
    \centering
    \begin{subfigure}[t]{0.9\linewidth}
        \input{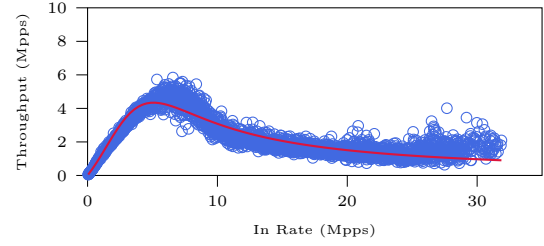}
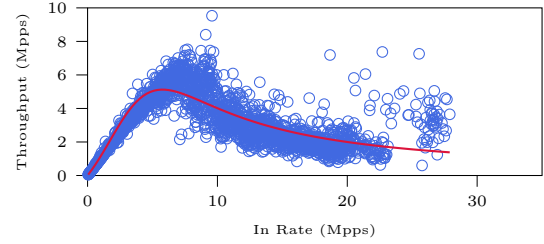
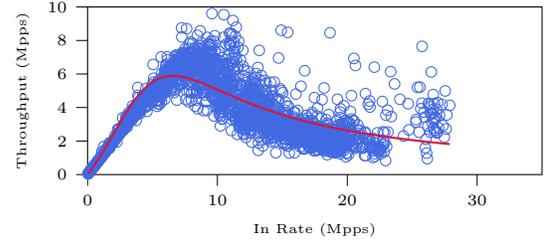
        \caption{3 PMTs}
    \end{subfigure}
    \begin{subfigure}[t]{0.9\linewidth}
        \input{figures/eval-results/usl-fitting/4}
        \caption{4 PMTs}
    \end{subfigure}
    \begin{subfigure}[t]{0.9\linewidth}
        \input{figures/eval-results/usl-fitting/5}
        \caption{5 PMTs}
    \end{subfigure}
    \caption{Scatter plot of collected data points for different numbers of PMTs: (a) 3 PMTs, (b) 4 PMTs, (c) 5 PMTs.}
    \label{fig:usl-vmt}
\end{figure}

As explained in Section~\ref{sec:opt}, we model the behavior of the throughput of each \gls{vmt} as a function of the input rate using the \gls{usl}. Figure~\ref{fig:usl-vmt} presents a scatter plot of some of the collected data points for three different values of \glspl{pmu} (3, 4, and 5), depicted in the first, second, and third subfigures, respectively. 

As the scatter plot shows, there is a clear \gls{usl}-like behavior. We employ a simple regression to estimate offline the \gls{usl} parameters and utilize them for the $S_{i}$ equation defined in Section~\ref{sec:opt}.

\section{Implementation Details}
\label{appendix:implementation}
In this section, we present the implementation of two key components of our system: the consistent hashing lookup and the action execution module. These components are integral to the efficient functioning of the key distribution and action processing pipeline in our system.

\noindent\textbf{Key Lookup with Consistent hashing.} The following listing demonstrates the implementation of the consistent hashing lookup function within a given \gls{vmt}, which is responsible for distributing keys across multiple \glspl{pmt} using a consistent hashing mechanism. The function takes the lookup request, calculates the appropriate hash value, and performs the lookup in the pre-allocated hash table (s\_lookup\_table).

\begin{lstlisting}[language=C++, basicstyle=\ttfamily\footnotesize,label={lst:consistent_lb}]
struct RequestT;
struct RequestD2T;
struct lb_entry_t;
template <int N, typename RequestT, typename RequestD2T, ...>
void consistent_lb(
    ......
    bool &sync_and_reset,
    stream<RequestT> &keys_in,
    stream<RequestD2T> &keys_out,
    lb_entry_t s_lookup_table[Power<2, N>::Value] // Passed as an argument for host access
) {
    #pragma HLS INTERFACE mode = s_axilite register port = sync_and_reset bundle=control
    #pragma HLS INTERFACE ap_ctrl_none port=return
    #pragma HLS INTERFACE s_axilite port=s_lookup_table bundle=control // Memory-mapped interface for the lookup table
    #pragma HLS PIPELINE II=1
    static bool prev_sync = false;

    // lookup table implementation (BRAM with a single port by default)
    #pragma HLS RESOURCE variable=s_lookup_table core=RAM_1P_BRAM

    if (!keys_in.empty()) {
        RequestT k = keys_in.read();
        RequestD2T m_key_out;
        m_key_out.data = k;
        m_key_out.last = 1;
        
        //byte level mask
        m_key_out.keep = -1;
        ap_uint<32> h = 
        xf::database
          ::details
          ::hashlookup3_core(k, h);
        ap_uint<N> index = h % Power<2,N>::Value;
        lb_entry_t tmp = s_lookup_table[index];
        #pragma HLS DISAGGREGATE variable=tmp
        //s_lookup_table entry stats update & flash them back
        .....
        m_key_out.dest = tmp.dest;
        keys_out.write(m_key_out);
    } else if (sync_and_reset && !prev_sync) {
        // Synchronize data stats upon reset
        .....
    }

    prev_sync = sync_and_reset;
}
\end{lstlisting}

\noindent\textbf{Action Execution Unit.} the following listing shows the implementation of the action execution module, which processes the results of the consistent hashing lookup and applies the corresponding actions to the \gls{phv}. This module operates immediately after the lookup stage and is responsible for executing actions based on the lookup results.
\begin{lstlisting}[language=C++, basicstyle=\ttfamily\footnotesize,label={lst:alu}]
struct phv_t;
struct action_reply_t;
void action_module(
    hls::stream<phv_t> &phv_in, // Input PHV stream from VMU
    hls::stream<phv_t> &phv_out, // Output PHV stream to next stage
    hls::stream<action_reply_t> &action_reply // Action reply stream
) {
    #pragma HLS PIPELINE II=1
    static hls::stream<phv_t> phv_fifo("phv_fifo"); 
    #pragma HLS STREAM variable=phv_fifo depth=D

    if (!action_reply.empty()) {
        action_reply_t reply = action_reply.read();

        if (!reply.valid_action) {
            // Miss case: explicit miss notification
            if (!phv_in.empty()) {
                phv_t phv = phv_in.read();
                phv_fifo.write(phv);
            }
        } else {
            // Valid action case
            if (!phv_fifo.empty()) {
                // Case 1: Apply action to a previously queued PHV from FIFO
                phv_t phv = phv_fifo.read();
                phv_t new_phv = apply_action(phv, reply.action);
                phv_out.write(new_phv);
            } else if (!phv_in.empty()) {
                // Case 2: Apply action to a PHV directly from the input stream
                phv_t phv = phv_in.read();
                phv_t new_phv = apply_action(phv, reply.action);
                phv_out.write(new_phv);
            }
        }
    }
}
\end{lstlisting}

%% file: figures/eval-results/ras-exec-time.tex
\begin{tikzpicture}
\pgfplotsset{every tick label/.append style={font=\scriptsize}}
\definecolor{darkgray176}{RGB}{176,176,176}
\definecolor{lightgray204}{RGB}{204,204,204}

\begin{axis}[
x grid style={lightgray204},
width=\linewidth,
height=0.6\columnwidth,
legend style={fill opacity=0.8, draw opacity=1, text opacity=1, draw=lightgray204},
tick align=outside,
xlabel={CFG Nodes},
xticklabel style={align=center, text width=1cm},
xmajorgrids,
ymajorgrids,
ymajorticks=true,
xmin=0, xmax=125.0,
xtick style={color=black},
y grid style={darkgray176},
ylabel={Avg Exec-Time (ms)},
ymin=0.0, ymax=6.5,
ytick style={color=black},
ylabel style={font=\fontsize{5}{5}\selectfont},
        xlabel style={font=\fontsize{5}{5}\selectfont},
        tick label style={font=\fontsize{6}{6}\selectfont}
]
\addplot [semithick, blue, mark=*, mark size=1.5, mark options={solid}]
table {%
	4 1.11053300940472
	10 1.14715894063314
	20 0.988044738769531
	30 1.34362979811065
	40 1.71713926354233
	50 2.03632821842116
	60 3.46461120916873
	70 3.79946231842041
	80 4.58708101389359
	90 4.90784158512038
	100 5.34380972385406
	110 5.84686795870463
	120 6.02834565298898
};
\draw[->,draw=black] (axis cs:7,3) -- (axis cs:6,1.2);
\draw (axis cs:8,3.1) node[
  scale=0.5,
  anchor=base,
  text=black,
  rotate=0.0
]{NetHCF.p4};

\draw[->,draw=black] (axis cs:22,2.8) -- (axis cs:10,1.2);
\draw (axis cs:25,3.0) node[
  scale=0.5,
  anchor=base,
  text=black,
  rotate=-30.0
]{UPF.p4};

\draw[->,draw=black] (axis cs:98,3) -- (axis cs:100,5.0);
\draw (axis cs:98,2.5) node[
  scale=0.5,
  anchor=base,
  text=black,
  rotate=0.0
]{Switch.p4};
\end{axis}

\end{tikzpicture}

%% file: figures/eval-results/usl-fitting/4.tex
\pgfplotstableread{
x       y
5 4.99462365591398
6 6.00609756097561
7 7
8 8.00694444444444
9 9
10 9.97637795275591
11 10.9811320754717
12 12
13 13
14 13.9850746268657
15 15.0298507462687
16 15.9857142857143
17 16.952380952381
18 18.0338983050847
19 18.9795918367347
20 19.9655172413793
21 20.9545454545455
22 21.9677419354839
23 22.8958333333333
24 23.9444444444444
25 25
26 26.2368421052632
27 27.12
28 28
29 29.0714285714286
30 29.9230769230769
31 30.9655172413793
32 31.8947368421053
33 32.9459459459459
34 34.2647058823529
35 34.5909090909091
36 36.0967741935484
37 36.9642857142857
38 37.7407407407407
39 39.04
40 39.4444444444444
41 40.4137931034483
42 41.9259259259259
43 42.4444444444444
44 43.7727272727273
45 44.0476190476191
46 46.2857142857143
47 47
48 48.04
49 48.96
50 49.8666666666667
51 50.4666666666667
52 58.6
53 52.0952380952381
54 54.5833333333333
55 55.9545454545455
56 55.5882352941176
57 55.2857142857143
58 58.2352941176471
59 58.6923076923077
60 59.8333333333333
61 72.2
62 60.8947368421053
63 63.0714285714286
64 63.6875
65 62.75
66 65.9
67 66.0714285714286
68 66.7142857142857
69 67.3
70 68.5909090909091
71 79.8888888888889
72 72.4444444444444
73 73.125
74 73.875
75 72.5652173913043
76 75.9285714285714
77 77.75
78 79
79 78.5294117647059
80 79.8
81 93.1111111111111
82 81.9090909090909
83 83
84 83.7777777777778
85 84
86 85.8
87 82.6
88 97.1428571428571
89 89.1818181818182
90 93.3636363636364
91 90.5454545454545
92 91.9
93 107.083333333333
94 93.125
95 95.2142857142857
96 117
97 96
98 97
99 101.545454545455
100 99.9090909090909
101 108.9
102 100.727272727273
103 102.1
104 104.363636363636
105 119.222222222222
106 107.083333333333
107 135.833333333333
108 103.375
109 109
110 119.857142857143
111 111.214285714286
112 113.4
113 116.909090909091
114 123.333333333333
115 115.125
116 114.714285714286
117 117.923076923077
118 120.545454545455
119 117.583333333333
120 116.833333333333
121 120.75
122 121.571428571429
123 125.272727272727
124 128.875
125 127.2
126 126
127 126.666666666667
128 128
129 130.1
130 163.833333333333
131 131
132 132.166666666667
133 133.1
134 134.230769230769
135 133.615384615385
136 136.142857142857
137 137.166666666667
138 203.285714285714
139 149.294117647059
140 143.444444444444
141 141.3
142 142.5
143 152.428571428571
144 144.625
145 145
146 155.466666666667
147 141.777777777778
148 144.333333333333
149 149.25
150 150.285714285714
151 151.571428571429
152 152
153 150.222222222222
154 154
155 168.545454545455
156 155.533333333333
157 154.769230769231
158 184.083333333333
159 162.363636363636
160 169.181818181818
161 161.615384615385
162 163.166666666667
163 163
164 162.375
165 165
166 162.5
167 184.7
168 175.4
169 151
170 194.5
171 170.375
172 172
173 176.166666666667
174 173.833333333333
175 174.777777777778
176 176.444444444444
177 185.111111111111
178 178.333333333333
179 190.7
180 177.857142857143
181 182.6
182 182.714285714286
183 182.692307692308
184 184.416666666667
185 185
186 186.75
187 186.142857142857
188 190.75
189 189.125
190 190.333333333333
191 191.5
192 192.25
193 191.125
194 193.3
195 195.444444444444
196 194.571428571429
197 196.5
198 197.888888888889
199 203.5
200 188.6
201 201.125
202 203.25
203 203
204 204.090909090909
205 205
206 199.785714285714
207 220.545454545455
208 208.538461538462
209 214.230769230769
210 212.444444444444
211 251.625
212 256
213 215.75
214 231.571428571429
215 218.4
216 215
217 210.25
218 215
219 261.888888888889
220 220.333333333333
221 222.5
222 234.9
223 227.1
224 222
225 225
226 194.142857142857
227 224.882352941176
228 227.714285714286
229 230.111111111111
230 229.5
231 234
232 248.555555555556
233 232.8
234 257.272727272727
235 238.625
236 255
237 198.4
238 236.5
239 222.846153846154
240 243
241 245.428571428571
242 328.090909090909
243 238.5
244 244.142857142857
245 246.375
246 244.714285714286
247 252.9
248 248.125
249 247.428571428571
250 252.133333333333
251 293.888888888889
252 273.166666666667
253 312.666666666667
254 245.375
255 254.846153846154
256 272.444444444444
257 254.75
258 257.857142857143
259 285.666666666667
260 262
261 310
262 274.125
263 263.285714285714
264 264.3
265 266.8
266 267.666666666667
267 267.75
268 288.5
269 269
270 304.666666666667
271 267.888888888889
272 260
273 261.454545454545
274 257.555555555556
275 275.461538461538
276 276.7
277 290.3
278 282.9
279 282.8
280 282.25
281 282.75
282 295.454545454545
283 284.090909090909
284 315.625
285 284.4
286 302.888888888889
287 287.75
288 289.166666666667
289 288.888888888889
290 289.714285714286
291 296.714285714286
292 294.916666666667
293 325
294 366.5
295 327.285714285714
296 282.333333333333
297 320.444444444444
298 287.25
299 304.166666666667
300 310.375
301 300.857142857143
302 305.25
303 314.090909090909
304 303.375
305 247.6
306 359
307 306.636363636364
308 355.444444444444
309 307.923076923077
310 297.583333333333
311 311.545454545455
312 311.125
313 311.166666666667
314 312.4375
315 313.416666666667
316 333.25
317 305.5
318 291.5
319 329.8
320 320.8
321 330.357142857143
322 355
323 324.555555555556
324 323.1
325 347.857142857143
326 326.363636363636
327 327.909090909091
328 343.368421052632
329 399.8
330 329
331 318.941176470588
332 331.25
333 366.727272727273
334 319.75
335 310.222222222222
336 335
337 336.2
338 366.076923076923
339 356.181818181818
340 382.888888888889
341 341
342 352.142857142857
343 342.666666666667
344 329.2
345 325.5
346 489.666666666667
347 350.9
348 346.3
349 342.428571428571
350 349.5
351 339.875
352 322.307692307692
353 365.636363636364
354 354.428571428571
355 343.222222222222
356 396.416666666667
357 356.428571428571
358 337
359 357.25
360 366.5
361 315.25
362 402
363 361.166666666667
364 348.222222222222
365 364.5
366 367
367 370.333333333333
368 365.5
369 352.5
370 409.222222222222
371 344.888888888889
372 365.083333333333
373 408.181818181818
374 340
375 358.357142857143
376 399.909090909091
377 377.5
378 385.6
379 395.666666666667
380 371.625
381 371.545454545455
382 365.142857142857
383 379.7
384 384.3
385 395.5
386 363.666666666667
387 544
388 412.833333333333
389 415.1
390 394.333333333333
391 392.285714285714
392 417.1
393 369.25
394 351
395 377.125
396 343
397 398
398 367.3
399 398.75
400 405.555555555556
401 446.555555555556
402 371.25
403 394.7
404 362.5
405 385.545454545455
406 408.571428571429
407 422.6
408 434.625
409 411.777777777778
410 471
411 446.333333333333
412 413.545454545455
413 415.375
414 427
415 414.625
416 418.833333333333
417 404.230769230769
418 410.666666666667
419 420.928571428571
420 379
421 364.75
422 477.285714285714
423 420
424 493.75
425 384.4
426 422.166666666667
427 429.166666666667
428 382.363636363636
429 441.25
430 389.461538461538
431 378.3
432 436.2
433 478.75
434 412.888888888889
435 408.375
436 434.428571428571
437 441.4375
438 436.214285714286
439 414.333333333333
440 406.6
441 490
442 426.461538461538
443 399.4
444 451.5
445 421.166666666667
446 436.625
447 405.4
448 451.625
449 489.375
450 450
451 427.111111111111
452 458.307692307692
453 410.666666666667
454 434.333333333333
455 547.545454545455
456 453.428571428571
457 414.5
458 480.428571428571
459 452.444444444444
460 459
461 495.857142857143
462 387.25
463 470.777777777778
464 444.1
465 461.857142857143
466 465.857142857143
467 445.2
468 460.285714285714
469 472.285714285714
470 355
471 470.833333333333
472 472.571428571429
473 450.083333333333
474 453.25
475 413.071428571429
476 473.888888888889
477 442.222222222222
478 493.818181818182
479 526.625
480 478.833333333333
481 477.142857142857
482 474.923076923077
483 481.555555555556
484 356.142857142857
485 488
486 529.833333333333
487 486
488 487
489 453.1
490 426.166666666667
491 491
492 456.631578947368
493 498.2
494 454.076923076923
495 467.714285714286
496 442.666666666667
497 495.9
498 497.888888888889
499 461.3
500 446.375
501 602.571428571429
502 497
503 428.9
504 505
505 460.25
506 468.909090909091
507 507.714285714286
508 490
509 337
510 492.333333333333
511 530.2
512 512.666666666667
513 436.777777777778
514 507.25
515 504.266666666667
516 495.6
517 515.888888888889
518 466.461538461538
519 447
520 584.833333333333
521 521.25
522 459.2
523 483.4
524 477.8
525 546.666666666667
526 544.133333333333
527 558.222222222222
528 482.333333333333
529 414.5
530 434.5
531 582.3
532 452
533 488.3
534 531.75
535 377.888888888889
536 472.454545454545
537 426.857142857143
538 490.571428571429
539 535
540 457.142857142857
541 494
542 400.555555555556
543 486.8
544 454
545 452.3
546 460.6
547 543.75
548 501.666666666667
549 510.545454545455
550 471.285714285714
551 502.375
552 522
553 423
554 476.5
555 452.25
556 451.5
557 594
558 569.571428571429
559 456.363636363636
560 445
561 449.285714285714
562 477.909090909091
563 547.555555555556
564 544.75
565 545.888888888889
566 584.125
567 625.333333333333
568 572.222222222222
569 387.2
570 548
571 552.5
572 445.333333333333
573 433.2
574 575.285714285714
575 572.666666666667
576 543.333333333333
577 472.6
578 474.166666666667
579 591
580 555.75
581 576.2
582 536.4
583 512.307692307692
584 524.625
585 483.166666666667
586 524.285714285714
587 543.625
588 427.5
589 583.125
590 581.142857142857
591 512.571428571429
592 523.625
593 434
594 454.25
595 505.111111111111
596 604.333333333333
597 512.833333333333
598 435.555555555556
599 486.833333333333
600 389.333333333333
601 583.666666666667
602 568
603 559
604 556.375
605 682
606 476.75
607 606.222222222222
608 504.666666666667
609 576.666666666667
610 499
611 578.333333333333
612 601.75
613 617
614 578.375
615 572.444444444444
616 454
617 515.142857142857
618 518.25
619 459
620 554.5
621 662
622 502
623 568.666666666667
624 536.75
625 565.909090909091
626 602.125
627 518.625
628 559.125
629 571.285714285714
630 593.230769230769
631 531.1
632 592.333333333333
633 632.5
634 568.285714285714
635 546.857142857143
636 582.6
637 609.571428571429
638 515.166666666667
639 550
640 550.7
641 598.333333333333
642 506.714285714286
643 474
644 525
645 471.666666666667
646 561.818181818182
647 590.2
648 535.5
649 611
650 616.666666666667
651 551
652 575.916666666667
653 569.666666666667
654 490.166666666667
655 563
656 572.083333333333
657 472.625
658 562.333333333333
659 588.875
660 674.4
661 506
662 615.666666666667
663 557.375
664 537.666666666667
665 551
666 651.4
667 355.5
668 570.666666666667
669 599.3
670 667.833333333333
671 573.076923076923
672 474
673 614.7
674 577.6
675 663.625
676 533.571428571429
677 603.2
678 602.5
679 566
680 650.285714285714
681 638
682 638.428571428571
683 635
684 544.875
685 544.545454545455
686 627.333333333333
687 453.75
688 655.571428571429
689 505.625
690 609
691 571.428571428571
692 566.9
693 605.857142857143
694 607
695 504.818181818182
696 532
697 701
698 490
699 614.272727272727
700 580.3
701 541.142857142857
702 608.363636363636
703 502.857142857143
704 657.111111111111
705 648.75
706 589.7
707 631.5
708 479.222222222222
709 654.375
710 215.5
711 592.444444444444
712 658.9
713 547.888888888889
714 555.545454545455
715 571.875
716 647.714285714286
717 615.714285714286
718 638.2
719 612.75
720 540
721 616.545454545455
722 533.727272727273
723 540.333333333333
724 245.5
725 541.625
726 652.181818181818
727 381
728 619.454545454545
729 642.333333333333
730 522.5
731 657.5
732 589.333333333333
733 603.25
734 681.555555555556
735 632
736 568.125
737 590.125
738 592.4
739 532.571428571429
740 597.272727272727
741 656.666666666667
742 587.833333333333
743 478.666666666667
744 705.25
745 586
746 587.363636363636
747 585.111111111111
748 599.166666666667
749 715.4
750 646
751 537.083333333333
752 747.25
753 625.25
754 578.125
755 440.75
756 397.428571428571
757 369.4
758 658.153846153846
759 580
760 505.916666666667
761 546.9
762 601
763 635.428571428571
764 540.363636363636
765 421.857142857143
766 630.5
767 590.4
768 468.666666666667
769 752.555555555556
770 457.5
771 422.8
772 665
773 611.125
774 660.666666666667
775 522.1
776 528.5
777 533.428571428571
778 579
779 250
780 450.222222222222
781 421.5
782 719.75
783 533
784 492.166666666667
785 490.833333333333
786 553.625
787 515.545454545455
788 547.111111111111
789 563.666666666667
790 590.25
791 713.6
792 492
793 618.8
794 539.1
795 402.142857142857
796 528.666666666667
797 544.4
798 374.5
799 691.285714285714
800 404
801 430.8
802 493.666666666667
803 491.454545454545
804 674.25
805 611.625
806 598.636363636364
807 389.222222222222
808 634.8
809 522
810 498.2
811 415.25
812 417.444444444444
813 602.333333333333
814 503.9
815 587.125
816 451.166666666667
817 427.6
818 517.166666666667
819 432.8
820 542.555555555556
821 525.142857142857
822 573.166666666667
823 612
824 466.666666666667
825 407.5
826 626
827 585.428571428571
828 287.666666666667
829 500.25
830 490
831 492.777777777778
832 319.6
833 318.2
834 552.142857142857
835 481.571428571429
836 407.666666666667
837 513.142857142857
838 588.4
839 389.4
840 487.166666666667
841 621.285714285714
842 364.5
843 439.4
844 346.4
845 529
846 533.25
847 494.777777777778
848 295.285714285714
849 380.545454545455
850 393.571428571429
851 456
852 424.571428571429
853 605.571428571429
854 487.833333333333
855 510.5
856 480.4
857 599.5
858 603.25
859 414.2
860 524.777777777778
861 469.666666666667
862 367.833333333333
863 423.5
864 368.25
865 516.666666666667
866 484.5
867 590.153846153846
868 630.9
869 620.5
870 469.428571428571
871 562.25
872 376.454545454545
873 460
874 744.666666666667
875 484.333333333333
876 388.5
877 355.666666666667
878 516.4
879 642.285714285714
880 511.333333333333
881 508.571428571429
882 469.666666666667
883 559.142857142857
884 605.4
885 556.833333333333
886 515.4
887 581.285714285714
888 467.333333333333
889 514
890 364.714285714286
891 472
892 399.8
893 377.5
894 657.333333333333
895 651
896 578
897 650.142857142857
898 692.285714285714
899 509.2
900 679.3
901 537.625
902 300.666666666667
903 308.333333333333
904 276.625
905 555
906 413.571428571429
907 286.666666666667
908 695.285714285714
909 453.125
910 840
911 228.666666666667
912 506.142857142857
913 520.8
914 559.25
915 343.166666666667
916 342.666666666667
917 489.666666666667
918 565.833333333333
919 330
920 249.5
921 387.2
922 611.375
923 505.2
924 472.166666666667
925 676.333333333333
926 552.571428571429
927 583
928 609.285714285714
929 538.7
930 624.125
931 562.9
932 463.833333333333
933 474.5
934 621.5
935 565.444444444444
936 620.307692307692
937 614.7
938 615.5625
939 579.833333333333
940 640
941 331.142857142857
942 445
943 572.8
944 745.375
945 545.6
946 692.133333333333
947 565.916666666667
948 570.333333333333
949 419.75
950 664.333333333333
951 286.857142857143
952 404.333333333333
953 674.714285714286
954 496.666666666667
955 423.8
956 581
957 414.333333333333
958 953
959 542.666666666667
960 398.8
961 606
962 328.666666666667
963 522.666666666667
964 452
965 505.636363636364
966 585.75
967 330.909090909091
968 403.5
969 381.6
970 736.555555555556
971 397.5
972 390.666666666667
973 304.875
974 413.5
975 440.111111111111
976 539
977 359.428571428571
978 291
979 272.75
980 571.5
981 310.142857142857
982 369.666666666667
983 346.5
984 416.285714285714
985 480.125
986 509.8
987 501.8
988 280.6
989 506.888888888889
990 375.25
991 403
992 497.25
993 600.142857142857
994 522
995 422
996 331.625
997 518.625
998 400.25
999 376.111111111111
1000 300.4
1001 483.571428571429
1002 405.272727272727
1003 493.75
1004 405.888888888889
1005 381.142857142857
1006 345.4
1007 427.333333333333
1008 412.833333333333
1009 372
1010 469.2
1011 257.333333333333
1012 281
1013 283.75
1014 270
1015 355.75
1016 391.75
1017 367.8
1018 272.25
1019 394
1020 414.333333333333
1021 349.333333333333
1022 309
1023 260.333333333333
1024 313.2
1025 220
1026 405.285714285714
1027 318.666666666667
1028 170.4
1029 331
1030 234.625
1031 327.25
1032 303
1033 296.6
1034 398.166666666667
1035 255.25
1036 437.8
1037 380.333333333333
1038 226.75
1039 331.454545454545
1040 326.8
1041 378.4
1042 360.428571428571
1043 232.555555555556
1044 387.111111111111
1045 375.857142857143
1046 285.111111111111
1047 397.666666666667
1048 306.2
1049 600.333333333333
1050 303.857142857143
1051 313.5
1052 316
1053 364.555555555556
1054 502.8
1055 232.5
1056 388.4
1057 293.375
1058 313
1059 252.333333333333
1060 365.5
1061 241.142857142857
1062 371.888888888889
1063 361.714285714286
1064 314.714285714286
1065 291.25
1066 424
1067 504.2
1068 340.888888888889
1069 384
1070 419.25
1071 278.142857142857
1072 447
1073 468
1074 220.333333333333
1075 290.666666666667
1076 394
1077 358.5
1078 330
1079 323.75
1080 353.833333333333
1081 291.333333333333
1082 327.333333333333
1083 424.6
1084 341.666666666667
1085 468
1086 273.333333333333
1087 364
1088 476
1089 300.166666666667
1090 237.285714285714
1091 421.555555555556
1092 259.833333333333
1093 384.166666666667
1094 345.833333333333
1095 290.909090909091
1096 354
1097 365.428571428571
1098 270.666666666667
1099 315.333333333333
1100 437.75
1101 302.5
1102 459.5
1103 388.5
1104 418
1105 405.4
1106 326.333333333333
1107 327.666666666667
1108 268.75
1109 336.75
1110 412.4
1111 425.285714285714
1112 326
1113 281.5
1114 379.777777777778
1115 490.666666666667
1116 315
1117 288
1118 267
1119 608.5
1120 357.25
1121 297
1122 318
1123 311.285714285714
1124 345.571428571429
1125 348.875
1126 386.833333333333
1127 319.571428571429
1128 406
1129 290.25
1130 377.2
1131 278
1132 338.8
1133 287.6
1134 409.5
1135 307
1136 247.333333333333
1137 459.666666666667
1138 418.5
1139 307
1140 345.4
1141 379.5
1142 378.333333333333
1143 468.333333333333
1144 242
1145 288
1146 543.75
1147 366.4
1148 227.666666666667
1149 315.333333333333
1150 265.8
1151 394.75
1152 264.285714285714
1153 317
1154 305.5
1155 329
1156 283.333333333333
1157 493.428571428571
1158 400.2
1159 318.25
1160 261
1161 307.5
1162 320.8
1163 320.333333333333
1164 222.5
1165 330.666666666667
1166 274.333333333333
1167 250.833333333333
1168 344.5
1169 349.166666666667
1170 388.571428571429
1171 248
1172 287.714285714286
1173 487
1174 474.333333333333
1175 219.8
1176 279.25
1177 215
1178 206
1179 305.625
1180 256.142857142857
1181 282.2
1182 299
1183 344
1184 373.666666666667
1185 332.5
1186 276.8
1187 354.666666666667
1188 334
1189 441.333333333333
1190 368
1191 201.333333333333
1192 165
1193 328
1194 272.833333333333
1195 317
1196 465.666666666667
1197 432.5
1198 289.4
1199 278.142857142857
1201 282
1202 264.6
1203 361.25
1204 227
1205 208
1206 390
1207 241.857142857143
1208 244
1209 188
1210 259.5
1211 411.25
1212 333
1213 226.5
1214 305
1215 210
1216 305.166666666667
1217 197
1218 243
1219 278.666666666667
1220 192.5
1221 326.75
1222 232.333333333333
1223 501.2
1224 271.5
1225 348
1226 349.5
1227 261.75
1228 262.555555555556
1229 199.5
1230 176
1231 350.666666666667
1232 285.75
1233 282
1234 184.8
1235 270.1
1236 291.5
1238 278.8
1239 182
1240 223.857142857143
1241 230.125
1242 144
1243 240.666666666667
1244 305.857142857143
1245 234
1246 319.090909090909
1247 482.5
1248 404.4
1249 228.333333333333
1250 277.75
1251 303
1252 205
1253 284.555555555556
1254 303.714285714286
1255 320.625
1256 272
1257 263.333333333333
1258 218.285714285714
1259 150
1260 186.4
1261 334
1262 223
1263 302
1264 181
1265 303
1266 331.625
1267 237.666666666667
1268 222
1269 228
1270 289.6
1271 300.111111111111
1272 413
1273 387
1274 265.333333333333
1275 169.5
1276 317.75
1277 316.857142857143
1278 227.6
1279 329.166666666667
1280 340.8
1281 361.125
1282 254
1283 337.4
1284 244.142857142857
1285 190.5
1286 217.5
1287 421
1288 268.2
1289 190
1290 373.333333333333
1291 161.5
1292 254.5
1293 201.25
1294 263
1295 246
1296 388
1297 280.5
1298 353
1299 284.8
1300 321.6
1301 330.142857142857
1302 244.375
1303 206.428571428571
1304 289.833333333333
1305 226.6
1306 232.5
1307 259
1308 273.8
1309 299.5
1310 327.2
1311 224
1312 229.571428571429
1313 299.666666666667
1314 197
1315 314.333333333333
1316 175
1317 240.666666666667
1318 197.75
1319 235.857142857143
1320 222.666666666667
1321 322.5
1322 421.666666666667
1323 254
1324 317.25
1325 243.8
1326 556.5
1327 226.5
1328 240.4
1329 262.833333333333
1330 184.75
1331 258.4
1332 327.833333333333
1333 245
1334 276.333333333333
1335 292.25
1336 201.833333333333
1337 387.75
1338 188.5
1339 258.6
1340 298.75
1341 287.25
1342 332.2
1343 315.555555555556
1344 332.833333333333
1345 141
1346 203.666666666667
1347 307.571428571429
1348 285.2
1349 276.571428571429
1350 235.5
1351 316.2
1352 206
1353 190
1354 256.5
1355 236.5
1356 144
1357 271.2
1358 340.8
1359 244.166666666667
1360 340
1361 205.666666666667
1362 329.5
1363 360.5
1364 271.833333333333
1365 367.333333333333
1366 197.666666666667
1367 242
1368 156
1369 326.75
1370 260.75
1371 291
1372 325.571428571429
1373 267
1374 242.5
1375 278.166666666667
1376 223.25
1377 323.8
1378 295.142857142857
1379 199.5
1380 313.285714285714
1381 262.6
1382 291.2
1383 309.571428571429
1384 417
1385 218.75
1386 173
1387 291.571428571429
1388 291.4
1389 257.5
1390 303.5
1391 207.833333333333
1392 304.6
1393 355.333333333333
1394 296.666666666667
1395 331.4
1396 329
1397 415.6
1398 265.6
1399 250
1400 278.166666666667
1401 232.5
1402 242
1403 334.2
1404 275
1405 350.5
1406 281.8
1407 339.285714285714
1408 253.571428571429
1409 267.666666666667
1410 246.142857142857
1411 288.666666666667
1412 242
1413 198
1414 225.333333333333
1415 270.833333333333
1416 258.666666666667
1417 316.2
1418 269.833333333333
1419 313.833333333333
1420 219
1421 228.6
1422 158.5
1423 248
1424 228.75
1425 279.333333333333
1426 235
1427 245
1428 298.875
1429 310.5
1430 306
1431 185
1432 269.285714285714
1433 175
1434 244.333333333333
1435 205
1436 260.833333333333
1437 204.25
1438 484.222222222222
1439 263.333333333333
1440 324.5
1441 256.5
1442 206.333333333333
1443 170.5
1444 288.5
1445 404.333333333333
1446 223.714285714286
1447 332.181818181818
1448 238.333333333333
1449 259.75
1450 142
1451 276.25
1452 171
1453 278
1454 176
1455 209.6
1456 179.166666666667
1457 238.666666666667
1458 278.333333333333
1459 319.5
1460 213.5
1461 285.777777777778
1462 230.285714285714
1463 235.666666666667
1464 212.666666666667
1465 268.4
1466 225.333333333333
1467 284
1468 241
1469 191
1470 245
1471 479
1472 255.5
1473 249
1474 171
1475 217.2
1476 260.4
1477 228.666666666667
1478 217.833333333333
1479 181.6
1480 167.5
1481 234
1482 262
1483 275.666666666667
1484 451.571428571429
1485 212.75
1486 214.333333333333
1487 260.333333333333
1488 273.142857142857
1489 300
1490 195.333333333333
1491 1923
1492 187
1493 269.5
1494 195
1495 236
1496 246.75
1497 201.25
1498 236.666666666667
1499 152
1500 231
1501 257
1502 141
1503 302
1504 225.666666666667
1505 199.5
1506 262.571428571429
1507 212.166666666667
1508 223.444444444444
1509 216.4
1510 157
1511 185.666666666667
1512 217.333333333333
1513 247.6
1514 203.333333333333
1515 296
1516 238.4
1517 180.25
1518 161
1519 209.75
1520 259
1521 294.333333333333
1522 247
1523 218
1524 325
1525 249
1526 167
1527 213
1528 216.25
1529 284.666666666667
1530 149.5
1531 223
1532 177.8
1533 230.285714285714
1534 234
1535 238
1536 221.666666666667
1537 237
1538 233
1539 251.571428571429
1540 181
1541 439
1542 247.666666666667
1543 241.333333333333
1544 119
1545 187.5
1546 212.333333333333
1547 167.5
1548 161.75
1549 212
1550 159.666666666667
1551 202.5
1552 231.333333333333
1553 203
1554 190.6
1555 205.428571428571
1556 199
1557 246.6
1558 210
1559 232.6
1560 158.666666666667
1561 252.333333333333
1562 198.25
1563 184
1564 235.666666666667
1565 206.25
1566 200.75
1567 202.8
1568 225.333333333333
1569 224
1570 229
1571 200.5
1572 226
1573 220
1574 176
1575 234.142857142857
1576 202.666666666667
1577 251
1578 219.2
1579 118
1580 179
1581 272
1582 222
1583 168.25
1584 170
1585 203.5
1586 331
1587 208.833333333333
1588 205
1589 159
1590 211.5
1591 219.5
1592 199.571428571429
1593 180
1594 130
1595 264.5
1596 304.75
1597 223
1598 199.571428571429
1599 238.8
1600 201.8
1601 262.4
1602 272
1603 185
1604 339.5
1605 215.428571428571
1606 281
1607 216.5
1608 151.25
1609 195.75
1610 187
1611 174.25
1612 84
1613 189.2
1614 226
1615 169
1616 167.75
1617 165.8
1618 238
1619 261
1620 245.5
1621 176.5
1622 212
1623 213.75
1624 137.333333333333
1625 172
1626 233.6
1627 168.75
1628 204
1629 201.25
1630 204.333333333333
1631 256.5
1632 222
1633 235.333333333333
1635 187.75
1636 208.333333333333
1637 221
1638 174.666666666667
1639 305
1640 176.8
1641 280.666666666667
1642 152
1643 259
1644 200
1646 200.6
1647 195
1648 205
1649 185
1650 205
1651 194.833333333333
1654 195
1655 265
1656 191
1657 228.5
1658 182.2
1659 191.666666666667
1660 269
1661 189.25
1662 244.25
1663 191
1664 153
1665 261
1666 190.5
1667 187.2
1668 181
1669 197.333333333333
1670 193.666666666667
1671 170.75
1672 262
1673 271.8
1674 195
1675 207
1676 155
1677 176.571428571429
1678 192
1679 144.2
1680 176.333333333333
1681 203
1682 154
1683 158.777777777778
1684 179
1685 236
1686 170.2
1687 192
1688 181.2
1689 198.333333333333
1690 199.666666666667
1691 195
1692 194.833333333333
1693 231.5
1694 229.75
1695 221.5
1696 202.5
1697 201.666666666667
1698 169.5
1699 167.666666666667
1700 219
1701 162.5
1702 134
1703 199.333333333333
1704 233
1705 175.5
1706 159
1707 461.75
1708 188.5
1709 254
1710 220.5
1711 203
1712 224
1713 187
1714 139
1715 231
1716 148.333333333333
1717 152.666666666667
1718 228
1719 237.666666666667
1720 198.333333333333
1721 223.5
1722 207
1723 222.25
1724 231.333333333333
1725 205
1726 199
1727 325
1728 206
1729 184.166666666667
1730 161.5
1731 167.2
1732 246
1733 196.5
1734 176
1735 173.5
1736 195
1737 204.333333333333
1738 161.7
1739 156.8
1740 227.666666666667
1741 182.5
1742 220
1743 211.125
1744 169
1745 144.75
1746 177.666666666667
1747 256
1748 186
1749 147.6
1750 226.428571428571
1751 229.333333333333
1752 154
1753 163
1754 220.5
1755 161.666666666667
1756 255.666666666667
1757 166.333333333333
1758 127.666666666667
1759 205.857142857143
1760 182.25
1761 226.666666666667
1762 159.666666666667
1763 186.375
1764 199.833333333333
1765 121.666666666667
1766 216.5
1767 152.833333333333
1768 205
1769 207.666666666667
1770 169
1771 315.6
1772 186
1773 138
1774 232
1775 230
1776 149.8
1777 184.75
1778 153.333333333333
1779 200.833333333333
1780 170
1781 135.5
1782 320
1784 201.666666666667
1785 130.5
1786 152
1787 195.6
1788 177.5
1789 203
1790 227.5
1791 193
1792 206
1793 165.333333333333
1794 142.4
1795 223.333333333333
1796 147.142857142857
1797 159.75
1798 208.666666666667
1799 177.5
1800 151
1801 212.666666666667
1802 217
1803 147.666666666667
1804 198.5
1805 205.8
1806 376.75
1807 185.333333333333
1808 140
1809 136.5
1810 158.666666666667
1811 268
1812 150.2
1813 150.5
1814 156.666666666667
1815 176
1816 151
1818 210
1819 101
1821 118
1822 425.666666666667
1823 170.2
1824 220
1825 149
1826 198.666666666667
1827 143.25
1828 184.4
1829 107
1830 226
1831 213
1832 155.5
1833 111.5
1834 184.5
1835 177.5
1836 130
1837 178.666666666667
1839 149.8
1840 125
1841 126.333333333333
1842 145
1843 139.666666666667
1844 186.5
1845 161
1846 313.5
1847 129
1848 136.5
1849 195
1850 253.6
1851 152.333333333333
1852 147.5
1853 168.333333333333
1854 122.333333333333
1855 104.5
1856 93
1857 167.333333333333
1858 302
1859 84
1860 164
1861 193.333333333333
1862 183
1863 89
1864 164.666666666667
1865 183
1866 206.25
1867 719
1868 245
1869 158.5
1870 171.333333333333
1871 195.25
1872 267
1873 175
1874 125
1875 173
1876 107.666666666667
1877 118
1879 127.333333333333
1880 214.5
1881 276
1882 128.5
1883 178
1884 178.75
1885 154.5
1886 225
1887 133.5
1888 182
1889 146
1891 189.666666666667
1893 107.5
1894 243
1895 255.666666666667
1896 140.666666666667
1897 177
1898 112.5
1899 168.5
1900 134
1901 232.5
1903 142.4
1904 194
1905 163.333333333333
1906 151
1907 148
1908 182
1909 180.5
1910 164
1911 102
1912 196
1914 267
1915 186
1916 159
1917 130
1918 209
1920 163
1921 203
1922 178
1923 152
1924 198
1925 116
1926 149.5
1927 121.333333333333
1928 79
1929 212
1930 98
1931 183
1932 136
1933 309
1934 187.333333333333
1935 255
1936 165.333333333333
1937 112
1938 185
1940 137
1941 139.5
1943 148
1944 187.5
1945 164
1946 131
1947 129.5
1948 178
1950 190
1951 105.5
1952 140
1953 261
1954 184
1955 136.5
1956 233
1957 179.5
1959 106
1960 160
1961 281
1962 247.5
1963 204.333333333333
1964 186
1965 251
1966 207.333333333333
1967 190
1968 175.6
1970 205
1971 285
1972 263.666666666667
1974 184
1975 141
1976 146
1977 277.5
1978 214
1980 100
1983 187.5
1984 181
1986 183
1987 219
1988 203
1989 217.5
1990 201
1991 236.333333333333
1992 158
1993 176
1994 246.666666666667
1995 177
1996 186
1997 269
1998 219
2000 184
2004 187
2005 242
2006 324
2007 202
2008 130
2010 164
2013 108
2014 184
2015 270
2016 149
2017 180
2019 160
2022 154
2023 114
2024 150
2025 222
2027 193.5
2028 232
2029 189
2030 172.5
2031 260
2032 269
2033 174
2037 144.5
2040 200
2045 203
2046 222.5
2051 582
2052 313
2054 206
2056 151
2059 213
2062 102
2063 507
2064 289
2068 185
2069 280
2078 187
2081 147.5
2086 134
2091 194
2099 218
2111 178
2116 171
2117 175
2119 200
2121 166
2128 181
2134 142
2137 216
2138 136
2139 131
2140 605
2142 173
2147 182
2153 155
2156 151
2161 236.5
2167 153
2172 389
2173 188
2174 126
2175 418
2180 158
2187 108.5
2191 134
2193 115
2196 95
2198 150
2202 102
2208 477
2211 177
2214 103
2215 160
2216 162
2221 199
2227 123
2232 154
2236 166
2243 130
2245 118
2249 142
2255 63
2256 148
2258 99
2266 200
2270 737
2276 195
2277 106
2278 227
2279 131
2284 128
2288 104
2289 105
2295 127
2297 392.5
2303 125
2310 458
2313 175
2334 371
2339 310
2364 400
2405 487
2433 504
2469 360
2474 362
2529 423
2530 352
2541 318
2553 726
2559 490
2566 306
2575 195
2576 60
2592 217
2596 287
2601 166
2606 345
2611 212
2616 199
2621 436
2624 274
2633 235
2635 448
2639 392
2640 328
2647 306
2649 372
2652 369
2658 198
2659 327
2662 299
2664 296
2665 334.333333333333
2666 309
2669 364.5
2670 505
2672 442
2680 290
2682 348
2688 316
2690 383
2693 124
2696 340
2703 222
2710 363.5
2729 313
2737 277
2747 234
2754 462
2761 268
2767 454
2789 365
}\DataScatter

\pgfplotstableread{
x       y
5 5.03025798196032
33.1212121212121 34.7031162582188
61.2424242424242 66.5961000563076
89.3636363636364 100.457580922182
117.484848484848 135.933087922474
145.606060606061 172.567005330526
173.727272727273 209.813336862829
201.848484848485 247.056182684944
229.969696969697 283.639123985119
258.090909090909 318.901097879542
286.212121212121 352.214962057786
314.333333333333 383.024192422832
342.454545454545 410.873287267191
370.575757575758 435.428475553764
398.69696969697 456.486984038519
426.818181818182 473.974974002742
454.939393939394 487.935864382213
483.060606060606 498.511798110775
511.181818181818 505.921373759828
539.30303030303 510.436538862909
567.424242424242 512.360925681937
595.545454545455 512.011132845653
623.666666666667 509.701704969917
651.787878787879 505.733953155428
679.909090909091 500.388337476835
708.030303030303 493.919891596143
736.151515151515 486.556074204431
764.272727272727 478.496437674215
792.393939393939 469.913569844476
820.515151515152 460.954858040495
848.636363636364 451.744723275545
876.757575757576 442.387064062437
904.878787878788 432.96772705191
933 423.556883924673
961.121212121212 414.211241305399
989.242424242424 404.976044929378
1017.36363636364 395.886863298663
1045.48484848485 386.971151984629
1073.60606060606 378.249609647999
1101.72727272727 369.737342460787
1129.84848484848 361.44485625843
1157.9696969697 353.378896425359
1186.09090909091 345.543154951263
1214.21212121212 337.938862802336
1242.33333333333 330.565284083639
1270.45454545455 323.420126658298
1298.57575757576 316.499882084329
1326.69696969697 309.800106018659
1354.81818181818 303.315648668443
1382.93939393939 297.040843463166
1411.06060606061 290.96966088167
1439.18181818182 285.095833289932
1467.30303030303 279.412955716554
1495.42424242424 273.914566698694
1523.54545454545 268.594212656296
1551.66666666667 263.445498681668
1579.78787878788 258.462128150573
1607.90909090909 253.637933157077
1636.0303030303 248.966897435988
1664.15151515152 244.443173153788
1692.27272727273 240.061092712702
1720.39393939394 235.815176515653
1748.51515151515 231.700137475849
1776.63636363636 227.710882918279
1804.75757575758 223.842514406959
1832.87878787879 220.09032593754
1861 216.449800856573
1889.12121212121 212.916607803842
1917.24242424242 209.486595920277
1945.36363636364 206.155789519317
1973.48484848485 202.920382382643
2001.60606060606 199.776731810558
2029.72727272727 196.721352532013
2057.84848484848 193.750910558362
2085.9696969697 190.862217047668
2114.09090909091 188.052222232192
2142.21212121212 185.318009449997
2170.33333333333 182.656789311996
2198.45454545455 180.065894027942
2226.57575757576 177.542771908423
2254.69696969697 175.084982054719
2282.81818181818 172.690189244161
2310.93939393939 170.356159015199
2339.06060606061 168.0807529537
2367.18181818182 165.861924179797
2395.30303030303 163.697713032928
2423.42424242424 161.586242951371
2451.54545454545 159.525716541582
2479.66666666667 157.514411831878
2507.78787878788 155.550678704479
2535.90909090909 153.632935499527
2564.0303030303 151.759665784502
2592.15151515152 149.929415282289
2620.27272727273 148.140788951144
2648.39393939394 146.392448209839
2676.51515151515 144.683108301348
2704.63636363636 143.011535788583
2732.75757575758 141.376546175836
2760.87878787879 139.777001649812
2789 138.211808934301
}\DataUSL

\begin{tikzpicture}

\definecolor{crimson}{RGB}{220,20,60}
\definecolor{darkgray176}{RGB}{176,176,176}
\definecolor{royalblue}{RGB}{65,105,225}

\begin{axis}[
width=\linewidth,
height=0.5\linewidth,
tick align=outside,
tick pos=left,
x grid style={darkgray176},
xlabel={In Rate (Mpps)},
xmin=0.0, xmax=35.0,
xtick style={color=black},
y grid style={darkgray176},
ylabel={Throughput (Mpps)},
ymin=0.0, ymax=10.0,
ytick style={color=black},
label style={font=\footnotesize},
tick label style={font=\scriptsize},
ylabel style={font=\fontsize{5}{5}\selectfont},
xlabel style={font=\fontsize{5}{5}\selectfont},
tick label style={font=\fontsize{6}{6}\selectfont}
]

\begin{pgfonlayer}{foreground}
\addplot [no markers, thick, crimson]
table [x expr=\thisrow{x}/100.0, y expr=\thisrow{y}/100.0] {\DataUSL};
\end{pgfonlayer}

\begin{pgfonlayer}{background}
\addplot [draw=royalblue, mark=o, only marks]
table [x expr=\thisrow{x}/100.0, y expr=\thisrow{y}/100.0] {\DataScatter};
\end{pgfonlayer}

\end{axis}

\end{tikzpicture}

%% file: figures/eval-results/usl-fitting/5.tex
\pgfplotstableread{
x       y
5 4.99462365591398
6 6.00609756097561
7 7.0062893081761
8 8
9 9
10 9.98425196850394
11 10.9905660377358
12 12
13 13
14 13.9850746268657
15 15.044776119403
16 15.9857142857143
17 16.9365079365079
18 18.0508474576271
19 19
20 19.9310344827586
21 21
22 22
23 22.8958333333333
24 23.9722222222222
25 24.9615384615385
26 26.2894736842105
27 27.08
28 27.9285714285714
29 29.0238095238095
30 29.8974358974359
31 30.7931034482759
32 31.8947368421053
33 32.972972972973
34 33.9117647058824
35 35.1363636363636
36 35.6774193548387
37 37.0357142857143
38 37.5555555555556
39 38.48
40 39.7407407407407
41 40.7586206896552
42 41.8888888888889
43 42.4814814814815
44 44.1818181818182
45 44.6666666666667
46 46.0952380952381
47 47.0909090909091
48 48.08
49 48.8
50 49.6666666666667
51 49.4666666666667
52 51.1333333333333
53 52.5238095238095
54 54.3333333333333
55 54.5909090909091
56 55.6470588235294
57 56.1428571428571
58 58.5294117647059
59 58.6153846153846
60 59.8333333333333
61 61.3333333333333
62 61.4210526315789
63 63
64 63.5
65 63.6666666666667
66 66
67 66.0714285714286
68 68.5714285714286
69 68.1
70 69
71 69.5555555555556
72 72.3333333333333
73 72.875
74 72.6875
75 72.1739130434783
76 76
77 77.25
78 78.2222222222222
79 78.1764705882353
80 79.4666666666667
81 80
82 81.9090909090909
83 83
84 84.4444444444444
85 84.7142857142857
86 85.4
87 86.8
88 87.7142857142857
89 88.9090909090909
90 88.0909090909091
91 90
92 92.3
93 96.0833333333333
94 93.75
95 95.0714285714286
96 96.1111111111111
97 96.4166666666667
98 98.5714285714286
99 98.8181818181818
100 99.9090909090909
101 102.2
102 100.727272727273
103 101.8
104 104.454545454545
105 102
106 105.666666666667
107 107.333333333333
108 104.75
109 109.166666666667
110 135.571428571429
111 110
112 113
113 136.636363636364
114 112.416666666667
115 115.25
116 114.571428571429
117 115.923076923077
118 118.272727272727
119 119
120 118.25
121 120
122 121.857142857143
123 119.545454545455
124 166.75
125 126.4
126 126.111111111111
127 127
128 127.076923076923
129 128.1
130 129.5
131 131.125
132 133.166666666667
133 133.2
134 133.461538461538
135 134
136 136.428571428571
137 136.583333333333
138 140.714285714286
139 141.588235294118
140 136.666666666667
141 140.8
142 144.5
143 142.357142857143
144 145
145 145.428571428571
146 143.933333333333
147 146
148 148.333333333333
149 149.25
150 150.714285714286
151 151.285714285714
152 151
153 151.333333333333
154 154
155 155.727272727273
156 156.2
157 156
158 158.583333333333
159 172.181818181818
160 159.818181818182
161 160
162 161.666666666667
163 162.571428571429
164 162.625
165 164.888888888889
166 165.75
167 172
168 168.1
169 156.555555555556
170 190.142857142857
171 170.375
172 172
173 169.833333333333
174 173.75
175 174.555555555556
176 177.111111111111
177 177.111111111111
178 178
179 179.9
180 191
181 181.2
182 183.142857142857
183 183
184 184.75
185 184.571428571429
186 186
187 186.571428571429
188 184.5
189 186.375
190 190.5
191 191.75
192 192.25
193 190.375
194 192.8
195 195.444444444444
196 195.428571428571
197 196.375
198 197.666666666667
199 198.4375
200 194.4
201 200.875
202 201.125
203 201.5
204 204.272727272727
205 204.857142857143
206 203.071428571429
207 207.090909090909
208 208.384615384615
209 209
210 211.333333333333
211 211.5
212 233.222222222222
213 211.75
214 215.142857142857
215 215.2
216 216.142857142857
217 215.5
218 215.0625
219 220.111111111111
220 220.777777777778
221 221.666666666667
222 228.6
223 225.7
224 223
225 225
226 195.285714285714
227 224.941176470588
228 228
229 230.555555555556
230 228.833333333333
231 234.333333333333
232 270.333333333333
233 232.4
234 233.727272727273
235 235.375
236 236.444444444444
237 248.4
238 237.2
239 241.384615384615
240 265.5
241 239.428571428571
242 272.545454545455
243 240
244 243
245 244.875
246 246
247 241.7
248 247.875
249 248.714285714286
250 249.933333333333
251 254.444444444444
252 257.25
253 253
254 250.375
255 255.846153846154
256 256.111111111111
257 258
258 257.285714285714
259 258.777777777778
260 260.5
261 260.714285714286
262 268.875
263 263.428571428571
264 263
265 265
266 267.333333333333
267 266.625
268 266.083333333333
269 267.75
270 338
271 269.888888888889
272 269.1
273 296
274 262.444444444444
275 274.692307692308
276 276.4
277 279.8
278 278.2
279 280.2
280 280.375
281 281.25
282 291
283 284.090909090909
284 286.125
285 285.4
286 286.777777777778
287 287.333333333333
288 288.333333333333
289 292.333333333333
290 289.857142857143
291 302.357142857143
292 317.166666666667
293 296
294 295
295 295.428571428571
296 335.166666666667
297 348.111111111111
298 285.5
299 303.25
300 298.5
301 301.428571428571
302 333.875
303 326.909090909091
304 304.75
305 283.6
306 305.666666666667
307 307.363636363636
308 343.555555555556
309 307.230769230769
310 317
311 312.272727272727
312 310.875
313 311.666666666667
314 311.3125
315 314.583333333333
316 342.5
317 316.625
318 370.5
319 313.466666666667
320 320.6
321 340.642857142857
322 361
323 322.666666666667
324 321.1
325 326.142857142857
326 326.181818181818
327 327.090909090909
328 328.789473684211
329 359.4
330 330.090909090909
331 341.588235294118
332 331
333 366.090909090909
334 334
335 329.444444444444
336 336.1
337 335.9
338 331
339 339.090909090909
340 344.444444444444
341 323.4
342 368.857142857143
343 341.333333333333
344 344.4
345 366
346 410
347 348
348 347.1
349 392.428571428571
350 349
351 342.875
352 344.846153846154
353 384.545454545455
354 354.142857142857
355 369.333333333333
356 398.083333333333
357 358.714285714286
358 344
359 357.5
360 361.375
361 359.5
362 359.142857142857
363 362.333333333333
364 376
365 363.666666666667
366 365.888888888889
367 368
368 384.5
369 359.9
370 406.333333333333
371 396.888888888889
372 383.333333333333
373 387.909090909091
374 323
375 364
376 373.818181818182
377 374
378 419.8
379 380.333333333333
380 359.375
381 392.090909090909
382 379.428571428571
383 396.1
384 383.8
385 399
386 421.083333333333
387 519.75
388 472.5
389 382.4
390 403.666666666667
391 391.571428571429
392 391.9
393 383.875
394 460.333333333333
395 371.875
396 370.2
397 398
398 382.4
399 398.5
400 374.666666666667
401 421.333333333333
402 395.375
403 389.5
404 364
405 380.545454545455
406 413.428571428571
407 422.1
408 408.625
409 433.444444444444
410 410.285714285714
411 419.777777777778
412 427
413 413
414 422.545454545455
415 416.125
416 416.5
417 424.076923076923
418 416
419 418.5
420 415.5
421 432.5
422 427
423 413.6
424 424.625
425 468.6
426 426.166666666667
427 432.666666666667
428 417.363636363636
429 428.375
430 430.615384615385
431 422.4
432 436.6
433 435.625
434 463.222222222222
435 399.5
436 434.428571428571
437 432.375
438 435.642857142857
439 401.666666666667
440 459.5
441 458.666666666667
442 441.230769230769
443 408.8
444 461.1
445 470.333333333333
446 402.375
447 455.3
448 447.375
449 513.25
450 450
451 436
452 439.384615384615
453 451.666666666667
454 447.777777777778
455 470.272727272727
456 452.142857142857
457 377
458 455.571428571429
459 458.333333333333
460 457.692307692308
461 415.857142857143
462 409.25
463 460.444444444444
464 464.9
465 444.857142857143
466 467
467 483.2
468 452.071428571429
469 472.714285714286
470 479.9
471 470.666666666667
472 473.571428571429
473 474.666666666667
474 468.125
475 462
476 476.555555555556
477 443.666666666667
478 478.909090909091
479 453.125
480 479.166666666667
481 472.5
482 480.615384615385
483 451.666666666667
484 451.142857142857
485 486.5
486 486.833333333333
487 486.333333333333
488 486.666666666667
489 471.9
490 462.333333333333
491 491.333333333333
492 490.736842105263
493 494.4
494 482
495 514.214285714286
496 429.333333333333
497 535.1
498 499
499 489.7
500 461.5
501 504.857142857143
502 501.933333333333
503 441.2
504 503.625
505 656.75
506 460.545454545455
507 507.714285714286
508 496.727272727273
509 574.333333333333
510 551.111111111111
511 507.666666666667
512 512.666666666667
513 477.666666666667
514 523.125
515 519.133333333333
516 495.8
517 499.666666666667
518 462.153846153846
519 493.666666666667
520 490.5
521 520.75
522 484.2
523 522.8
524 437.1
525 488.111111111111
526 523.533333333333
527 538.222222222222
528 491
529 473
530 492.25
531 525.9
532 545.25
533 520.4
534 533.5
535 407.333333333333
536 475.545454545455
537 484
538 442
539 538
540 518.428571428571
541 527.428571428571
542 422.333333333333
543 511.6
544 484.111111111111
545 443.6
546 470.4
547 561
548 520.777777777778
549 558.181818181818
550 506
551 540.75
552 520.444444444444
553 439.666666666667
554 419.25
555 506.75
556 551.166666666667
557 540
558 557.571428571429
559 463.727272727273
560 485.142857142857
561 495
562 480.727272727273
563 519.777777777778
564 533.25
565 564.888888888889
566 570
567 701.333333333333
568 545.888888888889
569 433.8
570 602.25
571 569.5
572 465
573 489
574 527
575 580
576 523.333333333333
577 521.2
578 512.5
579 563.555555555556
580 503.25
581 579.4
582 581
583 573.307692307692
584 540
585 495.833333333333
586 567
587 563.75
588 555.5
589 573.5
590 594.571428571429
591 508.857142857143
592 549.75
593 545
594 449
595 527.888888888889
596 626.166666666667
597 618.833333333333
598 489.111111111111
599 586.833333333333
600 614.666666666667
601 563
602 605.333333333333
603 605.166666666667
604 567.125
605 658.75
606 650.5
607 579.666666666667
608 498.833333333333
609 609.666666666667
610 520.25
611 569.166666666667
612 542.625
613 702.75
614 574.875
615 608.555555555556
616 685
617 573.571428571429
618 519.5
619 559.7
620 558.3
621 624.5
622 608
623 623.666666666667
624 595.5
625 618.181818181818
626 610.375
627 554.625
628 579.625
629 582.428571428571
630 625
631 647.6
632 632.666666666667
633 571.833333333333
634 615.714285714286
635 648.714285714286
636 627.6
637 626.571428571429
638 636.833333333333
639 615.166666666667
640 619.5
641 587.111111111111
642 503.285714285714
643 515
644 634.571428571429
645 633.833333333333
646 622.090909090909
647 613.2
648 612.666666666667
649 653
650 650.666666666667
651 629.8
652 609.416666666667
653 582.555555555556
654 557.5
655 627.636363636364
656 655.333333333333
657 524
658 550.833333333333
659 638.5
660 630.9
661 587
662 657.833333333333
663 625.875
664 625
665 616.8
666 634.6
667 489.125
668 624.5
669 702.4
670 671.5
671 675.153846153846
672 507.5
673 673.1
674 622.4
675 670.375
676 556.857142857143
677 668.2
678 619.166666666667
679 631
680 685.142857142857
681 641.222222222222
682 713
683 683.666666666667
684 665.5
685 644.727272727273
686 671.555555555556
687 444.25
688 684.857142857143
689 666.875
690 628.428571428571
691 518
692 665.5
693 633.285714285714
694 712.076923076923
695 535.727272727273
696 650
697 698.2
698 574.333333333333
699 653.181818181818
700 668.9
701 528.857142857143
702 670.909090909091
703 559.285714285714
704 661
705 618.625
706 633.9
707 674.25
708 687.888888888889
709 627.125
710 539
711 687.222222222222
712 635.3
713 570
714 599.636363636364
715 692.875
716 643.857142857143
717 668.428571428571
718 663.2
719 608
720 600.125
721 648.818181818182
722 639.636363636364
723 686.555555555556
724 875
725 564.875
726 821.090909090909
727 540.5
728 722
729 714.333333333333
730 654.5
731 728
732 597.166666666667
733 710.25
734 679.222222222222
735 708.75
736 619.75
737 696.75
738 648.6
739 549.142857142857
740 629.181818181818
741 666.333333333333
742 644.166666666667
743 601.333333333333
744 684.5
745 689.714285714286
746 724.909090909091
747 596.333333333333
748 719.333333333333
749 687.2
750 698.777777777778
751 600
752 750.5
753 673.125
754 714.625
755 642.875
756 447.857142857143
757 597.4
758 723.846153846154
759 636.666666666667
760 579.25
761 712.2
762 663.4
763 870.142857142857
764 697.363636363636
765 498
766 706.25
767 653.6
768 568
769 713.444444444444
770 607.25
771 602.2
772 680.428571428571
773 696.875
774 639.5
775 602.5
776 616.25
777 621
778 753.428571428571
779 434.666666666667
780 554.444444444444
781 538.1
782 780.25
783 651.444444444444
784 497.666666666667
785 661.166666666667
786 617.25
787 607.818181818182
788 683.444444444444
789 665.666666666667
790 592.625
791 730.8
792 648.375
793 790.2
794 732.8
795 537.285714285714
796 716.5
797 599.4
798 612.5
799 634
800 442.5
801 487
802 672.333333333333
803 628.818181818182
804 661.75
805 680.75
806 681
807 661.333333333333
808 781.8
809 682.25
810 624.1
811 730.5
812 586.222222222222
813 630.111111111111
814 615.5
815 595.875
816 500
817 634.8
818 658
819 735.9
820 645.222222222222
821 566
822 753.666666666667
823 699.375
824 609
825 581
826 584.25
827 699.714285714286
828 408.166666666667
829 591
830 568.142857142857
831 597
832 624
833 584
834 623.857142857143
835 586.428571428571
836 443.333333333333
837 681.428571428571
838 650.8
839 481.8
840 491
841 760.285714285714
842 505.166666666667
843 722.8
844 721.6
845 492
846 830.75
847 635.333333333333
848 393.142857142857
849 528.272727272727
850 588.857142857143
851 578.166666666667
852 532.285714285714
853 701
854 715.5
855 664.5
856 660.4
857 681.5
858 679.25
859 607
860 769.888888888889
861 751.5
862 578.5
863 635.4
864 531.75
865 703.5
866 634.166666666667
867 856.923076923077
868 696.2
869 672.666666666667
870 520.714285714286
871 675.5
872 590.545454545455
873 488.428571428571
874 787.666666666667
875 720.833333333333
876 534.75
877 511.333333333333
878 530.6
879 599
880 546.333333333333
881 742.428571428571
882 582.916666666667
883 633.571428571429
884 664.8
885 650.416666666667
886 592.6
887 757.857142857143
888 612.833333333333
889 547.666666666667
890 491.428571428571
891 536.666666666667
892 488.2
893 381.5
894 738
895 694.6
896 718.666666666667
897 708.571428571429
898 615.428571428571
899 625
900 695.6
901 561.625
902 540.333333333333
903 657
904 474.75
905 562
906 728.571428571429
907 516.333333333333
908 657.142857142857
909 568.125
910 672.2
911 435
912 637.285714285714
913 511.4
914 729.75
915 582.5
916 440.333333333333
917 593
918 679.833333333333
919 494
920 382.333333333333
921 589.6
922 818.25
923 692.3
924 536.5
925 672.833333333333
926 555.571428571429
927 611.285714285714
928 716.428571428571
929 730.8
930 774.875
931 665.6
932 568.916666666667
933 780.5
934 762.5
935 684.444444444444
936 667
937 628.8
938 726.5625
939 690
940 727.333333333333
941 555.714285714286
942 588
943 636
944 772.5
945 654.4
946 730.6
947 650.666666666667
948 663.5
949 586.75
950 794.833333333333
951 470.714285714286
952 673.166666666667
953 733.571428571429
954 572.444444444444
955 598.8
956 776.4
957 543.5
958 960.5
959 749.666666666667
960 466.2
961 616.25
962 544.444444444444
963 456
964 503
965 567.090909090909
966 688.75
967 502.727272727273
968 529.7
969 458.2
970 796
971 573.5
972 529.666666666667
973 350.125
974 688
975 587.444444444444
976 559
977 525.857142857143
978 518
979 734.25
980 828
981 405.428571428571
982 540.666666666667
983 444.833333333333
984 461.428571428571
985 660.5
986 491.8
987 552.8
988 554.2
989 691.222222222222
990 442.625
991 432.25
992 559.375
993 685.571428571429
994 670
995 412.666666666667
996 557.25
997 528.25
998 648
999 507.666666666667
1000 393.2
1001 493.571428571429
1002 488.818181818182
1003 663.5
1004 583.444444444444
1005 580.428571428571
1006 437.6
1007 420.333333333333
1008 642.666666666667
1009 624.428571428571
1010 518.4
1011 511.666666666667
1012 658
1013 404
1014 358.25
1015 370.125
1016 732.75
1017 453.2
1018 641.25
1019 678.5
1020 517.666666666667
1021 542.888888888889
1022 420.75
1023 290.666666666667
1024 348.8
1025 570.5
1026 585.857142857143
1027 408.833333333333
1028 394
1029 524.75
1030 379.75
1031 369.5
1032 649.5
1033 580.4
1034 696.5
1035 331.25
1036 644.8
1037 567
1038 333
1039 501.181818181818
1040 656.8
1041 657
1042 605.285714285714
1043 424.333333333333
1044 606.666666666667
1045 434.714285714286
1046 450.555555555556
1047 818.666666666667
1048 577.8
1049 953.666666666667
1050 441.857142857143
1051 646.5
1052 531.333333333333
1053 477.555555555556
1054 765.4
1055 404.166666666667
1056 487.1
1057 411.5
1058 491.333333333333
1059 513.666666666667
1060 458.666666666667
1061 472
1062 612.777777777778
1063 556.571428571429
1064 612.714285714286
1065 446
1066 778.25
1067 686.2
1068 504.111111111111
1069 459.5
1070 567.75
1071 663
1072 572.666666666667
1073 665
1074 462.555555555556
1075 512.666666666667
1076 484.666666666667
1077 523.25
1078 445.333333333333
1079 535.25
1080 442.833333333333
1081 438.666666666667
1082 564.777777777778
1083 519.4
1084 644.666666666667
1085 381
1086 359.333333333333
1087 711.333333333333
1088 399
1089 513
1090 584.142857142857
1091 779.666666666667
1092 512.5
1093 843
1094 448.833333333333
1095 548.181818181818
1096 658.5
1097 774.857142857143
1098 371.666666666667
1099 333
1100 514.5
1101 595
1102 472
1103 842.5
1104 796
1105 768.4
1106 465.666666666667
1107 524
1108 262.5
1109 452.5
1110 532
1111 693.142857142857
1112 661
1113 387.5
1114 555.111111111111
1115 835.666666666667
1116 354.5
1117 471.555555555556
1118 405.5
1119 919
1120 635
1121 348
1122 537.857142857143
1123 539.285714285714
1124 620.285714285714
1125 542.25
1126 671.833333333333
1127 692.285714285714
1128 789.166666666667
1129 567.375
1130 731.4
1131 432.25
1132 624
1133 400.8
1134 758.5
1135 364
1136 331
1137 824
1138 713
1139 589.714285714286
1140 528
1141 440.5
1142 324.666666666667
1143 455.333333333333
1144 363
1145 879
1146 689
1147 549.2
1148 412.666666666667
1149 373.833333333333
1150 390.8
1151 632.25
1152 475.428571428571
1153 478.25
1154 239.5
1155 448.222222222222
1156 426.666666666667
1157 646.285714285714
1158 758.6
1159 426
1160 387
1161 430.5
1162 358.4
1163 795.666666666667
1164 319
1165 471.333333333333
1166 337
1167 280.5
1168 294.5
1169 313.666666666667
1170 397.142857142857
1171 350.666666666667
1172 449.428571428571
1173 386
1174 525.666666666667
1175 314.2
1176 541.75
1177 284.5
1178 684
1179 415.375
1180 333.142857142857
1181 413.8
1182 354
1183 443
1184 449.666666666667
1185 528.5
1186 301.2
1187 342.666666666667
1188 341
1189 332.666666666667
1190 427.375
1191 375.666666666667
1192 302
1193 337
1194 271.5
1195 356
1196 741
1197 646.5
1198 467.8
1199 354.571428571429
1201 239.25
1202 246
1203 589.25
1204 364.666666666667
1205 312
1206 624
1207 344
1208 458.125
1209 287.666666666667
1210 274.5
1211 477
1212 328.833333333333
1213 274.5
1214 322.666666666667
1215 263
1216 290
1217 443.333333333333
1218 321.5
1219 508.333333333333
1220 221
1221 439.75
1222 349
1223 531.8
1224 358.666666666667
1225 548.5
1226 440
1227 438.5
1228 400.555555555556
1229 316.5
1230 311
1231 617.333333333333
1232 292.25
1233 481.333333333333
1234 342.4
1235 421.9
1236 578.5
1238 523.8
1239 373
1240 383.857142857143
1241 413.75
1242 286
1243 408.333333333333
1244 488.571428571429
1245 322.25
1246 364.272727272727
1247 524
1248 502
1249 305.333333333333
1250 333.5
1251 504.25
1252 266.666666666667
1253 401.222222222222
1254 526
1255 509.25
1256 248
1257 312.666666666667
1258 394.857142857143
1259 264
1260 432.8
1261 536
1262 515
1263 462.5
1264 194
1265 365.6
1266 499.5
1267 332.333333333333
1268 512
1269 419.5
1270 346.8
1271 460.222222222222
1272 570.666666666667
1273 546
1274 396.666666666667
1275 299.5
1276 324.75
1277 471.285714285714
1278 369.8
1279 484.833333333333
1280 500.2
1281 438.125
1282 263.5
1283 449.4
1284 499.428571428571
1285 292
1286 483.5
1287 556.5
1288 540.2
1289 308.25
1290 579
1291 286.5
1292 511.75
1293 359.25
1294 390.666666666667
1295 343.8
1296 370.5
1297 602.5
1298 417.5
1299 400.2
1300 461
1301 505.285714285714
1302 483.25
1303 341
1304 386.666666666667
1305 344.8
1306 356.5
1307 360.666666666667
1308 440
1309 361.833333333333
1310 354.2
1311 256.75
1312 638.714285714286
1313 480.666666666667
1314 145
1315 225.333333333333
1316 188.5
1317 280.666666666667
1318 361.5
1319 300.714285714286
1320 365
1321 494
1322 740.666666666667
1323 310.666666666667
1324 539
1325 309
1326 564
1327 343
1328 376.8
1329 386.5
1330 368.75
1331 421
1332 461.333333333333
1333 386.333333333333
1334 394.666666666667
1335 445.75
1336 312.333333333333
1337 590.75
1338 312
1339 430.8
1340 396.875
1341 379.5
1342 286.6
1343 346.888888888889
1344 435.5
1345 290
1346 361.833333333333
1347 375
1348 369.6
1349 313.857142857143
1350 171
1351 321.4
1352 355.666666666667
1353 282
1354 441.5
1355 339
1356 209
1357 449.2
1358 247.8
1359 345.5
1360 472.5
1361 245
1362 402.166666666667
1363 293
1364 347.166666666667
1365 562
1366 363.333333333333
1367 321.428571428571
1368 256.333333333333
1369 492.5
1370 259.75
1371 368.6
1372 409.428571428571
1373 501.4
1374 299.5
1375 544
1376 270.75
1377 379.8
1378 439.428571428571
1379 218
1380 382
1381 292
1382 405.2
1383 418.142857142857
1384 474
1385 353.25
1386 209.333333333333
1387 263
1388 465.4
1389 287
1390 283.5
1391 425
1392 391.4
1393 432
1394 365.333333333333
1395 309
1396 420.5
1397 448.3
1398 345
1399 248
1400 270.5
1401 406.166666666667
1402 233.2
1403 451.8
1404 358.5
1405 269
1406 309.8
1407 345
1408 261.714285714286
1409 550
1410 360
1411 402
1412 196
1413 220
1414 269
1415 316
1416 502.333333333333
1417 241.2
1418 464.833333333333
1419 293
1420 358
1421 312.6
1422 295.25
1423 469
1424 318.5
1425 420.166666666667
1426 314.8
1427 244.8
1428 313.5
1429 325.833333333333
1430 382.666666666667
1431 216.25
1432 292.571428571429
1433 294
1434 225.166666666667
1435 263.5
1436 413
1437 280
1438 551.777777777778
1439 354.5
1440 253.5
1441 328.833333333333
1442 197.666666666667
1443 313
1444 273.5
1445 406
1446 290.714285714286
1447 351.090909090909
1448 306.166666666667
1449 332.25
1450 130
1451 348.25
1452 290.75
1453 454
1454 261
1455 252.6
1456 307.333333333333
1457 295.666666666667
1458 355
1459 343.25
1460 268
1461 341.222222222222
1462 256
1463 293
1464 398.833333333333
1465 438.8
1466 312.666666666667
1467 229.666666666667
1468 234.666666666667
1469 265.285714285714
1470 241
1471 695.5
1472 447.5
1473 259
1474 299
1475 288.8
1476 284.2
1477 266.166666666667
1478 303.166666666667
1479 265.2
1480 328
1481 280.666666666667
1482 305
1483 273.666666666667
1484 511.285714285714
1485 324.25
1486 246.333333333333
1487 272
1488 276.285714285714
1489 346.2
1490 333.666666666667
1491 861
1492 371.5
1493 499.5
1494 339.333333333333
1495 291.666666666667
1496 277.75
1497 298.75
1498 324
1499 235
1500 326.166666666667
1501 310.25
1502 193.5
1503 488
1504 188
1505 288.5
1506 280.857142857143
1507 355.833333333333
1508 321.666666666667
1509 280.2
1510 213.5
1511 292.666666666667
1512 298.583333333333
1513 253.6
1514 309.666666666667
1515 292.375
1516 290.2
1517 211.75
1518 179
1519 287.625
1520 383.75
1521 369.333333333333
1522 266.75
1523 432.666666666667
1524 297.333333333333
1525 247
1526 282.333333333333
1527 267.25
1528 210.25
1529 266
1530 180
1531 348.333333333333
1532 233
1533 354.571428571429
1534 323.25
1535 409
1536 247.666666666667
1537 243
1538 255.666666666667
1539 321
1540 253.5
1541 848.5
1542 292.333333333333
1543 303.333333333333
1544 170.333333333333
1545 356
1546 233.666666666667
1547 216.25
1548 242.75
1549 406
1550 300.333333333333
1551 335.75
1552 268
1553 129
1554 305.6
1555 253.714285714286
1556 293
1557 257.4
1558 257.333333333333
1559 283.4
1560 289
1561 249.333333333333
1562 245.5
1563 251.333333333333
1564 204.666666666667
1565 246
1566 310.5
1567 258.4
1568 242
1569 280.666666666667
1570 215
1571 273.5
1572 253
1573 276.5
1574 297.5
1575 264.285714285714
1576 237.666666666667
1577 244.666666666667
1578 207.6
1579 257.5
1580 282.666666666667
1581 241
1582 289
1583 289.75
1584 196
1585 242
1586 256
1587 344
1588 372.5
1589 204
1590 226.5
1591 260
1592 333.714285714286
1593 187
1594 130.5
1595 308.5
1596 302.25
1597 371.2
1598 256.714285714286
1599 235.4
1600 259.4
1601 294.2
1602 287
1603 272
1604 480.25
1605 252.857142857143
1606 256
1607 270
1608 262.75
1609 292.75
1610 293.5
1611 238.75
1612 189
1613 298.8
1614 216
1615 274.25
1616 259.75
1617 275.8
1618 382.5
1619 306.857142857143
1620 248
1621 271
1622 213.666666666667
1623 275.5
1624 195
1625 259
1626 248.2
1627 208.5
1628 252.5
1629 236.5
1630 312.333333333333
1631 218.666666666667
1632 327
1633 280.333333333333
1635 272.75
1636 289.333333333333
1637 236.6
1638 165.333333333333
1639 306
1640 255.6
1641 193.666666666667
1642 229
1643 527
1644 223
1646 246
1647 297
1648 283
1649 306.5
1650 286
1651 241.166666666667
1654 230.666666666667
1655 329
1656 237.666666666667
1657 157.5
1658 282
1659 247.333333333333
1660 213.666666666667
1661 218.5
1662 280
1663 253
1664 208.75
1665 152
1666 219
1667 283.2
1668 206.4
1669 228.833333333333
1670 266
1671 303.25
1672 619
1673 306.2
1674 235
1675 269
1676 155
1677 227.285714285714
1678 312.5
1679 213.2
1680 242.333333333333
1681 196.75
1682 292.5
1683 222.777777777778
1684 212.5
1685 352.8
1686 217.8
1687 217
1688 289
1689 257.333333333333
1690 250.666666666667
1691 235
1692 231.666666666667
1693 184
1694 338.25
1695 270
1696 222.5
1697 199.666666666667
1698 186.5
1699 170
1700 248.666666666667
1701 266.333333333333
1702 204.666666666667
1703 192.666666666667
1704 163.5
1705 303.5
1706 216.6
1707 278.5
1708 225.833333333333
1709 247.5
1710 197.5
1711 202.2
1712 290.666666666667
1713 166
1714 158.333333333333
1715 222
1716 201.666666666667
1717 191
1718 264.6
1719 256
1720 198
1721 256.5
1722 211.166666666667
1723 282.75
1724 324
1725 201.833333333333
1726 276.2
1727 358
1728 217.333333333333
1729 217.833333333333
1730 370.5
1731 189
1732 321
1733 243
1734 202
1735 247.5
1736 276.833333333333
1737 259.5
1738 207
1739 163
1740 251.666666666667
1741 178
1742 183
1743 304.125
1744 142
1745 202.5
1746 218
1747 459
1748 198.8
1749 235
1750 279
1751 282.166666666667
1752 143
1753 167
1754 245.5
1755 217.666666666667
1756 257
1757 174
1758 185.666666666667
1759 209.714285714286
1760 217.125
1761 240.666666666667
1762 182.333333333333
1763 265.875
1764 225.5
1765 271.666666666667
1766 238
1767 205.333333333333
1768 348
1769 225
1770 145.5
1771 396.6
1772 290
1773 184.5
1774 148
1775 246
1776 218.2
1777 273
1778 230.666666666667
1779 283
1780 172
1781 172
1782 287
1784 267.333333333333
1785 170.5
1786 163.8
1787 194.6
1788 220.5
1789 205
1790 203
1791 287.5
1792 231.5
1793 182.333333333333
1794 205
1795 182.666666666667
1796 208.142857142857
1797 253.25
1798 174.666666666667
1799 259.625
1800 203.666666666667
1801 228.666666666667
1802 337
1803 245.666666666667
1804 240
1805 213.6
1806 446.5
1807 201
1808 178
1809 169.5
1810 204.333333333333
1811 146
1812 203.6
1813 263
1814 209.666666666667
1815 179
1816 159.5
1818 215
1819 200
1821 180.8
1822 517
1823 151.2
1824 271.5
1825 223
1826 227.333333333333
1827 173.75
1828 241.6
1829 140
1830 291
1831 263
1832 159
1833 167.5
1834 226
1835 267.5
1836 209
1837 328
1839 241.6
1840 195.5
1841 232.333333333333
1842 269.666666666667
1843 183.666666666667
1844 274
1845 119
1846 306
1847 167.333333333333
1848 141
1849 179.333333333333
1850 295.6
1851 215
1852 181
1853 226.666666666667
1854 150.333333333333
1855 201.5
1856 358
1857 222
1858 271
1859 269
1860 188
1861 174.833333333333
1862 167.333333333333
1863 159
1864 204.666666666667
1865 160
1866 181.25
1867 846
1868 206
1869 161
1870 177.333333333333
1871 201.25
1872 216
1873 197.75
1874 136
1875 169
1876 167.666666666667
1877 133
1879 188.333333333333
1880 177.5
1881 254
1882 147.5
1883 161
1884 163.5
1885 155.75
1886 239
1887 132
1888 225
1889 271.5
1891 223
1893 185.5
1894 164.5
1895 265.666666666667
1896 154.333333333333
1897 238
1898 139.5
1899 265
1900 148
1901 189.5
1903 234
1904 205
1905 210
1906 162
1907 143
1908 186
1909 240.5
1910 130
1911 137.666666666667
1912 210.5
1914 287
1915 203
1916 221
1917 253
1918 162
1920 165.333333333333
1921 172
1922 133.5
1923 324
1924 449
1925 182
1926 218.5
1927 149.666666666667
1928 142
1929 151
1930 165
1931 297
1932 191.666666666667
1933 239
1934 185.333333333333
1935 278
1936 226
1937 149
1938 196.25
1940 175
1941 165
1943 257
1944 373
1945 124
1946 340
1947 300
1948 392
1950 289.25
1951 139
1952 151.5
1953 209.5
1954 212.75
1955 196
1956 279
1957 244.75
1959 274
1960 267
1961 305.5
1962 299.5
1963 212.666666666667
1964 255.5
1965 162
1966 230.666666666667
1967 187
1968 249.4
1970 142
1971 259
1972 274.333333333333
1974 410
1975 200.666666666667
1976 211.5
1977 188.5
1978 177
1980 217
1983 188.5
1984 261
1986 207.666666666667
1987 188
1988 192
1989 322.5
1990 106
1991 268.666666666667
1992 203.5
1993 306.5
1994 305.666666666667
1995 176
1996 129
1997 214
1998 129
2000 211.5
2004 194
2005 311
2006 317
2007 193
2008 197
2010 119
2013 213
2014 282.5
2015 306
2016 228
2017 206
2019 306
2022 261
2023 263
2024 377
2025 194
2027 235.5
2028 242
2029 171
2030 203
2031 177
2032 471
2033 332
2037 186
2040 168
2045 167
2046 323
2051 693
2052 319
2054 241
2056 229
2059 235
2062 157
2063 651
2064 196
2068 210
2069 164
2078 247
2081 296.5
2086 261
2091 215
2099 266
2111 160
2116 234
2117 158
2119 183
2121 159
2128 168
2134 113
2137 254
2138 247
2139 211
2140 483
2142 113
2147 208
2153 184
2156 211
2161 419.5
2167 211
2172 439
2173 165
2174 204
2175 403
2180 178
2187 151
2191 169
2193 160
2196 217
2198 102
2202 142
2208 641
2211 124.5
2214 109
2215 156
2216 186
2221 150
2227 158
2232 249
2236 141
2243 167
2245 154
2249 143
2255 154
2256 149
2258 124
2266 165
2270 436
2276 255
2277 178
2278 203
2279 181
2284 136
2288 251
2289 190
2295 84
2297 354.5
2303 167
2310 384
2313 254
2334 614
2339 346
2364 541
2405 449
2433 230
2469 423
2474 267
2529 314
2530 328
2541 520
2553 253
2559 522
2566 393
2575 426
2576 765
2592 373
2596 343
2601 161
2606 211
2611 128
2616 95
2621 290
2624 424
2633 417
2635 404
2639 365
2640 252
2647 612
2649 362
2652 312
2658 363.333333333333
2659 388
2662 477
2664 325
2665 276.333333333333
2666 329
2669 264
2670 424
2672 340
2680 262
2682 341
2688 391
2690 248
2693 253
2696 429
2703 279
2710 354
2729 418
2737 292
2747 223
2754 499
2761 272
2767 356
2789 413
}\DataScatter

\pgfplotstableread{
x       y
5 5.02697929050895
33.1212121212121 34.5370433929567
61.2424242424242 66.0608755718823
89.3636363636364 99.4257037665993
117.484848484848 134.388426723956
145.606060606061 170.634458883088
173.727272727273 207.781079099084
201.848484848485 245.385910431703
229.969696969697 282.960647606397
258.090909090909 319.989499903597
286.212121212121 355.951131117772
314.333333333333 390.342288934468
342.454545454545 422.700958114236
370.575757575758 452.626840328674
398.69696969697 479.797281645296
426.818181818182 503.977376900664
454.939393939394 525.023752029662
483.060606060606 542.882304725097
511.181818181818 557.580828867489
539.30303030303 569.217866121263
567.424242424242 577.949289500425
595.545454545455 583.974056305877
623.666666666667 587.520336331156
651.787878787879 588.832903008829
679.909090909091 588.162340228187
708.030303030303 585.756316821594
736.151515151515 581.852942804165
764.272727272727 576.676055465558
792.393939393939 570.432184699241
820.515151515152 563.308903229296
848.636363636364 555.47426374956
876.757575757576 547.077047226143
904.878787878788 538.247583000913
933 529.098943182927
961.121212121212 519.728355328842
989.242424242424 510.218715169919
1017.36363636364 500.640113481382
1045.48484848485 491.051317638119
1073.60606060606 481.501169213721
1101.72727272727 472.02987479832
1129.84848484848 462.670178831018
1157.9696969697 453.448415474288
1186.09090909091 444.385442143777
1214.21212121212 435.497460887757
1242.33333333333 426.796735913464
1270.45454545455 418.292216602434
1298.57575757576 409.990075668285
1326.69696969697 401.894171932975
1354.81818181818 394.006446711782
1382.93939393939 386.327262132907
1411.06060606061 378.855688965745
1439.18181818182 371.589750754448
1467.30303030303 364.526630291062
1495.42424242424 357.662843740493
1523.54545454545 350.994387062015
1551.66666666667 344.516858765945
1579.78787878788 338.225562501048
1607.90909090909 332.115592486873
1636.0303030303 326.181904381957
1664.15151515152 320.419373809187
1692.27272727273 314.822844438456
1720.39393939394 309.387167249009
1748.51515151515 304.107232354465
1776.63636363636 298.977994567773
1804.75757575758 293.994493707011
1832.87878787879 289.15187049203
1861 284.445378753101
1889.12121212121 279.870394562735
1917.24242424242 275.422422808261
1945.36363636364 271.097101642949
1973.48484848485 266.890205185738
2001.60606060606 262.797644781943
2029.72727272727 258.815469088413
2057.84848484848 254.939863205055
2085.9696969697 251.167147039368
2114.09090909091 247.49377306074
2142.21212121212 243.916323575938
2170.33333333333 240.431507635704
2198.45454545455 237.036157664211
2226.57575757576 233.72722588774
2254.69696969697 230.501780625896
2282.81818181818 227.357002497702
2310.93939393939 224.29018058559
2339.06060606061 221.298708592483
2367.18181818182 218.380081020532
2395.30303030303 215.531889394528
2423.42424242424 212.751818548302
2451.54545454545 210.037642988538
2479.66666666667 207.387223347089
2507.78787878788 204.798502930188
2535.90909090909 202.269504370627
2564.0303030303 199.798326387074
2592.15151515152 197.383140653158
2620.27272727273 195.022188777626
2648.39393939394 192.713779395827
2676.51515151515 190.456285371936
2704.63636363636 188.248141110599
2732.75757575758 186.087839976171
2760.87878787879 183.973931817249
2789 181.905020593885
}\DataUSL

\begin{tikzpicture}

\definecolor{crimson}{RGB}{220,20,60}
\definecolor{darkgray176}{RGB}{176,176,176}
\definecolor{royalblue}{RGB}{65,105,225}

\begin{axis}[
width=\linewidth,
height=0.5\linewidth,
tick align=outside,
tick pos=left,
x grid style={darkgray176},
xlabel={In Rate (Mpps)},
xmin=0.0, xmax=35.0,
xtick style={color=black},
y grid style={darkgray176},
ylabel={Throughput (Mpps)},
ymin=0.0, ymax=10.0,
ytick style={color=black},
label style={font=\footnotesize},
tick label style={font=\scriptsize},
ylabel style={font=\fontsize{5}{5}\selectfont},
xlabel style={font=\fontsize{5}{5}\selectfont},
tick label style={font=\fontsize{6}{6}\selectfont}
]

\begin{pgfonlayer}{foreground}
\addplot [no markers, thick, crimson]
table [x expr=\thisrow{x}/100.0, y expr=\thisrow{y}/100.0] {\DataUSL};
\end{pgfonlayer}

\begin{pgfonlayer}{background}
\addplot [draw=royalblue, mark=o, only marks]
table [x expr=\thisrow{x}/100.0, y expr=\thisrow{y}/100.0] {\DataScatter};
\end{pgfonlayer}

\end{axis}

\end{tikzpicture}